\documentclass[prd,preprint,amsmath,nofootinbib,superscriptaddress]{revtex4}
\usepackage{graphicx}
\usepackage{bm}
\usepackage{epsfig}
\begin{document}
\title{TASI Lectures on Effective Field Theories} 
\author{Ira
Z. Rothstein\footnote{\uppercase{W}ork supported by \uppercase{DOE}
contracts \uppercase{DOE-ER}-40682-143 and
\uppercase{DEAC02-6CH03000}.}}
\address{Carnegie Mellon University \\
Dept. of Physics, \\ 
Pittsburgh PA  15213, USA}
\arabic{section}

\begin{abstract}{ These notes are a written version of a set of lectures given at TASI-02 on the
topic of effective field theories. They are meant as an introduction to some of the
latest techniques and applications in the field.}
\end{abstract}
\maketitle
\newpage
\tableofcontents
\newpage
\centerline{{\bf Prelude}}

It is a little known fact that the Pioneer spacecraft launched in the
mid-seventies contains a data repository with a summary of human
scientific knowledge. It is interesting to ask what would happen if an alien civilization found
this satellite and studied its contents? Let us furthermore suppose that while these aliens
were/are greatly advanced technologically, they are not terribly well
versed in basic research, mainly because of the alien government's 
short-sightedness \footnote {We know that without basic research
they probably would never have achieved space travel, but I'll invoke
suspension of disbelief at this point. } and  are thus  very interested
in the free knowledge that we are offering them. One of the pieces of
information in the data repository is a copy of the particle data
group booklet (the hand sized one, not the tome). The booklet is
handed over to the relevant scientists, and a massive project is
launched to decipher the information.  After a couple of years of
going over the data and a copy of Bjorken and Drell, which was also on
board, the aliens are up to speed and ready to test our
theories. They've come to the conclusion that we claim to have a well
defined theory of strong and electro-weak interactions up to the TeV
scale (let's not quibble over anachronisms), and that while we don't
know exactly what happens at the TeV scale there is something that
unitarizes the theory above that scale.  They also
conclude that our understanding of gravity is very limited. In
particular, while we know how to calculate quantum corrections at
energies which are small compared to the Planck scale, above the
Planck scale we have no idea what is going on.

After some  period of intense investigations, the aliens decide to test  the  theoretical predictions based on what they learned from us. A proposal is put forward by a prominent experimentalist for a high precision measurement of  the level splittings in Hydrogen. The proposal is summarily rejected based on a review written by a top
alien theorist. The rejection was based on the referee's argument that it is silly to try to calculate loop corrections in the human's field theory. The referee's justification for this statement is that given that gravitons run through the loops and the fact the integrals include momenta above the Planck scale, where we know the theory breaks down, there is no way that one can expect that the results will be correct. Moreover, even if gravitational couplings are suppressed by the Planck scale, there are power law divergences which can compensate for these suppressions, so it seems that any attempt at calculating loop corrections would be fruitless. 
Fortunately, for us,  there were some wise aliens working at the funding agency, and they said, ``let's for the moment ignore gravitational effects, and see if the data agrees with what we get
using just electro-magnetic effects". So the experiment was done and the theory did indeed meet with great success.  

So where did the referee go wrong? Well, to be fair, the referee was
not completely wrong.  There is a kernel of truth to its claims. It is
true that the theory breaks down and that we should indeed pause to
think about the results of any calculation which involves loop momenta
above the Planck scale (the scale where the theory is no longer
reliable).  Since we know we are not correctly portraying the physics at the
high scale,
whatever the result of the loop integration may
be, it will have to be treated in some way so as to remove the dependence on the UV physics. Indeed, the loop integrals will, in general, be
divergent and need to be regulated. But there is no preferred
regulation procedure \footnote {However, a prudent calculator would
choose to use a regulator which preserves the symmetries of the
action, though this may not always be possible.} and, therefore, the
numerical value of the integral is ambiguous. However, the point the
referee missed is that if we are not too ambitious, this is not a
problem. The short distance physics, which is not properly accounted for in the loop integrals, looks {\it
local} at low energies. Another way of saying this is, that if we
think of the loop integral in coordinate space, the part of the
integral from which we know we are getting nonsense can be contracted
to a point and therefore looks like an insertion of a local operator
into our Feynman diagram.  This means that all of the effects of the
unknown short distance physics can be mimicked by adding some new
operators into our Lagrangian (as we will see these operators may be
of arbitrarily high dimensions).  That is to say, all the effects
of the short distance physics can be accounted for by either shifting
the parameters in our renormalizable Lagrangian, or by adding higher
dimensional operators which are suppressed by the UV scale,
$M_{UV}$. Since we needed to perform measurements to fix those
parameters anyway we still have predictivity as long as we truncate
the Lagrangian at some order in powers of $1/M_{UV}$.  So we see now
in what way the referee was right. It's not that the unknown effects
of the short distance physics are parametrically small in any way,
indeed these effects are very important as they contribute at order
one to the parameters in the renormalizable part of the Lagrangian.
It's that all the effects go into making up the electron mass and all
the other low energy parameters of the theory \footnote{As we will
discuss at length in the text, the number of such low energy
parameters is dictated by the desired accuracy.}.  If you changed the
short distance physics it could, and would, change the values of the
low energy parameters. Thus, as long as we are not too ambitious, we
don't need to worry about the fact that our integrals diverge! Who
cares if they diverge, we didn't expect to get the answer right
anyway.
 So we should really call our field theories ``effective" field
theories. This moniker warns the consumer that it should be used with
caution, as at some scale the theory no longer makes any sense. What
that scale is depends upon the interactions one is trying to describe.

Let's for the moment consider a universe where the effects of short
distance physics could {\it not} be absorbed into low energy
parameters. We couldn't calculate {\it anything} without knowing the
complete theory of quantum gravity.
 While, this line of reasoning seems almost obvious now, thinking in
terms of effective field theories (EFT) only became generally accepted
in the last few decades. Indeed, from an EFT point of view,
renormalizability, which was a standard benchmark for acceptable
quantum field theories not that long ago\footnote{ As far as I know
the only text books prior to those by Peskin and Schroeder (95)
\cite{Peskin} and Weinberg\cite{Weinberg}(96), which espoused the
point of view described above was Georgi's book on weak
interactions\cite{Georgi}(84).}, is no longer relevant, unless we're
interested in theories which are supposed to correctly describe all the
short distance physics.  Does this mean that field theory as a
mathematical construct is incapable of being a ``complete" theory?
Certainly not. There are many field theories which are valid to
arbitrarily high scales. The classic examples being asymptotically
free theories such as QCD. These theories possess trivial UV fixed
points, which allow for the cut-off to be taken to infinity. In fact,
the necessary criteria for completeness is the existence of a UV fixed
point, and taking the cut-off to infinity is called taking the
``continuum limit", following the lattice terminology.  The reasoning
behind this criteria is that at the fixed point, the correlation
length $\xi$, (read inverse mass) diverges, and thus one can study physics at arbitrarily
large distances compared to the cut-off (the inverse of the lattice
spacing $a$), which is equivalent to saying that the cut-off can be
taken to infinity since lattice artifacts will be suppressed by powers
of $\xi / a $\cite{Wilson}.  Of course if we want a ``theory of
everything", it can't be asymptotically free since we know that
gravity should become stronger in the UV. Thus, any quantum field
theory description of gravity should possess a non-trivial UV fixed
point.  This possibility is very difficult, or perhaps impossible to
rule out.  Presently, the consensus is that it may well be that a local quantum field theory is not capable of describing physics at and
above the Planck scale. However, field theory is profoundly rich, and I
would guess it still possesses many tricks which have yet to be
revealed to us.
\section{Lecture I: The Big Picture}
This  lecture is meant to give a global overview of effective field theories.
The first part of the lecture in an introduction to the basic idea and
a tutorial on the matching procedure, all discussed within the context
of a simple toy scalar model. I then categorize the various types of
effective field theories into two main headings depending upon whether
the
theory admits a perturbative matching procedure. 
Most of the rest of the lecture is dedicated to some sample theories
where we can match.  I will assume throughout these lectures that the reader has
a basic understanding of renormalization.

\subsection{A Toy Model}

Let us now try to be more quantitative by starting with a simple toy
model. What we would like to illustrate is that all the effects of
ultra-violet (UV) physics may be absorbed into local counter-terms and
furthermore, that all of the  physical effects of the UV physics are
suppressed by powers of the UV scale\footnote{ 
This idea is usually referred to as ``decoupling'' as first
codified in \cite{decoup}.}. To do so we will work with a model of
two scalar fields , $\Phi_L$ and $\Phi_H$, with $m_H\gg m_L$, in four
dimensions.  The interaction Lagrangian will be given by
\begin{equation}
-{\it L}_{int}=\frac{\lambda_0}{4!} \Phi_L^4+\frac{\lambda_1}{2}\Phi_H \Phi_L^2+ \frac{\lambda_2}{4} \Phi_L^2 \Phi_H^2+\frac{\lambda_3}{4!}\Phi_H^4.
\end{equation}
It is typical, in most pedagogical treatises, to impose a discrete symmetry to forbid
tri-linear interactions, as such terms introduce dimensionful couplings, which complicate the systematics. However, such interactions will illuminate certain ideas which otherwise would not
arise until higher orders in the loop expansion.
The usual problems with tri-linear theories arise from issues of
vacuum stability, but for our purposes these issues will not be an obstruction.

I will assume that all the couplings, except for $m_L$, take on their ``natural values". Here I use the term ``natural"
in both the sense of Dirac and t'Hooft. That is, dimensionless
couplings should be of order one (Dirac) and  dimensionful couplings
should be of order of the largest scale in the problem, unless a
symmetry arises in the limit when the coupling vanishes
(t'Hooft). Thus, we take $\lambda_2\simeq 1$ and $\lambda_1\simeq
m_H$. Furthermore, since there is no symmetry protecting the light
scalar mass, choosing $m_L\ll m_H$ is unnatural, but this is just an
aesthetic issue with which we wont be bothered.

Now, according to the arguments in the previous section, it should be
possible to write down a local theory involving only the light degrees
of freedom which correctly reproduces the physics of the full theory
as long as we only consider external momenta much less than $m_H$.
The process of determining the low energy theory is called
``matching", and the strategy is quite simple. We build up our low
energy theory, which only involves the light fields, such that it
reproduces the full theory up to some orders in $(p,m_L)/m_H$ and
$\lambda_i$ where $p$ is the external momentum of the process under
consideration which we will take to be of order $m_L$.  Doing this is
quite simple in practice. The effective theory action will be given by
the full theory action with all heavy fields set to zero plus some new
terms which account for the effects of the heavy fields. The
additional terms are determined order by order in perturbation theory by calculating some amputated Greens
function in the full and effective theories and taking the difference
\begin{equation}
\delta L_{EFT}={\rm Full ~Theory~ Result}-{\rm Effective ~Theory~ Result.}
\end{equation}
In a sense what the left hand side of this equation is telling you is
exactly what the effective theory is missing.
Since the effective theory is designed to reproduce the IR physics of the full
theory, $ \delta L_{EFT}$ will have a well defined Taylor expansion around $p=0$.
Thus the holes in the effective theory can be plugged by adding local operators.
If this seems a little obscure at this point don't worry as
we will discuss many concrete examples of this matching procedure below.

Notice that matching each n-point function will generate distinct local operators in the effective theory, with higher n-point functions obviously generating higher dimensional operators. Thus the number of n-point function one needs to
calculate depends upon the desired accuracy. The more accuracy we want
the more operators we must keep, as each operator of successively higher dimension
will be suppressed by additional powers of the UV scale.
 
\begin{figure}[ht]
\centerline{\scalebox{0.6}{\includegraphics*[30, 280][530,590]{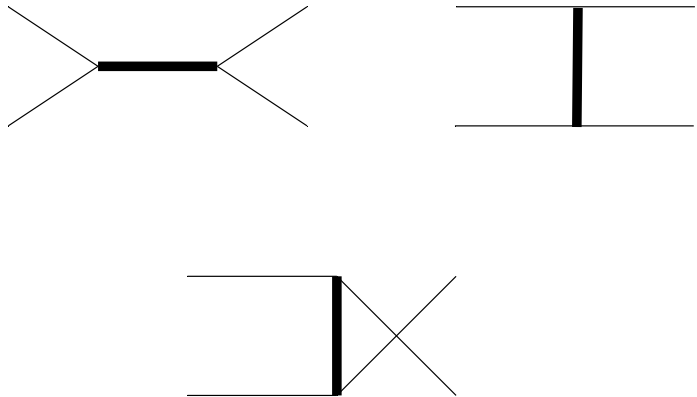}}}
\caption{Tree level contributions to the full theory four point
functions involving the heavy scalar (thick lines).\label{tree}}
\end{figure}

 Let's see how this works in our toy theory.
Consider the amputated four point function for the light fields.
At tree level, there are four diagrams which contribute to this process.
One is the usual contribution involving only the light fields gives $-i\lambda_0$, and the other three have intermediate heavy scalars as shown in figure (\ref{tree}). These latter contributions are given by
\begin{equation}
\label{fig1}
\textrm{Fig 1}=\frac{-i\lambda_1^2}{s-m_H^2}+s\rightarrow (t,u).
\end{equation}

After expanding (\ref{fig1}) we can see that the four point function can be reproduced, up to corrections of order $p^2/m_H^2$ by using
\begin{equation}
\label{loEFT}
\delta L_{int}=C_4^{(1,0)}\frac{\Phi_L^4}{4!}~~~~~
\end{equation}

Here I have established notation for the coefficients (called ``Wilson
coefficients"), the subscript counts the number of fields while the
superscripts denote the order in the expansion in coupling and
derivatives (more generally powers of $1/m_H)$, respectively.
We can now see clearly what was emphasized in the introduction. As far as a low energy observer is concerned this looks like a simple $\Phi^4$ theory with a coupling that we had to fix via experiment anyway. The low energy observer would be completely oblivious to the UV physics unless the experiment were sensitive enough to see the deviations from $\Phi^4$ theory due to $p/m_H$ corrections. If we wanted to, we could add additional operators to ensure that we correctly reproduce the higher n point functions as well. Suppose the low energy observer upgrades the machine, either by raising the energy or luminosity, to the point where  the data can no longer be fit by pure $\phi^4$ theory. We know exactly what to do to correctly reproduce the data, simply add the higher derivative interaction 
\begin{equation}
\label{deriv}
\delta L_{int}=\frac{C_4^{(1,2)}}{m_H^2}\frac{\Phi_L^2 \partial^2 \Phi_L^2}{4!}.
\end{equation}  
Other operators with two derivatives and four fields can all be put
into the form above via integration by parts and use of the equation
of motion. That this latter manipulation is allowed will be discussed
later.

As the energy of the experiment is further raised one needs to include higher and higher derivative interactions to reproduce the data. Notice that until we get to high enough energies to actually produce the heavy scalar we will be in the dark as to the true underlying theory.
This is  because in principle there will be an equivalence class
( called a "``universality class") of  theories which look the same in the infra-red. 
As we fit more and more of the couplings of the higher dimension operators we would be able to
eliminate theories, but removing all ambiguities would be difficult without including some other selection criteria.

So far, we've shown that to tree level we can reproduce the full theory by adjusting coefficients in an effective Lagrangian, this is called ``matching". 
This matching procedure can be carried out to arbitrary order in the couplings and $p/m_H$.
 Indeed, suppose we wish to match at higher orders in the couplings. The  matching procedure now takes on a new twist, but the idea is the same. We Taylor expand the full theory result  and compare it to the result in the effective theory. We then  modify the effective theory so as to get the answer right. I should mention at this point that in most theories the loop and coupling expansions coincide. But in our toy model, because of the $\lambda_1$ coupling this is no longer true.
Nonetheless we can overcome this issue by assuming that ($\lambda_1/m_H\equiv \hat{\lambda}_1$) is of order $\sqrt{\lambda_0}$. In this way all tree level contributions are of order $\lambda_0\simeq \lambda_2\simeq \lambda_3$.  This is why I labelled the Wilson coefficient above $C_4^{(1,2)}$.

\begin{figure}[ht]
\centerline{\scalebox{.6}{\includegraphics*[15, 273][540,590]{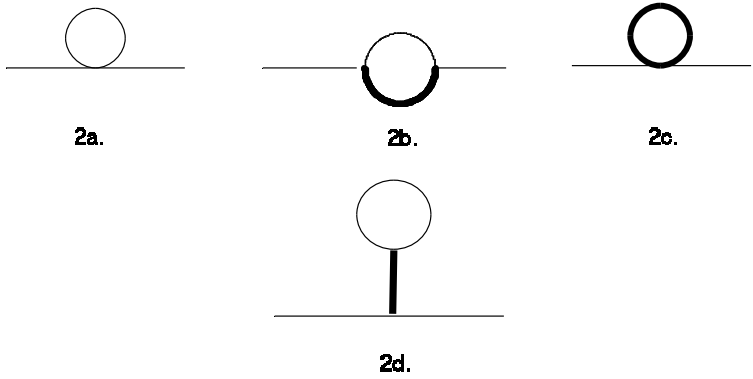}}}
\caption{One loop contributions to the full theory two point function. \label{2point}}
\end{figure}

Returning to our calculation, let us see if we can reproduce the full theory at next order in the couplings, beginning with the two point function. There are two  leading order (dimension four) operators which get
renormalized when calculating the two point function. The mass $m_L^2 \Phi_L^2$ and the kinetic energy
$\Phi_L \partial^2 \Phi_L$.
The one loop contributions to the two point function are shown in figure (\ref{2point}).
Diagram 2a. is the usual contribution found in $\Phi^4$ theory and is given by
\begin{equation}
\textrm{Fig 2a.}=\frac{i \lambda_0 m_L^2}{32 \pi^2}\left(\frac{1}{\overline{\epsilon}} -\ln(\frac{m_L^2}{\mu^2})+1 \right)
\end{equation} 
where I will use the notation
\begin{equation}
\frac{1}{\overline{\epsilon}}=\frac{1}{\epsilon}-\gamma+\ln(4\pi).
\end{equation}
The diagram in 2c. gives the same result as 2a., but with the replacement $(m_L,\lambda_0)\rightarrow (m_H,\lambda_2)$. Both of these contributions are momentum independent and thus only renormalize the mass.
Figure 2b, on the other hand, will have non-trivial momentum dependence.
We expect to have a cut beginning at $(m_H+m_L)^2$, and thus this
contribution will be analytic in the neighborhood $p^2 \simeq m_L^2$.
Doing the one-loop integration gives
\begin{equation}
\textrm{Fig 2b.}=\frac{i\lambda_1^2}{16
\pi^2}\left(\frac{1}{\overline{\epsilon}}-\ln(\frac{m_H^2}{
\mu^2})+1\right)
\end{equation}
where I have expanded in $1/m_H$, and kept only the leading term.
Finally, there is a contribution from the tadpole diagram 2d. which is given by
\begin{equation}
\textrm{Fig 2d.}=-\frac{i\lambda_1^2}{32 \pi^2}\frac{m_L^2}{m_H^2}\left(\frac{1}{\overline{\epsilon}}
-\ln(\frac{m_L^2}{ \mu^2})+1\right).
\end{equation}
Now let's calculate in the effective theory. We must note that to be consistent we must use the $O(\lambda)$ effective Lagrangian which we generated at tree level when we calculate on the effective theory side (the aforementioned twist). This is in contrast to the tree level matching where there is nothing to calculate on the effective theory side. At leading order in $1/m_H$, the only relevant operator is the $\Phi_L^4$ term and so we find that the effective theory  gives the same result as Fig 2a. except now the coupling is not $\lambda_0$ but $C_4^{(1,0)}$. 
To match we take the renormalized (here in $\overline{MS}$) full theory result expanded around small ($\hat{m}_L^2,\hat{p}^2=p^2/m_H^2$), and subtract the  renormalized effective theory
result, finding 
\begin{eqnarray}
C_m^{(1,0)}\!=\!\frac{1}{32 \pi^2}\left[\! \lambda_2 m_H^2\left(1- \ln(\frac{m_H^2}{ \mu^2})\right) -\lambda_1^2\left(2\ln(\frac{m_H^2}{\mu^2})\!-\!2\!     
 -2\frac{m_L^2}{m_H^2}(\ln\frac{{m_L^2}}{\mu^2}+1)\right)
 \!\right]. \nonumber \\
\end{eqnarray}
To insure that perturbation theory is well behaved, we should choose
$\mu$ to minimize the large logs. This scale is called the ``matching
scale".  Intuitively we would think that this scale should be the high
scale $m_H$. Given that the effective theory is supposed to reproduce
the infrared physics, we would expect that all the dependence on the
low energy parameter $m_L$ should be identical on both sides of the
matching equation, thus leaving all logs in the matching independent
of $m_L$.  However a glance at our results shows that choosing
$\mu=m_H$ does not get rid of all the large logs. Clearly we have done
something wrong.  The problem stems from the fact that we have kept some terms which
we had no business keeping. In particular, diagram 2d. is power suppressed relative (remember that
$\lambda_1$ scales as $m_H$)
to the other diagrams.  
 Thus,  to be consistent we need to expand
our full theory result for diagram 2b. to one more order.  So we make the replacement
\begin{equation}
\textrm{Fig 2b.}\rightarrow \textrm{(8)}+i\frac{\lambda_1^2}{16 \pi^2}(\frac{m_L^2}{m_H^2}\ln\frac{m_L^2}{m_H^2}).
\end{equation}
The log term will exactly kill off all of the large logs in $C_m^{(1,0)}$, leaving
\begin{equation} 
C_m^{(1,0)}=\frac{1}{16\pi^2} \lambda_1^2(1+\frac{m_L^2}{m_H^2})+\lambda_2 \frac{m_H^2}{32\pi^2}.
\end{equation}
Notice that this correction wants to pull up the light scalar mass to the scale $m_H$ since $\lambda_1\simeq m_H$.
 This is just what we called ``unnatural" in the sense of t'Hooft, since
as $m_L$ is taken to zero we do not enhance the symmetry. 
Keeping $m_L\ll m_H$ will require a fine tuning of the counter-terms at each order in
perturbation. In the $\overline{MS}$ scheme this wont show up until we try to relate the $\overline{MS}$ mass to the physical mass. That is, there will be a relation of the form
\begin{equation}
m_L^{2 ~(phys)}=m_L^{2~ \overline{MS}}+ O(\lambda/(32\pi^2))(m_H^2),
\end{equation}
and we will need to delicately choose $m_L^{2~ \overline{MS}}$ to
cancel the large $O(m_H^2)$ piece. 
 
 The momentum dependent terms will simply shift the field normalization in the effective theory with $C_p^{(1,0)}=\frac{\lambda_1^2}{32\pi^2 m_H^2}$. If we wish to go to higher orders in the $1/m_H$ expansion we will generate the operator
\begin{equation} 
O_2= C_2^{(1,2)}\Phi_L\frac{\nabla^4}{m_H^2}\Phi_L.
\end{equation}
\vskip.3in
{\bf Exercise 1.1}: Calculate $C_2^{(1,2)}$
\vskip.3in

\begin{figure}[ht]
\centerline{\scalebox{.6}{\includegraphics*[33, 120][590,650]{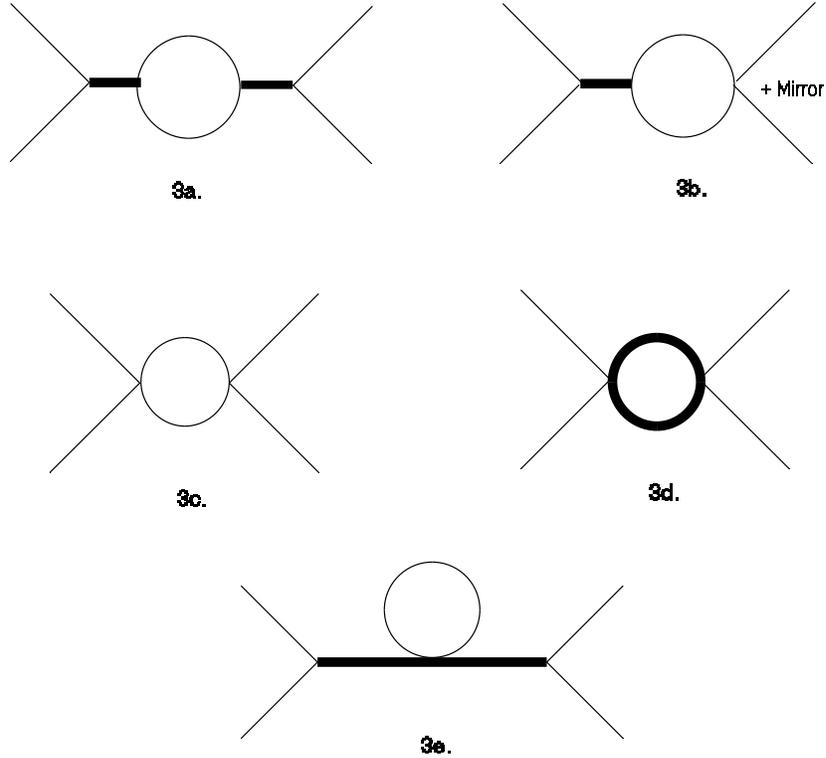}}}
\caption{A partial list of the one loop contributions to the full theory four point functions. \label{4point}}
\end{figure}

Now let us consider the four point function.  A partial list of the relevant full theory
diagrams are shown in figure 3.The diagrams with external leg
corrections will automatically be reproduced by the effective theory,
since we already matched the two point function.  In the effective theory, there will be only one
diagram, and it is identical to 3c. except that at each vertex we use the
coupling $C_4^{(1,0)}$. By considering the combination of couplings
coming from $(C_4^{(1,0)})^2$, we can begin to see that we will indeed
reproduce 3a, 3b and 3c.  However, we can also see that this diagram
will not be able to reproduce the log of $m_L$ from figure 3e. To
match this log we will have to first reproduce the six point function
at tree level. Then diagram 3e. will be reproduced by tying
together two external legs.  You can see how this will happen by
noting that heavy field propagators can be contracted to a point in
the full theory graph. Figure 3d will have no partner in the effective theory and will contribute ( in the s channel), after renormalization and expansion in $\hat{s}\equiv \frac{s}{m_H^2}$, 
\begin{equation}
\textrm{Fig 3d.}=-\frac{i\lambda_2^2}{16 \pi^2}\left(\ln(\frac{m_H^2}{ \mu^2})-\frac{\hat{s}}{6}+\cdot \cdot \cdot \right).
\end{equation}
We see that since we are matching at the scale $\mu=m_H$, this diagram will not contribute
to $C_4^{(2,0)}$, at least in the $\overline{MS}$ scheme. The other channels give similar
contributions.

\vskip.3in
{\bf{Exercise 1.2}} First draw all the remaining diagrams which contribute to the four point function in the full theory. Then show that all the dependence on $\ln(m_L)$ cancels in the matching for the four point function. Hint: By working at zero external momentum the calculation greatly simplifies.
\vskip.3in

What happens if we go to first order in $p^2/m_H^2$, staying at one loop? 
At this order  when we calculate in
the effective theory, we will need to include the graph depicted in figure 4, where the circle represents an insertion of the operator (\ref{deriv}). 
The integral for this Feynman diagram will have a contribution which is quadratically divergent 
\begin{equation}
I_{\Lambda^2}= \int \frac{d^4p}{(2\pi)^4} \frac{p^2}{(p^2-m_H^2+i\epsilon)((p_1+p_2+p)^2-m_H^2+i \epsilon)}.
\end{equation}
\begin{figure}[ht]
\centerline{\scalebox{.6}{\includegraphics*[16, 550][290,710]{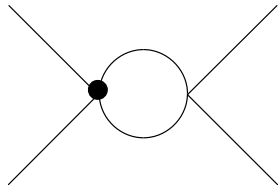}}}
\caption{One loop contributions to the effective theory four point functions, with the dark circle representing an insertion of the derivative operator (\ref{deriv}). \label{4pointeft}}
\end{figure}

 Now, suppose we regulate this divergence with a cut-off, then this
contribution will scale as $\Lambda^2/m_H^2$. What order is this in
our momentum expansion? If it's order one, it would seem that we have
lost all hope of calculating in a systematic fashion, since no matter
how many powers of $m_H$ may suppress a given operator, it may give an
order one contribution if we stick it into a multi-loop diagram. This
would be a disaster. However, this reasoning is misleading. The fact
that something can't be right can be immediately seen by working with
a different regulator. Suppose we work in dimensional
regularization. In this scheme, there are no power divergence as they
are automatically set to zero \cite{Sterman}. It had better not be
that our answer depends on our choice of regulator, otherwise we're
clearly lost. So what's going on? Which regulator is giving us the
correct result? The answer is, both. The difference between the two is
``pure counter-term", which means the discrepancy in the two results
can be accounted for by a shift in the couplings of local
operators. Since the couplings are \textit{measured}, and not
predicted, there is no physical distinction between the two
results. The only difference is in the bookkeeping, that is, the
counter-terms in the two schemes differ but that's it.  The nice thing
about dimensional regularization is that we don't have to worry about
mixing orders in $1/m_H$. We can see that if we use a cut-off and take
it to be of order $m_H$, then operators which are formally suppressed
will ``mix up" with operators of lower dimension\footnote{ I am using
term ``mix" in the technical sense of operator mixing.}.  But if we
stick to dimensional regularization, then powers of $1/m_H$ are
explicit, as they only appear in coefficients of operators (modulo the
caveat about $\lambda_1$ in our toy theory), so we can read off the
order of a loop diagram immediately. Thus, in general it is wiser to
always work in ``dim. reg." when doing effective field theory
calculations.  This is not to say that the cut-off doesn't have its
time and place. In particular the cut-off makes quadratic divergences
explicit, whereas in dim. reg. they show up as poles in lower
dimensions.

Another interesting point which arises from the calculation of the diagrams in figure 4 is the fact that there are divergences which can only be subtracted using counter-terms for operators with dimensions greater than four. The effective theory is non-renormalizable, an anathema in ancient times. That is, there are new UV divergences in the effective theory that arise in each order of perturbation theory. These divergences were not there in  the full theory.
This should not surprise us, since the effective theory is only built to mock up the IR of the full theory, not the UV.  Thus the UV divergences  will not cancel in the matching. This isn't a problem, we simply choose a scheme, and subtract. In the effective field theory approach we should be careful with out choice of scheme. If we're not careful we could mess up our power counting by introducing a scale. So we should choose a nice simple mass independent scheme like
$\overline{MS}$. I will have more to say about this later on in these lectures.
Notice that the matching coefficient will clearly be prescription dependent, but any such prescription dependence will
drop out in low energy predictions.
\vskip.2in

\textbf{Exercise 1.3} Calculate the anomalous dimensions of the
dimension six operator $\Phi_L^2 \partial^2 \Phi_L^2$ to one loop. The additional UV
divergences which arise in the effective theory correspond to IR
divergences in the full theory.
This is easily understood from the fact that we have taken $m_H$ to infinity in the
effective theory. So the ``IR'' divergence $\ln(m_H)$ now looks UV. 
 Thus effective field theory allows us to resum
IR logs in a simple way using standard renormalization group techniques. We will
see an example of this in the coming sections.
\vskip.3in
\subsection{Matching on or off the mass shell?}
Before moving on I'd like to discuss one more technical aspect of
matching. In particular, it's important to understand that it doesn't
matter whether or not one chooses to match with external states
on-shell or off-shell. The reason is that putting external lines
off-shell only changes the IR behavior of the theory, and since the
effective theory is designed to reproduce the non-analytic structure
of the full theory, all the information regarding the virtuality of
the external lines cancels in the matching. This can greatly reduce
the amount of work one has to do to match.  For instance, if one is
not interested in matching operators with derivatives, then one could
choose the external momenta to vanish identically, greatly simplifying
the necessary integrals. Also if one matches off-shell one can induce
operators in the effective theory which vanish by the leading order
equations of motion.  For example, in our toy model we could induce
$\Phi_L^3 \partial^2 \Phi_L$. However, it is quite simple to show that we
can eliminate this operator via a field redefinition, which are known
to leave S-matrix elements unchanged due to the equivalence
theorem\cite{equiv}. The proof \cite{GeorgiI} is inductive.  Suppose
we have set all operators of order $(p/m_H)^n$ which vanish by the
leading order equations of motion (by leading order I'm referring to
the momentum expansion) to zero. If there exists an operator at order
$(p^2/m_H^2)^{n+1}$ of the form $F(\Phi_L,\Phi_H)\partial^2 \Phi_L$, then we
make the field redefinition
\begin{equation}
\delta \Phi_L=-F(\Phi_L,\Phi_H)
\end{equation}
so that variation of the leading order Lagrangian
\begin{equation}
\delta L^{LO}=(\delta \Phi_L) \partial^2 \Phi_L
\end{equation}
exactly cancels the offending term. Notice that since $F$ is order $n+1$ in the derivative expansion the lower order operators remain unchanged but we do induce a change in the subleading set of operators, i.e. at order $n+2$. However, since in an effective
theory we generate all possible operators consistent with the symmetries, all
this redefinition will do is to shift the value of some matching coefficients.

\subsection{Non-Decoupling and Wess-Zumino Terms}

Let us consider the case of  a chiral gauge theory. These are theories 
for which a fermion mass term is disallowed by gauge invariance. We know
that if the theory is to be unitary and gauge invariant, the particle
content must be such that all the gauge anomalies arising from
triangle diagrams sum to zero. The classic example of such a theory is the standard
model itself, where  the fermions get their masses from the Higgs Yukawa
couplings. Taking the fermion masses large while holding the symmetry
breaking scale fixed corresponds to the limit of a large Yukawa coupling $\lambda$. Since
the coupling is proportional to the mass, we might expect that low
energy observables (below the scale of symmetry breaking) could be
enhanced by a large mass. The classic example of this
``non-decoupling'' arises  in $K-\bar{K}$ mixing. After integrating
out the top quark one generates a term in the $\Delta S=2$ Hamiltonian
\begin{equation}
H_t=m_t^2 \frac{G_F^2}{16 \pi^2}\eta_2 O_4,
\end{equation} 
where $O_4$ is a four quark operator. A similar effect shows up in the
$\rho$ parameter, which is a measure of  the ratio of charged to neutral
currents. These effects would disappear in the limit where the bottom
and top quark become degenerate, the former vanishes as a consequence
of the GIM mechanism.  Both  of these effects arise due to $SU(2)$ breaking.  If we took the limit where both
the bottom and the top quark masses become equal and large, then we
would expect that the effects of this heavy pair could all be absorbed
into $SU(2)$ symmetric operators.

\begin{figure}[ht]
\centerline{\scalebox{0.5}{\includegraphics*[116, 400][430,530]{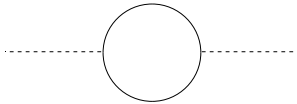}}}
\caption{A diagram which contributes to a dimension six operator which is suppressed by $1/v^2$. \label{dim6}}
\end{figure}

The thoughtful reader might be confused by this last statement if they are
familiar with anomalies.  Suppose that one of the $SU(2)_L$ doublets
is much heavier than the weak scale, and we wish to integrate it
out. The low energy theory would then have anomalous matter content
and would apparently no longer be gauge invariant.  Actually things
can look even worse. If we start with an even number of $SU(2)_L$
doublets and integrate out one such pair, then naively, the partition function
would vanish due to the non-perturbative anomaly \cite{Witten}.
\footnote{This anomaly arises as a consequence of the fact that $\pi_4(SU(2))=Z_2$
and a topologically non-trivial gauge transformation multiply the
partition function by a factor of -1 for each $SU(2)_L$ doublet.}. 
Now it should not come as no surprise that at ``low energies'' we do
not have manifest gauge invariance. Nor should we be worried about
this fact, since the lack of
gauge invariance will only manifest itself in a pernicious way (a
breakdown of unitarity) at energies near the masses of the gauge
bosons. 
Nonetheless we should be able to add a tower of higher
dimensional operators which will allow us to calculate sensibly at energies
above the masses of the gauge bosons.
The question is, what scale suppresses the effects of the fermions
in these higher dimensional operators? 
The answer is the symmetry breaking scale $v$, and not the mass of
the fermion.    Integrating out a heavy
fermion will generate a set of higher dimensional operators such as
those arising from the graph in figure (\ref{dim6}).
This graph will generate a whole string of operators once we expand it in momentum. Consider the contribution to the dimension 6 operators, which are suppressed by $1/M_{UV}^2$.  Just from counting vertices and dimensional analysis we see that it must scale as
$1/M_{UV}^2\propto\lambda^2/m^2=1/v^2$. Thus we can't write down a
local 
effective theory which
is valid in the region $v<E<\lambda v$, at least with these
variables. The trouble all starts from the fact that fermions are
getting their relatively large masses from dimensionless couplings. 
Even so, we should be able to calculate in a reliable fashion up
to the
scale $v$ even with our low energy theory which appears anomalous.

Imagine we're taking the limit $g\ll 1$ so that there is a large
hierarchy between $M_W$ and $v$, and we wish to calculate in this window.
  How can we recover a sensible theory
 without the heavy fermion pair? The answer is via a
``Wess-Zumino'' term \cite{WZ}.  In the present context the generation
of this term was worked out in a series of papers by D'Hoker and
Farhi, which I will only briefly summarize. The serious student may
wish to spend some time looking at the original work \cite{DF}.

If we allow for  $\lambda\gg 1$,  then the heavy fermions will be strongly coupled, but this should not stop
us from matching onto an effective theory. We know that we will not be
able to match perturbatively. However, the large $\lambda v$ limit
lends itself naturally to a derivative expansion which will keep our
matching under control.  We will simplify the problem by considering a
toy Yukawa theory with $SU(N)_L\times SU(N)_R$ symmetry. The field
content is a left (right) handed fermion in the fundamental and a
Higgs in the bi-fundamental, which transform as
\begin{equation}
\psi_L\rightarrow g_L\psi_L,~~~~~\psi_R\rightarrow g_R \psi_R,~~~~~\Phi\rightarrow g_L \Phi g^\dagger_R,
\end{equation}
$g_L$ and $g_R$ are independent elements of $SU(N)_L$ and $SU(N)_R$ respectively.
The addition of gauge fields complicates things only slightly.
The action for the  model is given by
\begin{equation}
\textsl{L}=i\overline{\psi}_L\partial \!\!\!\slash\psi_L+
\overline{\psi}_R i\partial \!\!\!\slash\psi_R-\lambda \overline{\psi}_L\Phi \psi_R-\lambda \overline{\psi}_R\Phi^\dagger \psi_L,
\end{equation}
where the Higgs doublet, $\Phi$ is written as $\Phi=v U$, with $U$
being a unitary matrix. That is we have frozen the modulus of the
Higgs field to simplify the problem. Allowing the modulus to fluctuate
wont change our conclusions.

Now we would like to match on to an effective theory without the heavy
fermions. The matching involves only calculations in the full theory
since the are no graphs in the effective theory. If we were working in
the standard model and wanted only to integrate out the top quark,
this would no longer be true.  Thus matching corresponds to formally
do the path integral over the fermions then expanding the effective
action
\begin{equation}
W(U)=-i \ln \int d[\psi_L] d[\psi_R]exp\left(i\int d^4x \textsl{L}\right)
\end{equation}
in powers of $1/(\lambda v)$, keeping only the terms which do not
vanish in the $\lambda\rightarrow \infty$ limit. Here I will only
sketch how this calculation is done.  The trick is to perform a change
of variables to absorb the Higgs field into a gauge field.  This is
accomplished by the redefinition
\begin{equation}
N_L=U^\dagger \psi_L,~~~ N_R=\psi_R,
\end{equation}
which results in 
\begin{equation}
\textsl{L}=\overline{N}_L(i\partial \!\!\!\slash-A\!\!\!\slash_L) N_L+\overline{N}_R i\partial \!\!\!\slash N_R-\lambda v(\overline{N}_L N_R+\overline{N}_R N_L),
\end{equation}
where $A_L=U^\dagger \partial^\mu U$. The advantage of using this form is that a derivative expansion corresponds to an expansion in the number of external fields.

We may write the new effective action defined in terms of an integral over the $N$ fields, as
$W^\prime$ with 
\begin{equation}
W=W^\prime-i J(U)
\end{equation}
since \begin{equation}
[d\psi_L]=J(U) [dN_L].
\end{equation}
Where $J(U)$ is the Jacobian for the change of variables (23).
$W^\prime$ can be calculated using the usual Feynman diagram techniques, and, in general, 
contains many interesting pieces, such as the famous ``Goldstone-Wilzcek'' current 
\cite{GW} which can endow solitons with fermion number. However,  in this model, all of these terms will be invariant under the anomalous
symmetry, so we will not concern ourselves with them at present.
We will be mostly concerned with the Wess-Zumino term which  resides in the Jacobian\cite{WZ}. This term can not be written as the four dimensional  integral of an $SU(2)\times SU(2)$ invariant density but  can be written as an invariant five dimensional integral \cite{Witten2}
\begin{eqnarray}
\Gamma_{WZ}=\frac{-i}{240 \pi^2}\int d^4x \int_0^1 dx^5 \epsilon^{abcde}
Tr\left[\tilde{U}^\dagger\partial_a \tilde{U} \tilde{U}^\dagger \partial_b
\tilde{U}\tilde{U}^\dagger\partial_c\tilde{U}^\dagger\tilde{U}\partial_d\tilde{U}\tilde{U}^\dagger\partial_e\tilde{U}\right].\nonumber \\
\end{eqnarray}
$\tilde{U}$ is an arbitrary (differentiable) extension of $U$ into the $x^5$ direction which need only to satisfy
\begin{equation}
\tilde{U}(x^5=0,x^\mu)=1 ~~;~ \tilde{U}(x^5=1,x^\mu)=U(x^\mu).
\end{equation} 
The reason that the interpolation is arbitrary is that the 
integrand is invariant under general coordinate transformations. Moreover, 
under a small variation of $\tilde{U}$, the integral is a total derivative
whose value  is fixed in terms of $U(x^\mu)$.
\vskip.2in
{\bf Exercise 1.4} Prove that under  small but arbitrary variations in $\tilde{U}$, the Wess-Zumino term changes by a total derivative, and thus depends only on the vale of $\tilde{U}$ in four dimensional space-time.
\vskip.2in

 If we were to include the gauge fields, then a variation of the WZ term would
exactly reproduce the usual $\epsilon^{\mu \nu \rho \sigma} F_{\mu \nu}F_{\rho \sigma}$ contribution to the anomaly.
It would also account for the non-perturbative anomaly \cite{Witten} in the case where the
gauge group is $SU(2)$. The result for the gauged theory can be found in \cite{DF}, and is slightly
more complicated then what I have described above. The non-local WZ term can be
expanded in powers of $v$ and the resulting theory will give well defined predictions in this
standard model-like theory up to the scale $v$.

\subsection{Types of EFTs}
I will separate  EFTs into two types. Those for which the underlying UV physics is known and the matching can be done perturbatively, which I call ``Type I", and those for which it is not possible to match, either because the UV physics is unknown, or because matching is non-perturbative,  which  I will call ``Type II". 
Some examples of type II theories  for which the underlying theory is unknown are
\begin{itemize}
\item The Standard Model
\item Einstein Gravity
\item Any higher dimensional gauge theory \footnote{There are some higher dimensional SUSY field theories which are conjectured to have UV fixed points\cite{Seiberg}. In these cases the theory is not effective, in that it is valid up to arbitrarily large momenta.}.
\end{itemize}
Whereas  for the following effective theories 
\begin{itemize}
\item Chiral perturbation theory: Describes low energy pion interactions.
\item Nucleon Effective theory: Describes the low energy interaction of nucleons \cite{nuc}.
\end{itemize}
the underlying theory  is just QCD, but the matching coefficients are not calculable, at least perturbatively.

Examples of Type I theories are
\begin{itemize}
\item Four Fermi Theory:
For low energy weak interactions. In this case the underlying theory is the standard model, which in itself is a type II theory.
\item Heavy Quark Effective Theory: Describes mesons with one heavy quark\cite{MW}.
\item Non-Relativistic Effective Theories  (NRQED, NRQCD): For non-relativistic bound states such as onia or atomic physics \cite{BL,lmr}.
\item Soft-Collinear Effective Theory (SCET): For large light-like
interactions such as those in heavy to light quark transitions
\cite{SCETI},  high energy jet processes\cite{SCETII} or quarkonia
decays \cite{SCETNRQCD}.
\end{itemize}

For type I theories, one may ask the question: ``Why bother with an effective theory when we know the complete theory?". The answer is that in general the full theory can be quite complicated and going to an effective theory simplifies matters greatly. In particular, going to an effective theory can manifest approximate symmetries that are hidden in the full theory and, as we know, increased symmetry means increased predictive power. Furthermore, when the full theory contains several disparate scales $(m_l,m_h)$ perturbation theory can be poorly behaved as typically one generates terms of the form $\alpha \ln(m_l/m_h)$. When these logs become large, they need to be resummed in a systematic fashion in order to keep perturbation theory under control. Working within an effective theory simplifies the
summation of these logs.
The classic example of such a resummation arises in $B$ meson decays, which I will now explore.

\subsection{EFT and Summing Logs}
 Let us consider the decay of a free $b$ quark. This calculation is discussed in the text by Peskin and Schroeder \cite{Peskin} using slightly different language, but I will sketch it here for the sake of completeness . I will also simplify it some by ignoring operator mixing.  At first sight there is no reason to believe that free quark decay has anything to do with the actual meson decay, but we'll come back to that point later.

The tree level decay rate is proportional to the usual muon decay calculation, while
 at one loop we have diagrams such as those in figure 6a. 
\begin{figure}[ht]
\centerline{\scalebox{0.8}{\includegraphics*[40, 375][480,590]{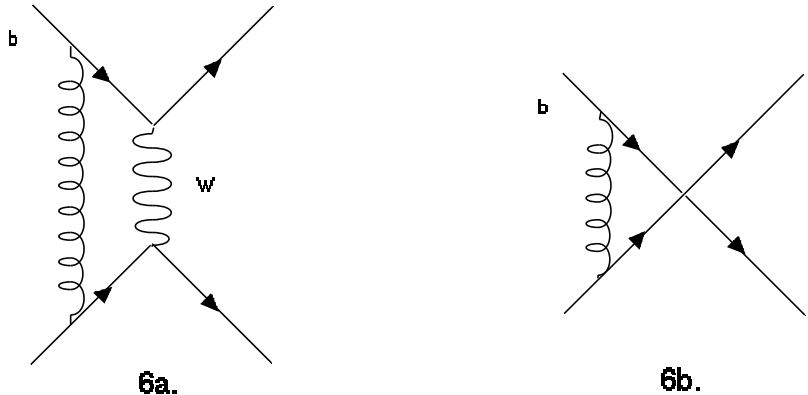}}}
\caption{One loop correction to heavy quark decay. \label{Hdecay}}
\end{figure}
This  correction is finite and contains a large log of the form
\begin{equation}
\label{1L}
\textrm{Diag. 6a}=\lambda_1 \frac{\alpha_s(\mu)}{4 \pi} \ln(m_W^2/m_b^2),
\end{equation}
The diagrams which do not traverse the $W$ propagator do not contribute large logs and will not be of interest to us. 
If we are to save perturbation theory we must resum these logs. Indeed, the perturbative series for a generic Greens function of a theory which contains a large disparity of scales whose ratio is $\delta$, will contain the series
\begin{equation}
\label{LL}
G_{log}=\sum_{n=1}^\infty (\frac{\alpha_s}{4\pi})^n \sum_{m=0}^n C_m^{(n)} \ln(\delta)^m.
\end{equation}
If we take $(\frac{\alpha_s}{4\pi} \ln(\delta))$ to be order one, then we must resum all terms of the form
$(\frac{\alpha_s}{4\pi}\ln(\delta))^n$. This is called the ``leading log (LL)" approximation. This resummation as well as the ``next to leading log (NLL) ($(\frac{\alpha_s}{4\pi}\ln(\delta))^{n-1}$)" and so forth are easily performed once we work within the framework of EFT. 

The way EFT makes the resummation of the logs  simple is by converting the problem to one with which we are more familiar. We know how to sum logs arising from UV divergences using the renormalization group (RG), so all we need to do is to figure out how to relate the logs in (\ref{LL}) to RG logs.
The logs in (\ref{LL}) can be thought of as IR logs in the $m_b$ goes to zero limit, or UV logs  if we take $M_W$ to infinity. But this second limit is exactly the limit we take in constructing
 the   EFT. In the four-Fermi theory $M_W$ is taken formally to infinity, usually keeping only the first non-trivial term in the $1/M_W$ expansion. Let us see how the four Fermi theory reproduces the log in (\ref{1L}).
The loop correction in the effective theory corresponding to those in figure 6a is shown in figure 6b, and is given by
\begin{equation}
\textrm{Figure 6b}=\lambda_1 \frac{\alpha_s(\mu)}{4
\pi}\left(\frac{1}{\bar{\epsilon}}- \ln(m_b^2/\mu^2)\right)+\kappa.
\end{equation}

Note that without doing any calculation we could have guessed the coefficient of the log in the effective theory from the full theory result. Since, as we discussed in the context of our toy model, the effective theory must reproduce the IR physics of the full theory, 
so the logs of $m_b$ must be identical in both theories.
Using this result we may match onto the effective theory. The Wilson coefficient of the four-Fermi operator is 
\begin{equation}
\label{4Fmatch}
C_{4F}(\mu)=G_F\left(1+\lambda_1 \frac{\alpha_s(\mu)}{4 \pi}\left( \ln(M_W^2/m_b^2)+\ln(m_b^2/\mu^2)+ \rho\right)\right)
\end{equation}
Notice that we work in a mass independent prescription (usually $\overline{MS})$ where we subtract only the pole and possibly some dimensionless constants. Thus, the constant in (\ref{4Fmatch}) is {\it scheme dependent}. However, this scheme dependence is order $O(\alpha)$, which is sub-leading
in the LL approximation. To cancel this $O(\alpha)$ scheme dependence we would need to go to the NLL, where the shift in the two loop anomalous dimension, generated by shifting schemes, will account for  the change in the matching coefficient. Details of higher order calculations can be found in \cite{buras}.
\vskip.3in
{\bf Exercise 1.5} Prove that the anomalous dimension is scheme independent only up to one loop. Also show that the beta function is
scheme independent up to two loops. Assume that the coupling in the new scheme is analytic
in the coupling in the old scheme, that is $g\prime(g)=a_1g+a_2g^2+.~.~.~$
\vskip.3in

 Now in order to match we must make sure that series for $C_{4F}$ is converging. Thus we should  choose $\mu$ to be order $M_W$, in which case $C_{4F}= G_F(1+\lambda_1 \frac{\alpha_s(\mu)}{4 \pi}\rho)$. In making this choice  we have removed the large log from the matching, but if we are not careful the logs can still reappear.
To see this we must consider the observable of interest. 
Using the optical theorem we may write the $B$ meson decay rate as \cite{MW}
\begin{equation}
\label{OPE}
\Gamma=Im \int d^4x \frac{i}{m_B}\mid C_{4F}(\mu)\mid^2 \langle B \mid T(O_{4F}^\dagger(x) O_{4F}(0) \mid B\rangle _\mu.
\end{equation}
Any $\mu$ dependence in the Wilson coefficient must be cancelled by the $\mu$ dependence of the matrix element.
If we choose $\mu$ to be order $M_W$, then the matrix element will
also be 
renormalized at this high scale. This would clearly be a blunder, since the physics responsible for the decay is all taking place at the scale $m_b$. This blunder will reprise the log, since it will now show up in the matrix element as $\ln(m_b/\mu)$. To truly vanquish the log we want to renormalize the four Fermi operator at the lower scale $m_b$.
But we know how to do this. We use  the fact that the bare  operator, $O_B= Z O_R$, where $Z$ includes both the operator and wave function renormalization, is $\mu$ independent
\begin{equation}
\label{RGE}
\left( \mu \frac{\partial}{\partial\mu}+\beta(g)\frac{\partial}{\partial g}+\gamma_{O_{4F}}\right)O_{4F}^R=0,
\end{equation}
then since $Z=Z_q^2/Z_{O_{4F}}$
\begin{equation}
\gamma_{O_{4F}}=\mu \frac{\partial}{\partial \mu}\left(-\delta_{O_{4F}}+ 4 \times \frac{1}{2} \delta_2\right)
\end{equation}
Where $\delta_{O_{4F}}=-\frac{1}{\epsilon} \lambda_1 \frac{\alpha_s}{4\pi}$ is the counter-term necessary to renormalize the four Fermi operator, and $\delta_2$ is the fermionic
wave function counter-term
\begin{equation}
\delta_2=\frac{4}{3}\frac{\alpha}{4\pi}\frac{1}{\epsilon}.
\end{equation}
The solution to the RG equation for $O_{4f}^R$ is 
\begin{equation}
O_{4f}^R(\mu)=O_{4f}^R(\mu_0)\exp\left(\int^{g(\mu)}_{g(\mu_0)} \frac{dg}{g} \frac{\gamma_{O_{4F}}(g)}{\beta(g)}\right),
\end{equation}
leaving
\begin{equation}
O_{4F}(m_W)=O_{4F}(m_b)\left[ \frac{\ln(m_W^2/\Lambda^2)}{\ln(m_b^2/\Lambda^2)}\right]^{(A_1/(2b_0))},
\end{equation}
where $b_0$ is the leading order beta function coefficient $11-2/3
n_f$, and $\gamma_{O_{4F}}=\frac{\alpha_s}{4\pi}
(A_1=2\lambda_1-\frac{16}{3})$.  We are now free to evaluate the time
order product, renormalized at the low scale $m_b$ without fear of
generating large logs.  Note that we can not claim to be accurate at order
$\alpha_s$. We can achieve this level of accuracy only if we run at two
loops. Since at order $\alpha_s$ we must keep all terms of the form
$\frac{\alpha_s}{4\pi}^n \ln(\delta)^{n-1}$. Thus keeping the constant
in the matching really bought us nothing. The systematics follow the
rule that to achieve accuracy at order $\alpha^n$, you must match at
order $\alpha^n$ and run at order $n+1$.

 We have
successfully removed that large and irrelevant scale $M_W$ from our
calculation. Doing so enabled us to easily sum the large logs. But we
are not quite done removing scales. In the decay process itself there
are two relevant scales, namely $m_Q$ and $\Lambda_{QCD}$, the scale
of hadronic physics. So we would expect that in our matrix elements we
could induce logs of the ratio of these scales. But this is not a
large log, so should we bother integrating out the scale $m_Q$? The
answer is, of course, yes. Because if you recall in the introduction
to this section we said that effective field theories do much more
than just sum logs. They make approximate symmetries manifest.

Just as the scale $M_W$ was irrelevant for the decay process, so is $m_Q$. Since $m_Q\gg \Lambda_{QCD}$, the quark effectively acts as a static color source. As far as the light degrees of freedom are concerned, the heavy quark is a brick wall. If you increased its mass, the effects on the bound state would be essentially nil. Thus  in the limit where the quark mass is taken to infinity,  a new symmetry flavor symmetry emerges. If we have more than one quark with $m_Q\gg\Lambda$, then to leading order in $\Lambda/m_Q$, the quark decay is flavor independent. 
Furthermore, the spin of the heavy quark should decouple since the color magnetic dipole moment scales inversely with the mass, so we would expect the emergence of a spin symmetry as well in the large mass limit.

\subsection{Heavy Quark Effective Theory}
In this section I will briefly introduce HQET. I wont go into detail as this is now textbook material \cite{MW}. But some of the concepts inherent in this theory will be necessary for the second lecture, so I will briefly recap the highlights and refer the reader to \cite{MW} for more details.
 
\begin{figure}[ht]
\centerline{\scalebox{0.8}{\includegraphics*[140, 400][410,550]{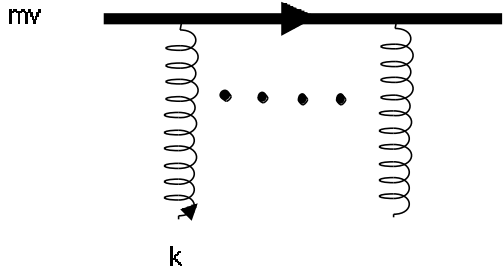}}}
\caption{Scattering of a heavy quark off of gluons with momenta of order $\Lambda$. \label{Hscat}}
\end{figure}

What we would like to do is eliminate the scale $m_Q$ from the theory. However, we obviously need to keep the heavy quark field in the theory, so how do we keep the field and lose the mass?
We begin by studying the dynamics of a heavy meson decay. As I said above the heavy quark should be static up to corrections of order $\Lambda/m_Q$.
We may write the heavy quark momentum as 
\begin{equation}
p_Q^\mu=m_q v^\mu+k_\mu~,
\end{equation}
where $v^\mu$ is the quarks' 4-velocity ($(1,\vec{0})$ in the rest frame) and $k^\mu$ is what's known as the ``residual momentum" which is of order $\Lambda$. By studying the propagation of a heavy quark in a hadronic environment, It is easy to illustrate that the heavy quarks only  role is to be a static color source 
Consider a heavy quark sitting inside a hadron being constantly bombarded by ``soft" gluons(i.e. gluons with momentum of order $\Lambda$), as shown in figure (\ref{Hscat}). In the large mass limit we may simplify the intermediate propagators as follows
\begin{equation}
\frac{\gamma_\mu(m_Q v\!\!\! \slash+k\!\!\!\slash+m_Q)\gamma_\nu}{(m_Qv+k)^2-m_Q^2} \approx
\frac{\gamma_\mu(1+v\!\!\!\slash)\gamma_\nu}{2v\cdot k}+~O(k/m_Q),
\end{equation}
so that the dependence on the heavy quark mass has disappeared. We expect the mass only to appear at sub-leading order in the mass expansion, and in phase space integrals.
Furthermore, we also have 
\begin{equation}
v\!\!\!\slash u(b)=u(b)+~O(k/m_Q)
\end{equation}
so we may make the replacement $\gamma_\mu \rightarrow v_\mu$ and the spin structure becomes trivial. In the real world where we have effectively two heavy quarks with $m_Q\gg\Lambda$, charm and bottom\footnote{ The top quark decays too quickly to be relevant to this discussion.}, 
we can see that in the large mass limit we should generate an $SU(2)_{flavor} \times SU(2)_{spin}$
symmetry, and this symmetry should be manifest in our effective theory.

We can determine the action for the effective field theory by matching. We consider the full theory propagator, again splitting the momentum into a large and small pieces, and  expand in the residual momentum
\begin{equation}
G^{(2)}(p)=\frac{i}{m_Q v\!\!\! \slash+k\!\!\!\slash-m_Q+i\epsilon}\approx \frac{i(1+v\!\!\!\slash)}{2 v\cdot k+i \epsilon}.
\end{equation}
Now we need to find an  action which has the proper quadratic piece to reproduce this two point function.
We do this in two steps: First we notice that the numerator of the effective propagator is a projector onto the large components of the heavy quark spinor. So we should split up the heavy quark spinor into ``large" and ``small" components. Then, to remove the heavy quark mass yet retain the field itself, we simply extract the mass dependence via a field redefinition \cite{GeorgiII},
\begin{equation}
q(x)=e^{-im v\cdot x}\left(h_v(x)+H_v(x)\right),
\end{equation}
where
\begin{eqnarray}
h_v(x)&=& e^{imv\cdot x} \frac{1}{2}(1+v\!\!\!\slash )q(x) \nonumber \\
H_v(x)&=& e^{imv\cdot x} \frac{1}{2}(1-v\!\!\!\slash )q(x) .
\end{eqnarray} 
Here is where HQET differs from the usual effective theories where we
remove entire fields from the action. The fields have {\it labels}
$v$. A field with a label $v$, destroys a particle\footnote{If
we were interested in anti-particles we would chose an opposite sign
in the re-phasing.} whose four momentum is fixed to be within
$\Lambda$ of $m v_\mu$. That is, the field itself only has Fourier
components less than $\Lambda$. This is in accordance with our
physical expectations.  Once we integrated out the hard modes, the
only modes left have momenta of order $\Lambda$ whose effect is only
to jiggle the heavy quark. So we write the momentum of the heavy quark
as $p=m_Q v+k$, where $k$ is called the ``residual momentum" and is of
order $\Lambda$. Thus, derivatives acting on this rescaled field are
of order $\Lambda$.

We may now substitute this parameterization of the full QCD field into
the Lagrangian to arrive at
\begin{equation}
{\it L }_{QCD}=\bar{q}(iD\!\!\!\slash -m_Q)q~=~\sum_v
\bar{h}_v(iv\cdot D)h_v~+~ O\left(\frac{\Lambda}{m_Q}\right).
\end{equation}
Notice the sum over velocities. It is needed because we should allow
for arbitrary four velocities of the heavy quark, and remember that
once we fix a label the momentum is fixed up to some small
an amount of $O(\Lambda)$. Technically, this means that there is a super-selection
rule, which is defined by the statement that states with
four-velocities which differ by more than $\Lambda$ exist in different
Hilbert spaces. So to allow for differing velocities we have to
``integrate in" the different sectors. If we are only interested in
processes with no weak decays this point becomes moot since we can go
to a frame where the heavy quark is static. However,
if we want to look at for instance a $b$ to $c$ transition then a
large "external" momentum is injected which drastically changes the
velocity of the heavy quark. In the effective theory this is described
by the flavor changing current, whose labels reflect the connection
between different Hilbert spaces
\begin{equation}
J(x)=\sum_{v,v^\prime}~\bar{h}^c_v~ \Gamma~ h_{v^\prime}^b(x)
\end{equation}
These type of label changing operators will play an important role in the next lecture.

 We performed a Taylor
expansion of our tree level propagator, should we expect this
replacement to work even in loop diagrams? The answer is of course,
yes, and the reason is (not to flog a dead horse) that the part of
the loop integral for which the Taylor expansion is ill suited comes
from UV physics, which is local and can always be absorbed into
counter terms.  Technically, this can be see in a simple fashion. Take
any full theory integral, $I$ and write it as
\begin{equation}
I=(I-I_{EFT})+I_{EFT}
\end{equation}
If $I$ is finite, then Taylor expanding and integrating commute and
there is no work to be done, the EFT will reproduce the full theory to
any desired order in the inverse mass.  If the integral diverges, then
all we need to do is differentiate it enough times with respect to the
external momentum to make it finite and then perform the Taylor
expansion. The value of the full integral will  differ from the
expanded integral by some polynomial in the external momentum, i.e. a
counter-term. Using this type of argument inductively an all orders
proof has been provided both for four Fermi theory \cite{WittenEFT}
and HQET \cite{Grinstein}.
HQET is particularly simple because there is only one relevant IR region, $k \sim \Lambda_{QCD}$.
So we expect that Taylor expanding around small loop momenta will correctly reproduce
the IR physics. Other theories may have more relevant IR momentum regions as will be discussed later on in these lectures.
\subsection{The Method of Regions}
\label{regions}
This line of reasoning can greatly simplify matching calculations.
Suppose we wish to match onto to HQET at one loop. The standard recipe
is, take the full theory result, expanded in powers of $k/m_Q$, and
subtract the value of the corresponding effective theory result
\footnote{In addition one needs to worry about how the states are
normalized in the two theories.}. But there is actually a way to make
life a lot less complicated by utilizing some of the magic of
dimensional regularization and asymptotic expansions\cite{AE}. Instead
of first calculating in the full theory and then expanding, one can
Taylor expand at the level of the integrand, assuming that the loop
integrals are dominated by momenta much larger than the external
momenta. By working in this way we will miss out on non-analytic
pieces, but that's OK since we know that they cancel in the matching.
How do we know that this method correctly reproduces the matching
coefficient?  One way to think about it is in terms of the ``method of
regions"\cite{MoR}. The idea is that in dimensional regularization the
integral will receive contributions from momenta of order the heavy
quark mass (``hard modes") as well as momenta of order the residual
momenta (``soft modes"). This is true even for divergent integrals
since the contribution from modes much larger than the quark mass lead
to scaleless integrals which vanish. Let us see how this works in a
simple one loop example.  Consider the two point function integral
\begin{equation}
I=\int
\frac{d^nq}{(2\pi)^n}\frac{1}{(q^2+i\epsilon)}
\frac{1}{((p_Q+k-q)^2-m_Q^2+i\epsilon)},
\end{equation}
$k$ is the residual momentum of order $\Lambda_{QCD}$.  To extract the
hard contribution we assume $q$ is of order $m_Q$ and Taylor expand
the integrand up to order $1/m_Q$
\begin{equation}
I_h=\int
\frac{d^nq}{(2\pi)^n}\frac{1}{(q^2+i\epsilon)}\frac{1}{(q^2-2p\cdot
q+i\epsilon)}(1-\frac{2(p-q)\cdot k)}{q^2-2p\cdot q+i\epsilon}).
\end{equation}
Whereas the soft part of the integral is found by assuming the loop momentum is order $k$. Keeping only the leading order piece in this region we find
\begin{equation}
I_s=\int \frac{d^nq}{(2\pi)^n}\frac{1}{(q^2+i\epsilon)}\frac{1}{2p\cdot(k-q)+i\epsilon},
\end{equation}
which is exactly the HQET contribution. So if we can show that the sum of
these two contributions is equal to $I$, then we've shown that the
hard piece is exactly what we would call the matching.  The full
integral can be performed using text-book methods
\begin{equation}
\label{full}
I=\frac{i}{16\pi^2}\left(\frac{1}{\bar{\epsilon}}-\ln(\frac{m_q^2}{\mu^2})+2+\left( \frac{1}{\hat{p}^2}-1\right)\ln(1-\hat{p}^2)\right),
\end{equation}
where $\hat{p}^2=1+\frac{2p_Q\cdot k}{m_Q^2}$.
The hard piece is given by
\begin{equation}
\label{hard}
I_h=\frac{i}{16\pi^2}\left(\frac{1}{\bar{\epsilon}}+2-\ln(\frac{m_Q^2}{
\mu^2})+\frac{k\cdot
p_q}{m_Q^2}\left(-\frac{1}{\bar{\epsilon}}+\ln\frac{m_Q^2}{\mu^2}
-2\right)\right)
\end{equation}
The soft integral has denominators of differing dimensions, so it's
helpful to use the identity
\begin{equation}
\frac{1}{a^n b^m}=2^n
\frac{\Gamma[n+m]}{\Gamma[n]\Gamma[m]}\int_0^\infty d\lambda
\frac{\lambda^{m-1}}{(a+2 b \lambda)^{n+m}}.
\end{equation}
Then following standard techniques we arrive at
\begin{equation}
\label{soft}
I_s=\frac{i}{16 \pi^2}\frac{k\cdot p_q}{m_Q^2}\left(\frac{1}{\bar{\epsilon}}+2-2\ln(\frac{-2v\cdot k}{\mu})\right).
\end{equation}
Expanding the full theory result (\ref{full}) and comparing it to the
sum of (\ref{soft}) and (\ref{hard}), we find agreement between the
two. Notice that both $I_h$ and $I_s$ have spurious divergences which
cancel each other.  For more complicated theories, such as $NRQCD$,
there are more regions to worry about, and one must be sure that they are
all accounted for in order to correctly reproduce the full theory
result. Furthermore, one must be careful when matching at two loops
using the above trick. Doing so would necessitate keeping $\epsilon$
dependence in the effective theory action\cite{BSS,MSI}.

\vskip.2in {\bf Exercise 1.6} In the last example I have expanded in
$\epsilon$ before expanding in $1/m_q$.  Show that the equality holds
true even if you don't expand in $\epsilon$ by calculating the full
result exactly and then expanding in $1/m_Q$.
\vskip.2in

The method of ``Asymptotic Expansions" has been proven to work to all
orders for both the large momentum and large mass limits for Euclidean
momenta.  However, there are no known counter-examples to two loops
in Minkowski space.

\subsection{A Caveat About Scaleless Integrals}
Suppose we chose to match exactly at zero external momentum. We are free to do
this if we are not interested in corrections which are sub-leading $p$. 
It is often the case that at this point the integrals in the effective theory will
be scaleless and thus vanish (this is true in HQET as well as NRQCD).
 However, this does not mean that we can conclude that the
effective theory operator under consideration has no anomalous dimension.
To calculate the running in the effective theory we need to
account for the fact that the scaleless integral $I_{EFT}$ is giving zero
because of a cancellation
\begin{equation}
I_{EFT}=C~(\frac{1}{\epsilon_{UV}}-\frac{1}{\epsilon_{IR}}).
\end{equation}
Given any logarithmically divergent scaleless integral 
we can always perform this split algebraically
\begin{eqnarray}
\int {d^dk \over{(2\pi)^d}} \frac{1}{k^4}&=&\int \frac{d^dk}{(2\pi)^2}\left( \frac{1}{(k^2)(k^2-m^2)}-
\frac{m^2}{(k^4) (k^2-m^2)}\right)\nonumber \\ &=&\frac{i}{16\pi^2}\left(\frac{1}{\epsilon_{UV}}-\frac{1}{\epsilon_{IR}}\right).
\end{eqnarray}
However, if and only if, we know that the full theory has no IR
poles, can we  correctly conclude that the anomalous dimension in the
effective theory vanishes when we have scaleless integrals.
When the full theory has no IR poles, the effective theory has no IR poles and since
in the effective theory (for scaleless integrals) there is also a one to one correspondence
of UV and IR poles we may conclude that the effective theory has no UV poles.
This is an important point: We
can't ignore dimensionless integrals in dimensional regularization in the effective theory unless we
are sure that we have correctly accounted for all of the IR poles in the full theory.

\subsection{Power Counting}
\label{HQETPC}
In most effective field theories  the power counting is as simple as
keeping track of powers of the mass. However, this will no longer be
true in more complicated EFTs. Power counting is trivial in HQET, but I want to go
through it pedantically, since such an analysis will bear fruit later.
Once we've removed the heavy quark mass, there is only one
relevant scale, $\Lambda$. All of the fields have support only over
regions of order $1/\Lambda$ so  we should expect that dimensional
analysis will yield the correct power counting. Let's see if this is true. The leading order
action,
\begin{equation}
S^{HQET}=\int d^4x ~\sum_v \bar{h}_v(iv\cdot D)h_v,
\end{equation}
 should scale as $O(1)$. 
Each field scales as $\Lambda^{3/2}$ while the derivative scales as
$\Lambda$. Finally, since the fields only have support over region of
size $\Lambda$, $d^4x\propto \Lambda^{-4}$,  the leading order
action does indeed scale properly. In general, there may be more than
one scale involved and the scaling of the fields is instead fixed
according to the leading order action.  All other operators in the
action should scale as positive powers of $\Lambda$, since if there
were operators with negative powers, power counting will be jeopardized.

\subsection{The End of Our Calculation}
We now have all the ingredients to finish our inclusive B decay
calculation. We started with a theory with three scales, $M_W$, $m_b$ and
$\Lambda$. We eliminated $M_W$, ran down to $m_b$  and are now have the
power, via $HQET$, to remove $m_b$. Removing this scale is crucial, even if we are not interested in the enhanced symmetry. To see this, we have to figure out
how to evaluate the time ordered product (\ref{OPE}). I will only sketch this final part of the
calculation since it's is explained in details in \cite{MW}.
The main point is that the  non-local operator product can be simplified using an operator product expansion 
(OPE) much as in  deep inelastic scattering (DIS) \cite{DIS} (though the use of the OPE in the present context is not rigorously justified as it is in DIS). The
TOP can be expanded into a set of local operators, schematically 
\begin{equation} 
\int d^4x e^{-iq\cdot x} O(x) O(0)\simeq C_1(q^2,\mu)+C_2(q^2,\mu) \frac{O_2(\mu)}{q^2}+C_3(q^2,\mu)\frac{O_3(\mu)}{q^4}+ . ~. ~.
\end{equation}
where $q$ here is the large scale in the process (for us  $q\sim m_b$).
The utility of the expansion is predicated on our ability to truncate
it\footnote{In DIS the expansion is in ``twist=Dimensions-Spin'',
whereas in B decays its purely dimensional, unless one probes certain
end-point spectra.}.  As long as when we take the matrix elements of
the $RHS$, we don't introduce a scale of order $q$ upstairs, we can
safely ignore terms of higher order in $1/q^2$. Now we can see that in
the case of $B$ decays we will run into exactly this problem since we
will be taking matrix elements between $B$ states. So that while each
term in the expansion may be down by powers of $q$, i.e. $m_b$, taking
the matrix element will generate expectation values which scale with
positive powers of $m_b$. The solution to this problem is simply to go
to the effective theory, where the scale $m_b$ is gone and all matrix
elements scale only as $\Lambda$. How the OPE is performed in HQET is
discussed in \cite{MW}.
\subsection{EFT and $\overline{MS}$}
I want to emphasize the fact that using a mass independent prescription is crucial for this, as well as most other, EFT calculations. In fact, EFT and mass independent prescriptions go usually go hand in hand (lattice effective theories are exceptions to this rule). 
To see the need for this marriage,  suppose we're trying to calculate the running of a gauge coupling. Our prescription can not affect any physical prediction, but a poor choice of prescription can lead to an ill behaved perturbative series.
Let's compare and contrast the contributions  of a massive fermion to the QED vacuum polarization in  the $\overline{MS}$ and momentum subtraction $(ms)$ schemes. 
\begin{equation}
\Pi_{\mu \nu}^{\overline{MS}}=\frac{-ie^2}{2\pi^2}(p_\mu p_\nu -p^2
g_{\mu \nu} )\int_0^1 dx ~x(1-x)\ln\left[\frac{m^2-p^2
x(1-x)}{\mu^2}\right],
\end{equation}
\begin{equation}
\Pi_{\mu \nu}^{{ms}}=\frac{-ie^2}{2\pi^2}(p_\mu p_\nu -p^2 g_{\mu \nu} )\int_0^1 dx ~x(1-x)\ln\left[\frac{m^2-p^2 x(1-x)}{m^2+M^2 x(1-x)}\right],
\end{equation}
where $M$ is the scale at which we have chosen to define the coupling, so that in the $ms$ scheme
$\Pi(-M^2)=0$.
For $p^2\ll m^2$ we may choose $\mu, M\simeq m$ and neither scheme will produce large logs.
The problem arises when we have two fermions with very disparate masses. Indeed, suppose that in addition to the fermion of mass $m$, we also a massless charged fermion in the theory.
Then we can see that in the $\overline{MS}$ scheme, there is no choice of $\mu$ for which there are no large logs. On the other hand in the $(ms)$ scheme we can choose $M^2\simeq -p^2$ and no large logs will remain. The reason for this difference is that the $ms$ scheme is a ``decoupling scheme" while $\overline{MS}$ is not. That is, in the $\overline{MS}$ massive particles do not decouple in the low energy limit. That is exactly why we should work in an EFT, where we remove the heavy particles by hand. This is the small price we have to pay in order the easily sum the logs.
Indeed, in the $ms$ scheme we will need to calculate  $\alpha_s(-M^2)$, given its value at some other momenta at which it was measured $M_0$, and solving the RG equations in this scheme is more difficult. I should also say that even if you work in a mass independent scheme you don't {\it have to} use an EFT,  it's just that without EFT calculations become more cumbersome since the low energy observables will receive contributions from all the heavy fields which  re-introduce large logs. But in the end if you are careful you should always get the same result. Using an EFT will just make your life {\it a lot} easier.
 

\subsection{The Force Law in Gauge Theories}

When working in an effective theory, or any theory for that matter, it is easy to get confused if one does not work with physical quantities. Coupling are in general scheme
dependent and have no physical meaning, thus it is always best to have in mind some
measurable observable when doing a calculation. Often in these lectures instead of talking about
measuring a coupling, I will instead talk about measuring the force between static
sources. In this way we can be assured that our result is in no way ambiguous.
 This force or potential is intimately related to what we normally call the
coupling, but is a measurable quantity and thus scheme independent. 

We begin by showing that the static potential can be extracted by calculating the expectation value
 \begin{equation}\label{Z}
Z[J]=\langle 0 \mid T(e^{-i e \int d^4x e j^\mu(x) A_\mu(x)}\mid 0 \rangle
 \end{equation}
 for a stationary charge distribution which turns off adiabatically at $\pm \infty$.
then using standard arguments (see \cite{Peskin})
\begin{equation}
\lim_{T\rightarrow \infty} Z[J]\propto e^{i E(J) T},
\end{equation}  
where $E(J)$ is the vacuum energy in the presence of the source $J$.
It is left to the reader to show that $Z[J]$ is nothing but the expectation value of
the gauge invariant time-like Wilson loop
\begin{equation}
W(r,T)\equiv \langle 0 \mid Tr\left(P(e^{\oint A_\mu dx^\mu})\right)\mid 0 \rangle
\end{equation}
where $P$ stands for path ordering.

For a free Abelian theory, i.e. no matter, things simplify tremendously since the action is
quadratic in the gauge field, and hence the functional integral 
can be done exactly, resulting in 
\begin{equation}
Z[J] \propto e^{-\frac{ie^2}{2} \int d^4x d^4y J^\mu(x) \Delta_{\mu \nu}(x,y) J^\nu(y)}
\end{equation}
where $\Delta_{\mu \nu}$ is the  photon propagator.
We will be interested in the energy of two static sources, of opposite charge, separated by a spatial
distance $r$. So that 
\begin{equation}J_\mu(x)=e \delta_{\mu 0} \delta^3(\vec{x}-\vec{r})-e \delta_{\mu 0} \delta^3(\vec{x})
\end{equation}
then it is straightforward to show that 
\begin{equation}
E(r)=  \frac{-e^2}{4 \pi r}.
\end{equation}
Diagrammatically, one can arrive at result by using Wicks' theorem on (\ref{Z}).
At leading order in the coupling there is one diagram corresponding to one photon
exchange, resulting in the integral
\begin{equation}
I_2=\frac{1}{4\pi^2}\int \frac{dt_1 dt_2}{(t_1-t_2)^2-r^2} 
\end{equation}
At next order there are two Wick contractions, but these sum to give the
square of $I_2$ with a combinatoric factor of $1/2$. One can then see that
subsequent Wick contractions will just fill out the power series expansion of the
exponential\cite{Fischler}.

Now when we include matter, the only new diagrams which arise correspond to corrections
to the photon propagator. Thus all the logarithmic corrections to Coulomb's law will
arise from loops in the photonic two point function. Resumming the vacuum polarization
graphs results in the corrected form of the potential
\begin{equation}
E(r)= \frac{-e^2_{eff}(r)}{4 \pi r}
\end{equation}
where 
\begin{equation}
e^2(r)_{eff}=\frac{e^2(\mu)}{(1+\frac{e^2(\mu)}{6 \pi^2} \ln(r/\mu))}.
\end{equation}

When we go to QCD, things get slightly more complicated but, at least up to two loops,
the same reasoning goes through \cite{Fischler,Kogut}. The potential is given by the
logarithmically corrected Coulomb potential with the appropriate beta function.
However, at three loops one encounters an infra-red divergence\cite{ADM}
arising from the fact that there can be intermediate color octet states that
live for long times. These colored states can interact with long
wavelength gluons since they have non-vanishing dipole moments. 
However, this should  not bother us. The vacuum matrix element 
contains long distance physics, and thus it need not be infra-red finite in perturbation theory.

\section{Lecture II: Non-Relativistic Effective Theories.} 
\subsection{The Relevant Scales and the Free Action}
In the previous lecture we displayed the utility of encapsulating the
UV physics in an effective field theory. In our toy model we
completely removed a heavy field, while in HQET we kept the field but
removed the mass. In both of these cases the non-analytic momentum
dependence came from rather simple kinematic configurations. In the
toy model all the cuts were generated from intermediate light particles
going on shell. In HQET the cuts come from either
gluons or heavy quark lines going on shell with $k_\mu\simeq
\Lambda$.  But there are certain kinematic situations where the
momenta configurations which determine the cuts of a Greens function
can be more complicated. Consider for instance a non-relativistic
bound state such as Hydrogen or quarkonia\footnote{For the quarkonia
case we will assume that the bound state is Coulombic such that
$mv^2>>\Lambda$.This hierarchy most probably does not hold for the
case of the $J/\psi$ and may not for the $\Upsilon$ either. Attempts
at alternate power countings which arise when this criteria are not
met can be found in \cite{NRQCDc}.}. In this case the on-shell external
lines in our Greens functions will obey a non-relativistic
dispersion relation, leading to the propagator
\begin{equation}
G^{(2)}(E,\vec{p})=\frac{i}{E-\frac{\vec{p}^2}{2m}+i\epsilon},
\end{equation}
and the cuts will come from a momenta region where $E\simeq
\frac{\vec{p}^2}{2m}$.  Notice that, as opposed to the case of our toy
model and HQET where there was only one relevant low energy scale,
$m_L$ and $\Lambda$, respectively, a non-relativistic theory
contains two scales namely $E\simeq mv^2$ and $p\simeq mv$. That is,
the typical size of the bound state is the Bohr radius $r_B\simeq
1/(mv)$ and the typical time scale is the Rydberg $R\simeq 1/mv^2$.
The existence of these two scales complicates matters.  As in our
previous cases we'd like to eliminate all the large scales from the
theory, leaving a nice simple theory with only one low energy scale.
Unfortunately, as we shall see this is not possible due to the fact that the
two scales $mv$ and $mv^2$ are correlated, and must exist simultaneously.

Suppose we try to treat the  scale $mv$ in the same way that we treated the
mass in HQET, i.e. by rescaling the quark fields. This would indeed 
remove it from the theory, and all derivatives would scale homogeneously  as $mv^2$. 
However, since spatial fluctuations on the scale $mv$ are crucial to the dynamics of the bound state, we know that this can't be the end of the story as far as this scale is concerned. In fact,
as we shall see, the Coulomb potential is built from the exchange of
gluons with three momentum of order $mv$. So if we eliminate the modes with momenta of order
$mv$ by re-phasing the field we had better figure out how to put (integrate)
them 
back in. At first, this seems
like a silly thing to do, but we will see that this formal
manipulation will simplify  the construction of the effective theory since
all spatial derivatives will scale homogeneously. If we did not do this re-phasing we would not have  manifest power counting, as we shall see.

But we are jumping the gun a little. Let us treat one scale at a time from top
down.
We begin by first eliminating the mass from theory as we did in HQET
by a simple rescaling of the field
\begin{equation}
\label{exp}
Q=e^{-im_Q t}(\psi+\chi),
\end{equation}
where 
\begin{eqnarray}
\label{2spins}
\psi=e^{im_Qt}\frac{1}{2}(1+\gamma_0)Q \\
\chi=e^{im_Qt}\frac{1}{2}(1-\gamma_0)Q 
\end{eqnarray}
are the large and small two spinors of the full Dirac quark spinor, respectively.
As opposed to HQET, there is now a preferred frame, namely the center of mass frame. So we will assume that we are working in this frame from now on. Thus when we rescale the field we won't sum over all four velocities, as we did in HQET. Also  I won't bother labelling the fields by their four velocity, as there is only one relevant label velocity $(1,\vec{0})$.
If we wanted to allow for weak decays we would have to include a sum over  
four-momenta labels as we did in HQET.

\begin{figure}[ht]
\centerline{\scalebox{0.35}{\includegraphics*[0,0][600,750]{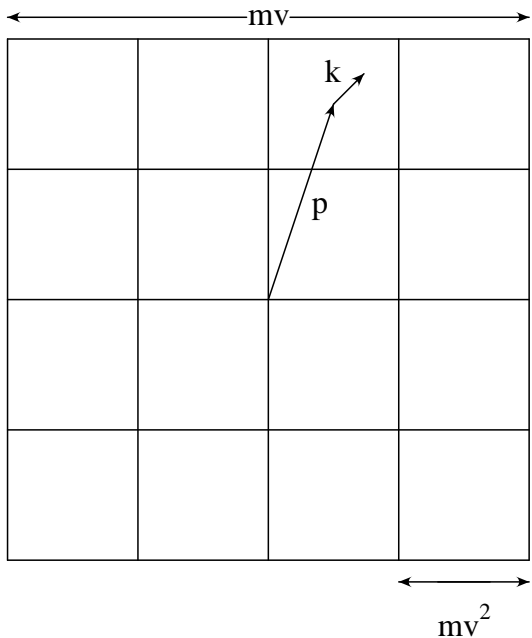}}}
\caption{This diagram depicts how the momenta are divided by the
labelling process. The largest label $v_\mu$ has been frozen to be
$(1,0,0,0)$. The three momenta are split into a large label $\vec{p}$
of order $mv$, and a small residual momenta of order $mv^2$ depicted
by the smallest boxes.}\label{boxes}
\end{figure}

Now we can separate the next largest scale, $mv$, from the other scales in the theory by writing  
\begin{equation}
\label{fieldexp}
\psi=\sum_{\bf{\vec{p}}}e^{-i\vec{p}\cdot \vec{x}}\psi_{\bf{\vec{p}}}(x)
\end{equation}
where the label on the field fixes the three momentum to be order $m\vec{v}$.
Spatial derivatives acting  on $\psi_{\bf{\vec{p}}}(x)$ are of order the residual three momenta $mv^2$.
This way of splitting up momenta into a large label piece and a residual
momentum is depicted in figure (\ref{boxes}).

The free piece of the action can then be read off by substituting 
the expansions (\ref{exp}) and (\ref{fieldexp}) into the full QCD
action, solving for the small component $\chi$ in terms of the large
component $\psi$, and keeping only the leading order terms in an
expansion in the small parameter $v$,
\begin{equation}
\label{LONRQCD}
L_0=
\sum_{\bf{\vec{p}}}\psi^\dagger_{\bf{\vec{p}}} (i\partial_0-\frac{\bf{\vec{p}}^2}{2m})\psi_{\bf{\vec{p}}}~+~(\psi\rightarrow \zeta).
\end{equation}
Notice that we have dropped all spatial derivative terms, such as 
${\bf \vec{p}}\cdot \vec{\partial}$,  as they are
higher order in $v$.  Remember that the $\bf{\vec{p}}$ which appears in the
action is a C number and not an operator.  $\zeta$ is the {\it large}
component of the anti-quark field, which arises when we re-phase with
the opposite sign in (60), and its action on the Fock space is to destroy an anti-quark.

\begin{figure}[ht]
\centerline{\scalebox{0.6}{\includegraphics*[87, 400][470,600]{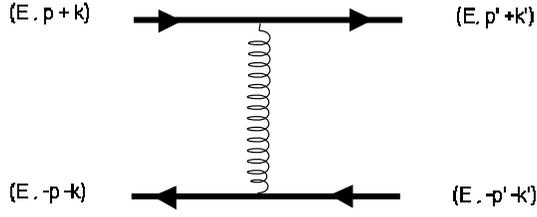}}}
\caption{
Full theory matching which generates the Coulomb interaction. The momenta are split up into a large piece ($\vec{p}$) which is order $mv$ and a residual piece $k$ which is of order
$mv^2$. The large piece corresponds to the label of the field.}\label{Coulfig}
\end{figure}

\subsection{Interactions}

The interactions can be fixed by a  matching calculations, though some interactions 
are more simply read off by gauge invariance, as we shall see. 
Let's start
by considering the t-channel exchange of a gluon between a quark and anti-quark
in the full theory as shown in figure (\ref{Coulfig}). Since the initial and final state quarks are on shell, the gluon momentum must
scale as $(mv^2,mv)$.
We split the external three
momenta into a piece of order $mv$, corresponding to the label of the
field, and a residual momenta $k$, which if of order $mv^2$.  We then
expand the full theory result in $v$ and write the full theory spinors
in terms of the two spinors (\ref{2spins}). The leading order piece
generates the operator
\begin{equation}
\label{Coulomb} 
L_C=g^2\sum_{\bf{\vec{p}}_i}\psi^\dagger_{\bf{\vec{p}}^\prime} 
T^a\psi_{{\bf \vec{p}}}
\frac{1}{(\bf{\vec{p}}-\bf{\vec{p}}^\prime)^2}
\zeta^\dagger_{-\bf{\vec{p}}^\prime}\overline{T}^a
\zeta_{-\bf{\vec{p}}}
\end{equation}
This operator is nothing but the Coulomb potential. It is called a ``potential" because
the interaction is non-local in space but instantaneous in time. 
Notice that $L_C$ includes a sum over all possible
labels. This is in stark contrast to HQET where an external current is
needed to change labels. Here in NRQCD, there are terms in the action
which themselves change labels. What we have done seems a little
perverse at first. We have integrated out modes (via a field rescaling) only
to integrate them back in by summing over labels. However, we will see that
there is a method to the madness.

Now let us power count to see if the operator (\ref{Coulomb}) we matched onto should
appear in the leading order Lagrangian. To do so we first need to
determine how the fields scale. The easiest way to do this is to use
the fact that the kinetic piece should be order one in our power
counting. So, following the line of reasoning developed in section
\ref{HQETPC}, we first power count the measure in the action.  The
quark field has support over $1/v^2$ in the time direction\footnote{All units are scaled to $m$.} and order
$1/v$ in the spatial directions so we conclude that $d^4x \sim v^{-5}$
and $\psi \sim v^{3/2}$. In this way the kinetic piece
(\ref{LONRQCD}) scales as $v^0$. Now let us consider the operator
(\ref{Coulomb}). The measure again scales the same way as in the
kinetic piece since the operator contains only heavy quarks, so that
\begin{equation}
O_C\sim v^{-5} (v^{3/2})^4 v^{-2} \sim v^{-1}. 
\end{equation}
Thus $O_C$ certainly should be treated as a leading order piece in
the Lagrangian.  Indeed, the fact that it actually is enhanced
relative to kinetic piece is worrisome as it could lead us to the conclusion that there is no sensible $v\rightarrow 0$ limit, which is the limit in which the leading order action is exact. We will come back to this issue later. For now we will
conclude that the Coulomb potential will have to be treated non-perturbatively
in our theory. The operator can be interpreted in terms of the exchange of
a non-propagating Coulomb gluon. If we wish we can re-write this operator
in terms of an auxiliary field (i.e. a field without a kinetic term)\cite{LS}
\begin{equation}
\label{coulomb}
L_c=\sum -g(\psi^\dagger_{\vec{\bf{p}}}\psi_{\vec{\bf{p}^\prime}}+
\chi^\dagger_{\vec{\bf{p}}}\chi_{\vec{\bf{p}^\prime}})
A_{0_{\vec{\bf{p}}^{\prime  \prime}}}+{1\over 2}(\vec{\bf{\nabla}}A_{0_{\vec{\bf{p}}}})^2.
\end{equation}
Writing things in this way will make our relevant degrees of freedom more explicit when we discuss NRQCD in terms of the method of regions as discussed in (\ref{regions}). But this is just a formal manipulation. In general, ``off shell'' modes 
will not appear in the Lagrangian.

\begin{figure}[ht]
\centerline{\scalebox{0.9}{\includegraphics*[37,490][270,560]{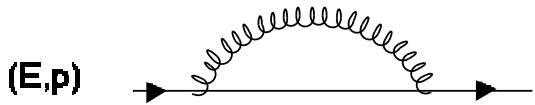}}}
\caption{Self energy diagram is NRQCD. The labelling technique automatically implements the dipole expansion.}\label{seNRQCD}
\end{figure}
\subsection{The Multipole Expansion and Ultra-Soft Modes}
What interactions are there in the one quark sector? We should allow
for all interactions which leave the fields on-shell to within order
$mv^2$, which is the scale of their residual momentum, as these modes will certainly
generate  non-analyticity in our amplitudes. 
One obvious
interaction we can have is of the usual form $\psi^\dagger A \psi$, if
$A$ is a gluon which carries momentum scaling as $(mv^2,mv^2)$.  In
fact, gauge invariance seems to lead us to this type of interaction,
since we would expect that the $\partial_0$ in (\ref{LONRQCD}) should
be elevated to a covariant derivative. Let's see if this new
interaction does indeed to correspond to $(mv^2,mv^2)$ gluons.  The
self energy diagram in figure (\ref{seNRQCD}) is given by
\begin{equation}
\label{NRQCDSE}
i\Sigma = g_s^2 C_F \int {d^dg \over{(2\pi)^d}}\frac{1}{(g^2+i\epsilon)}\frac{1}{(E-g_0-\frac{\bf{\vec{p}^2}}{2m} +i \epsilon)}.
\end{equation}
Now we will show that the gluon running through the loop does have momentum which scales as $(mv^2,mv^2)$.
The important point to note here is that the fermion propagator does not carry any of the three momentum of the gluon. The $\bf{\vec{p}}^2$ piece is just
the label of the external line. This is an implementation of a dipole expansion
\cite{Labelle,GR}
\begin{equation}
\label{dipole}
A^a_0(t,{\bf{\vec{x}}})=A^a_0(t,{\bf{\vec{0}}})+{\bf{\vec{x}}\cdot \vec{\nabla}}
A^a_0(t,\bf{\vec{0}})+ .~.~.~
\end{equation}
\vskip.2in
{\bf Exercise 2.1}
Working with a coordinate space dipole expanded Lagrangian reproduce
equation (\ref{NRQCDSE}) for the self-energy. 
\vskip.2in 
It is interesting that the act of separating the scales via
the re-labelling technique has automatically generated a dipole
expansion. This demonstrates a persistent fact in effective field
theories. If you have properly separated the scales, then each diagram
in the effective theory should scale homogeneously in the expansion
parameter. In order for this to be true any given integral should be dominated
by just one region of momenta.
Let us see how this happens in our self-energy
diagram. Doing
the integral by contours we see that the $g_0\sim \mid
{\bf{\vec{g}}}\mid \sim mv^2$, there are no other relevant regions.
Note that if we had not dipole
expanded then the integral would have received a contribution where
$g$ was of order $m$, and the integral would not scale
homogeneously in $v$. This would spoil the power counting in the theory
\cite{GR}. Gluons which have momenta of order $(mv^2,mv^2)$ are called
``Ultra-Soft (US)'', and are on shell-modes which have the usual $F^2$ kinetic
term in the Lagrangian. These modes are sometimes also referred to as ``radiation'' or``dynamical'' gluons which can be thought of as contributing to 
higher Fock states  in the onia. 
Note, however, that the $A_0$ US mode can be eliminated from the leading order action
by redefining the quark field in the following fashion
\begin{equation}
\psi^\prime(x)=P e^{ig \int_{-\infty}^t A_0(n_\mu \lambda+{\bf \vec{x}}) d\lambda} \psi(x),
\end{equation}
where P stands for path ordering and $n_\mu$ is the time like unit vector $(1,\vec{0})$. The $A_0$ US gluons will then reappear in the action at higher order in $v$.
Thus without doing any calculations we can conclude that the interactions due to $A_0$ US
modes must vanish at leading order.

There are additional US spatially polarized gluons ($\vec{A}$)
which play an important role at sub-leading order in $v$.
We can determine the coupling of these gluons in a very simple way using what
is known as {\it reparameterization invariance}\cite{ML} (RPI).
The idea behind RPI is quite simple, yet its ramifications are 
powerful. The point is that when we pulled out the three momenta 
from the heavy quark field, there was an ambiguity \footnote{A similar ambiguity exists in the HQET rescaling. See \cite{MW} for a discussion.}. We could just 
as 
easily rescaled by $e^({imv+k})$ where $k$ is order $mv^2$.
Thus our action should be independent under small shifts in the label.
This just means that every time we see a ${\bf \vec{p}}$ it should be accompanied by a ${\bf \vec{\partial}}$. Then by gauge invariance we know that we should
make the replacement
\begin{equation}
{\bf \vec{p}} \rightarrow {\bf \vec{p}}+i{\bf \vec{D}}.
\end{equation}
everywhere we see 
${\bf \vec{p}}$. 
Notice that this form has to hold to all order in perturbation theory.
This is quite a useful piece of information, as it tells us that many
operators in our action will have vanishing anomalous dimension.

The net effect of this substitution is the generation of subleading
terms in the dipole expansion (\ref{dipole}). For instance, the kinetic term
${\bf \vec{p}^2}/(2m)$ will generate the subleading dipole
interaction $\psi^\dagger ({\bf \vec{p}\cdot \vec{A}}/m) \psi$.   Note that
${\bf{\vec{D}}}=\vec{\partial}-ig \vec{A}$, scales as $v^2$, since
when the derivative acts on the heavy quark field it brings down a
factor of the {\it residual momentum} which is of order $v^2$, and, as
you can prove for yourself, the ultra-soft gluon field scales as $v^2$ as well.
\begin{figure}[ht]
\centerline{\scalebox{0.7}{\includegraphics*[130,390][315,550]{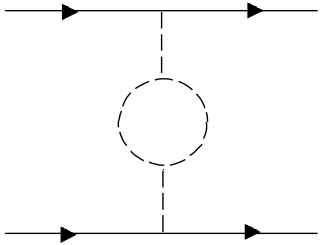}}}
\caption{A graph which one might have hoped would induce a running
potential, which does not exist in the effective theory.}\label{coulSE}
\end{figure}
\subsection{Soft Modes}
So now we have shown that there are two types of gluonic modes which
contribute to the non-analytic behavior of the low energy theory: potential gluons (which are not
dynamical), whose momenta scale as $(mv^2,mv)$, and  ``ultra-soft'' (US) gluons
whose momenta scale as ($mv^2,mv^2)$. But this can not be the end of
the story.  In particular, we know that the Coulomb
potential should run even once we are below the scale $m$, due to the
non-Abelian nature of the theory. We might expect that a diagram such
as (\ref{coulSE}) could account for the  running of the coupling/potential. 
However,
this diagram does not exist in the effective theory, since the Coulomb
gluon is not a propagating degree of freedom (it is simple to show that this diagram vanishes). The Coulomb gluon
sometimes called an ``off-shell" mode because it can never
satisfy the gluon dispersion relation $g_0^2={\bf\vec{g}}^2$, and thus never appears as an external state. 
To generate a running Coulomb potential we must
include a new set of propagating degrees of freedom\cite{MoR,Griesshammer},
namely, ``soft gluons", whose momenta scale as $(mv,mv)$.

\begin{figure}[ht]
\centerline{\scalebox{0.65}{\includegraphics*[100,390][570,700]{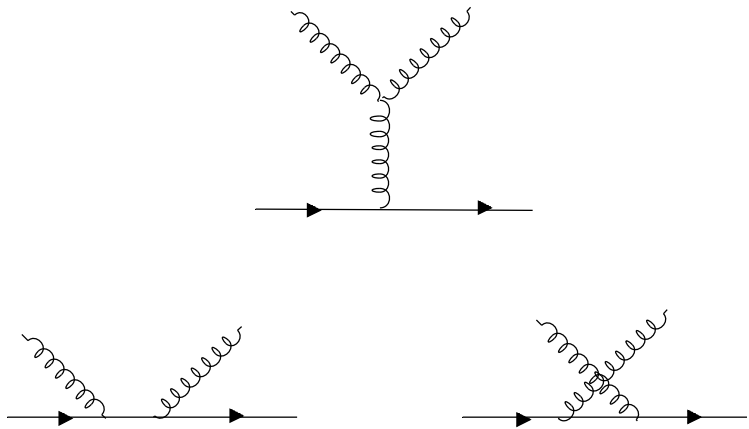}}}\caption{Leading
order matching graphs between heavy quarks and ``soft'' gluonic modes with
momenta of order $(mv,mv)$. Not shown are the similar diagrams with ghost fields,}\label{compton}
\end{figure}

How could such degrees of freedom interact with the heavy quarks?  We
notice that because their energy scales as $mv$, any such interaction
would throw the heavy quark off-shell by an amount of order $mv$. 
So we know that we need at
least two interactions.  The simplest such interactions are shown in
figure ($\ref{compton}$).
Now since these modes have a ``large'' component, in both their energy {\it and} momentum, we need to rescale the field so that all derivatives acting on the field scale as the lowest scale in the theory, $mv^2$. 
So as in the case of the heavy quark field we write
\begin{equation}
A^a_\mu(x)=\sum_q e^{-iq\cdot x}A^a_{q \mu}(x).
\end{equation}
Notice the field is now labelled by a four momentum. Now we match the
Compton graphs in figure (\ref{compton}) onto the effective theory. To
do so we split the external momenta of the gluon fields into a label
piece and a small residual momentum
\begin{equation}
Q_\mu=q_\mu+k_\mu~,
\end{equation}
and expand the diagram in powers of $v$. Keeping only the leading
order result we find
\begin{eqnarray}\label{softint}
L_{soft}^{int} &=& - 4 \pi \alpha_s \sum_{q,q^\prime\mathbf
  p,p^\prime}\Biggl\{{1\over q^0} \psi_{\bf {\vec{p}^\prime}} ^\dagger
\left[A^0_{q^\prime},A^0_q \right] \psi_{\bf \vec{p}} \nonumber \\ &&
+ {g^{\nu 0} \left(q^\prime-p+p^\prime\right)^\mu - g^{\mu 0}
  \left(q-p+p^\prime\right)^\nu + g^{\mu\nu} \left(q-q^\prime \right)
  \over \left( {\bf\vec{p}}^\prime-{\bf \vec{p}} \right)^2} \psi_{\bf{\vec{p}^\prime}}
^\dagger \left[A^\nu_{q^\prime},A^\mu_q \right] \psi_{\bf \vec{p}}
\nonumber \\ &&\qquad\qquad +
\frac{-q_0}{(\bf{\vec{p}^\prime-\vec{p}})^2}
\psi^\dagger_{\bf \vec{p}}
[\bar{c}_q^\prime,c_q]\psi_{\bf \vec{p}}\Biggr\}
+ \psi \leftrightarrow \chi,\ T
\leftrightarrow \bar T
\end{eqnarray}
Where I have also included the contribution from the ghosts $c_q$, which
are not shown in the figure.
As we will see, this interaction will be responsible for the running
of the Coulomb potential. Note that it vanishes for the Abelian case,
which is consistent with the fact that there is no change in the force
law below the scale of the electron mass in QED.  
This cancellation would not arise if we did not drop the $i \epsilon$ in the denominator.
Keeping the $i \epsilon$ would mean including a contribution where the gluons are potential and not soft. To see this write the intermediate fermion propagator as
\begin{equation}
\frac{1}{q_0+i\epsilon}=P(\frac{1}{q_0})-i \pi \delta(q_0)
\end{equation}
The delta function forces the gluon to have vanishing energy, i.e. it becomes a potential gluon \cite{MoR}.
In the effective theory this potential contribution is accounted for by operators which describe
the scattering of soft modes off of a four fermion potential (see \cite{MSII} for details).

Furthermore, to
include the effects of the ``massless'' quarks we will need to include
a soft fermion field for each such flavor.  So now we have two
propagating gluonic degrees of freedom, the soft $(mv,mv)$ and
ultra-soft (US) $(mv^2,mv^2)$. The kinetic terms
for these fields are 
\begin{eqnarray}\label{gaugekinetic}
L^{K.E} &=&  
 -{1\over 4}F^{\mu\nu}F_{\mu \nu} + \sum_{p} \mid p^\mu A^\nu_p -
 p^\nu A^\mu_p\mid^2 
\end{eqnarray}
Here the fields strengths are composed of purely US gluons $igF^{\mu
\nu}=[D^\mu, D^\nu]$. While the second term in (\ref{gaugekinetic})
generates the propagator for the soft gluons. We may treat the US field
as a background gauge field. Given the relative time scales of these two modes, the US modes  are frozen,  as far as the soft gluons are concerned
Thus in a background field formalism (see \cite{Peskin} for a review),
we may gauge fix the US fields and the soft fields independently.
The US gauge invariance is manifest while the soft gauge invariance
is only seen when combined with re-parameterization invariance.
Note that there will also be soft (labelled) ghost fields as well. The full soft
Lagrangian, including ghosts, can be found in \cite{MSI}.

\begin{figure}[ht]
\centerline{\scalebox{0.65}{\includegraphics*[65,462][400,750]{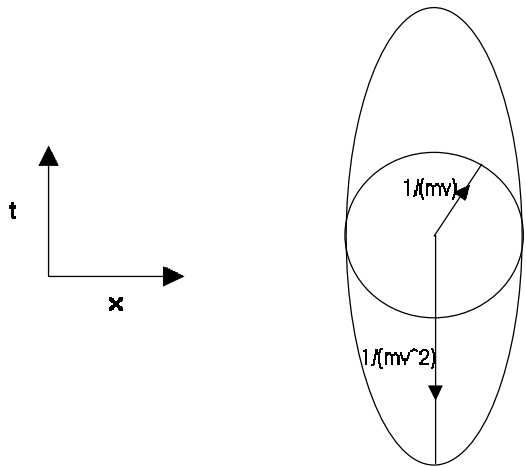}}}
\caption{The soft gluons and heavy quarks have differing
spatio-temporal extent, The overlap regions, which determines the
scaling of the measure in the action is of order $(v,v)$}\label{overlaP}
\end{figure}

We are now prepared to
see if the soft interaction Lagrangian is leading order in $v$. Each
soft gluon field scales as $v$, as can be read off from
(\ref{gaugekinetic}), while each fermion field gives a factor of
$v^{3/2}$. How does the measure scale? The fermion has
(temporal,spatial) support on the scale $(1/(v^2),1/(v))$, whereas
the gauge field has support only on scales of order $1/v$. Thus, the 
measure scales as $1/(v)^4$, since outside this volume the integrand
vanishes as depicted in (\ref{overlaP}).  Thus the soft interaction
Lagrangian $L_{soft}^{int}$ scales as $v^0$ and is indeed leading
order. However, since this interaction is also order $\alpha_s$ we need not treat it
non-perturbatively.

\subsection{Calculating Loops in NRQCD}
\begin{figure}[ht]
\label{onelopcoul}
\centerline{\scalebox{0.65}{\includegraphics*[100,375][450,540]{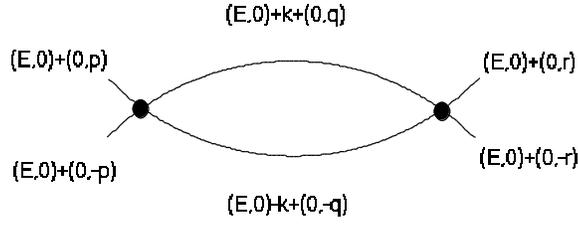}}}\caption{A
one loop graph in the effective theory with two insertions of the
coulomb interaction (solid dot).}
\end{figure}

\begin{figure}[ht]
\label{softex}
\centerline{\scalebox{0.65}{\includegraphics*[100,375][490,540]{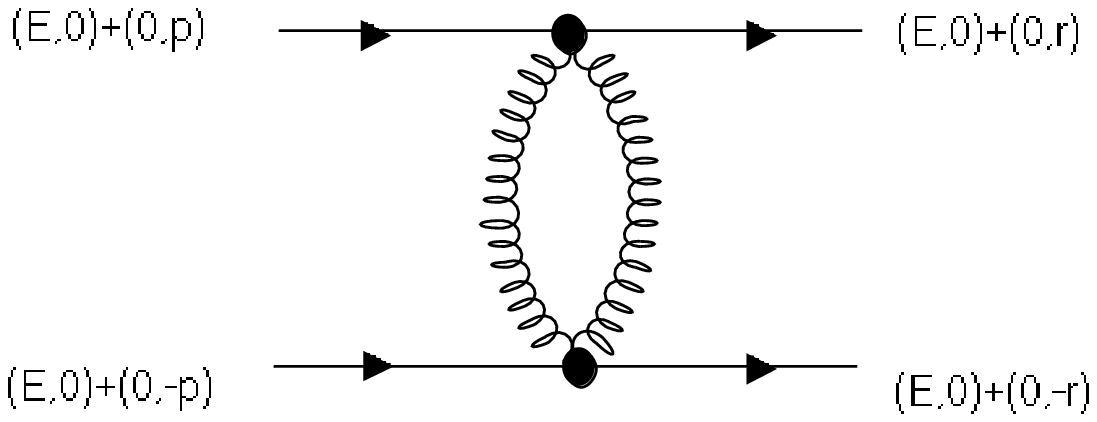}}}\caption{A one loop graph in the effective theory with two insertions of the
soft interaction (solid dot).}
\end{figure}

Now let us take stock in what we have built. So far we have a leading
order action which contains  the kinetic pieces for the
fermions and gluons (\ref{LONRQCD}) and (\ref{gaugekinetic}), as well as  two leading order interactions (\ref{coulomb}) and (\ref{softint}).
If we want to match beyond leading order in $\alpha$, we need to know
how to calculate using our effective theory. Calculating loop graphs 
seems rather complicated given that we will have loops with
sums over labels. However, it will turn out in the end to be quite
simple. As a first example suppose we're interested in a loop consisting of
two insertions of $L_C$, as shown in figure (\ref{onelopcoul}).
The external fermions have energy $E$ and momenta $\pm {\bf
\vec{p}},{\bf \vec{r}}$. In the effective theory, the external
fermions have soft momentum (labels)  $\pm {\bf \vec{p}},\pm {\bf
\vec{r}}$, and the ultra-soft momentum $(E,0)$. The intermediate
fermions in the effective theory have soft momentum $\pm q$, and
ultra-soft momentum $(E,0)\pm k$. The graph in the effective theory is
proportional to the integral
\begin{equation}\label{nrqed:5}
I_{Coul}= \sum_{\bf \vec{q}} \int d^4 k {1 \over {\bf
(\vec{p}-\vec{q})^2}} \ {1 \over {\bf (\vec{r}-\vec{q})^2}} \ {1 \over
k^0+E - {\bf \vec{q}}^2/2m+i \epsilon}\ {1 \over -k^0+E - {\bf q}^2/2m+i\epsilon}.
\end{equation}
There is an integral over the ultra-soft energy $k^0$, the ultra-soft
momentum $\mathbf k$, and a sum over the soft momentum $\mathbf q$.
Summing over ${\bf \vec{q}}$ and integrating over ${\bf \vec{k}}$ is
equivalent to integrating over the entire momentum space. Thus one can
replace Eq.~(\ref{nrqed:5}) by
\begin{equation}\label{nrqed:6}
I_{Coul}=\int d^4 q {1\over ({\bf \vec{p}}-{\bf \vec{q}})^2}~ {1\over ({\bf
\vec{r}}-{\bf \vec{q}})^2}~ {1 \over q^0+E - {{\bf \vec{q}}}^2/2m+i\epsilon}~ {1
\over -q^0+E - {{\bf \vec{q}}}^2/2m+i\epsilon},
\end{equation}
where 
\begin{equation}\label{nrqed:11}
dk^0 \to d q^0,\qquad \sum_{\mathbf q} \int d{\mathbf k} \to 
\int d{\mathbf q}.
\end{equation}
Soft loops are treated in a similar fashion. Consider the diagram with the insertion of
two soft interactions as shown in figure (\ref{softex}).
It gives a contribution of the form
\begin{equation}\label{nrqcd:8}
I_{soft}= \sum_q \int d^4 k {1\over {\bf (\vec{p}-\vec{r})^2}}\
{1 \over {\bf (\vec{p}-\vec{r})^2}}\ {1 \over \left(q^0\right)^2-{\bf \vec{q}}^2}\
{1 \over \left(q^0\right)^2-\left({\bf \vec{q}+\vec{p}-\vec{r}}\right)^2},
\end{equation}
where I have used the last interaction in Eq.~(\ref{softint}) for the soft
vertices. The sum on $q$ is over a four-vector. As in the case of the
potential loops, we make the replacement 
\begin{equation}\label{nrqed:11s}
\sum_{q} \int d^4{ k} \to 
\int d^4{q},
\end{equation}
where the replacement must be done for all four components of $q$, since soft
gluons carry a four-vector label $q$.
There is a much simpler way to arrive at the effective theory integrand.
Just take the full theory graphs, assume the proper scaling for the
gluon loop momentum, i.e. soft, potential or US, and just keep the leading order term in 
each propagator. 
\subsection{Matching at Subleading Orders in $v$}
Let us now consider to matching at higher orders in $v$, but still at tree level.
We can guess the first correction
to the kinetic energy from the dispersion relation
\begin{equation}
L^{(2)}_{kin}=\sum_{\bf{\vec{p}}}\psi^\dagger_{\bf{\vec{p}}} (\frac{\bf{\vec{p}}^4}{8m^3})\psi_{\bf{\vec{p}}}~+~(\psi\rightarrow \chi).
\end{equation}
Formally this piece can be derived simply by keeping the sub-leading
terms in the expansion of the QCD action in terms of the
non-relativistic fields (\ref{fieldexp}). The more relevant correction
comes from higher order terms in the multipole expansion,
\begin{equation}
\label{dip}
L^{(1)}_{mp}=\sum_{\bf{\vec{p}}}\psi^\dagger_{\bf{\vec{p}}} (\frac{\bf{\vec{p}}\cdot {i \bf \vec{D}}}{m})\psi_{\bf{\vec{p}}}~+~(\psi\rightarrow \chi).
\end{equation}
The ultra-soft gluon field inside $ {\bf \vec{D}}$ scales as $v^2$, as does the derivative since it brings down
a factor of the residual momentum of the heavy quark.
Now let's consider the corrections to the quark/anti-quark potential. These are derived by expanding the
full theory graph in figure (\ref{Coulfig}) to higher orders in $v$.

{\bf Exercise 2.2} Show that, at order $\alpha_s v$ the subleading
potentials generated by T-channel gluon exchange are given by

\begin{eqnarray} \label{V1}
 V^{(1)} &=&  (T^A \otimes \overline{T}^A) 
\left[ 
 { {V}_r\: {\bf \vec{p}^2 } \over  m^2\,
 {\bf \vec{k}}^2}
 + { V_{hf} \over m^2}\, {\bf \vec{S}}^2 
+ {{  V}_{so}
 \over m^2}\, \Lambda({\bf \vec{p}^\prime ,\vec{p}}) + { V_t \over  m^2}\,
 {T}({\bf \vec{k}})
\right] 
\end{eqnarray}
where ${\bf \vec{k}=\vec{p}-\vec{p}^\prime}$ and 
\begin{eqnarray}
\bf \vec{S} &=& { {\bf \vec{\sigma}_1 + \vec{\sigma}_2} \over 2},
 \qquad \Lambda({\bf \vec{p}^\prime, \vec{p} }) = -i {\bf \vec{S} \cdot ( \vec{p}^\prime
 \times \vec{p}) \over  {\bf k}^2 },\qquad
 {T}({\bf \vec{k}}) = {\bf \vec{\sigma}_1 \cdot \vec{\sigma}_2} - {3\, {\bf \vec{k}
 \cdot \vec{\sigma}_1}\,  {\bf \vec{k} \cdot \vec{\sigma}_2} \over {\bf k}^2} \nonumber
\end{eqnarray}
and ${\bf \vec{\sigma}}_1/2$ and ${\bf \vec{\sigma}}_2/2$ are the spin-operators on the quark and
anti-quark.   $V_{hf}$ and $V_{so}$ are the familiar hyperfine and spin-orbit potentials respectively.
The matching coefficients are given by
\begin{eqnarray}  \label{bc}
 { V}_r &=& 4 \pi \alpha_s\,,\qquad 
 { V}_{hf} = -\frac{4 \pi \alpha_s}{3}\,, \qquad \nonumber\\
 { V}_{so} &=& -6 \pi \alpha_s \,,\qquad 
 { V}_t = -\frac{\pi \alpha_s}{3}.
\end{eqnarray}
 
\begin{figure}[ht]
\centerline{\scalebox{0.75}{\includegraphics*[60,250][360,425]{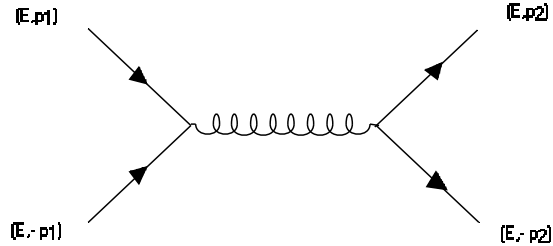}}}\caption{Tree level full theory diagram contributing to the potential.}\label{annihl}
\end{figure}

The annihilation diagram in Fig.~\ref{annihl} generates terms of order
$\alpha_s v$ in the potential.  Using Fierz  identities and
charge conjugation, the contribution of this diagram  can be transformed into the  basis in
Eq.~(\ref{V1}) and give additional contributions to the matching. Only ${
V}_{hf}$ receives a non-zero contributions at  tree level from annihilation:
\begin{eqnarray} \label{bc2}
 V_{hf,a} = {1 \over N_c}\: \pi\, \alpha_s \,, \qquad
\end{eqnarray}
where the subscript $a$ stands for annihilation.
In addition this diagram generates the a hyper-fine interaction with a different color structure 
\begin{equation}
\delta V^{(1)}=(1\times 1) 
{ V_{hf}^{(1)} \over m^2}\: {\bf S}^2
\end{equation}
with $V_{hf}^{(1)}=\alpha \pi {N_c^2-1 \over{2N_c}}$. The color factor arises from the
Fierz rearrangement.

\subsection{Matching at Subleading Orders in $\alpha$}
Let's now consider matching beyond tree level. But before we plunge
into a calculation we must address an intricacy of NRQCD which I
touched upon previously. In particular, we noticed that the Coulomb
potential scales as $1/v$. On the face of it, this presents a serious
problem to the theory for the following reason. Suppose we wish to
calculate the matrix element of some operator in the effective theory,
 $\langle \psi\mid  O \mid \psi \rangle$. at order in $v^k$. In
this case we have
\begin{equation}
\label{example}
\langle \psi\mid O \mid \psi \rangle=\int d^4x_1 ...d^4x_n\langle \psi \mid T (O H_{1}(x_2)....H_{n}(x_n)exp({i\int d^4x L_C(x)})\mid \psi \rangle
\end{equation}
where $H_{i}$ are terms in the sub-leading Hamiltonian such that
their net order in $v$ is $v^k$.  Given that the contribution from the
Coulomb potential is of order $1/v$ and must be treated non-perturbatively, it appears that  the time ordered product will not have a definite scaling in $v$. However, we are saved by the fact
that while the Coulomb potential is of order $1/v$ it is also of order
$\alpha$.  Furthermore, in a Coulombic state the Virial theorem tells
us that
\begin{equation}
{mv^2}\simeq {\alpha_s(1/r \simeq mv)\over{r}}
\end{equation} 
so that $\alpha_s(mv) \simeq v$. \footnote{If we are not in a Coulombic state, for instance we could be looking
at threshold production, then we must have $\alpha \leq v$.}

So when we calculate matrix elements we should treat the Coulomb potential
to be order one.  However, when we match, we must account for the fact
that the Coulomb potential can enhance a sub-leading operator, at the
cost of a power of $\alpha_s$.  For instance, if we want to match at
order $\alpha_s^2$, then we have to include  contributions of the
form
\begin{equation}
\langle L^{(2)} L_c^2 \rangle \propto \alpha_s^2 
\end{equation}
where $L^{(2)}$ is an order $v^2 \alpha_s^0$ piece of the Lagrangian.

As a practical example let us consider the matching of a production
current at order $v^0$. This matching calculation is relevant
to top-quark threshold production\ \cite{HMST}.  At tree level we
match by expanding the full QCD current in terms of the NRQCD fields,
\begin{equation}
 \bar{q}\gamma^i q= \sum_{\bf \vec{p}}\psi^\dagger_{\bf \vec{p}}
 \sigma^i \chi^\star_{\bf -\vec{p}}+ ...
\end{equation}
This operator produces a quark/anti-quark pair in a relative $^3S_1$ state.

{\bf Exercise 2.3} Show that at order $v^2$ one generates the operators
\begin{equation}
O_1^i= \sum_{\bf \vec{p}}\psi^\dagger_{\bf \vec{p}}({\bf \vec{p} \cdot
\vec{\sigma}} ){\bf p^i} \chi^\star_{\bf -\vec{p}}
\end{equation}
\begin{equation}
O_2^i= \sum_{\bf \vec{p}}\psi^\dagger_{\bf \vec{p}} ({\bf \vec{p}^2}) \sigma^i
\chi^\star_{\bf -\vec{p}}
\end{equation}

\begin{figure}[ht]
\centerline{\scalebox{0.75}{\includegraphics*[60,400][430,615]{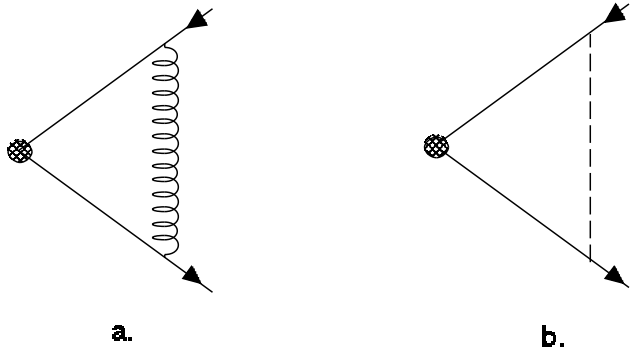}}}\caption{One loop matching diagrams in the full and effective theory. In the effective theory there are additional diagrams involving temporal US gluons which vanish. }\label{oneloopcurrmatch}
\end{figure}

At one loop we must calculate the graphs shown in
(\ref{oneloopcurrmatch}).  Since the full theory current is
conserved, it has vanishing anomalous dimension once the wave function
renormalization is included. 
We can simplify the matching calculation by working at threshold. In this way
the effective theory diagram vanishes, since it is scaleless. 
Calculating the full theory diagram gives 
\begin{equation}
\rm{Figa} =\psi^\dagger \sigma^i \chi^\star \frac{i \alpha C_F}{ 4 \pi}\left[\frac{1}{\bar{\epsilon}_{UV}}+\frac{2}{\bar{\epsilon}_{IR}}-3\ln(m^2/\mu^2)-9\right], \end{equation}
where I have explicitly separated the UV from the IR divergences.
Adding the contribution from the wave function renormalization, all the UV and IR poles
cancel. 
 We then  find the
matching coefficient to be
\begin{equation}
C=(1-2C_F\frac{\alpha}{\pi}).
\end{equation}
You should be able to show for yourself that this short cut to calculating the
matching coefficient is identical to calculating the hard piece of the graph using the
method of regions.

Notice that at one loop the current has no anomalous dimension.
We can conclude this from the fact that all of the IR poles in the
full theory cancelled (see section 1.8.2)
In general we would not expect this since in the effective theory
the current does not generate a symmetry. In fact we will see that at two loops 
we will generate an anomalous dimension.

\begin{figure}[ht]
\centerline{\scalebox{0.75}{\includegraphics*[20,120][600,550]{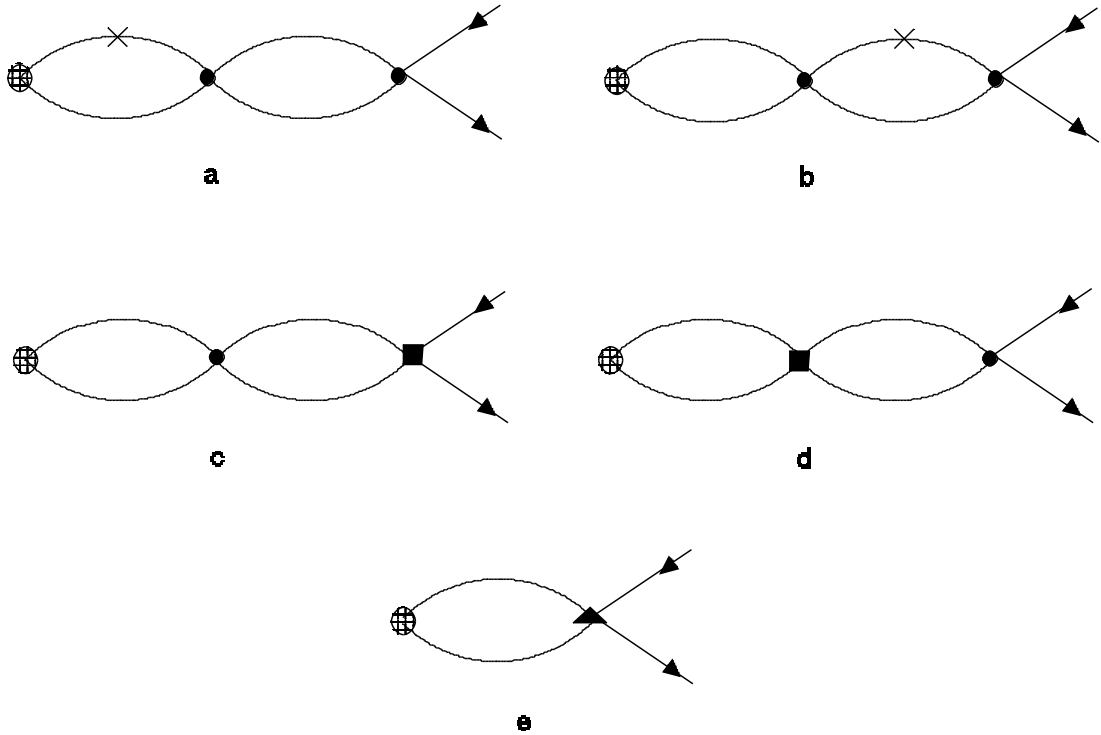}}}\caption{Diagrams
contributing to the matching of the current at order $\alpha^2
v^0$. The large hatched circle is the current insertion. The small
filled circle is a Coulomb potential while the square represents
insertions of order $v^1$ potential. The cross corresponds to the insertion of 
the $p^4/(8m^3)$ operator. The triangle is the order $v^0$
potential.}\label{twolooprun}
\end{figure}

  At
two loops one must account for the possible Coulomb enhancement
discussed previously in this section. So if we concerned with
contributions of order $\alpha_s^2 v^0$, we must include all possible
operator insertion such that the net power is $v^0$. This means we
must also include any order $v^0 \alpha_s^2$ potentials which might be
generated. At this order the full  theory generates the
$v^0$ potential
 \begin{equation}\label{Vk}
   V^{(0)}= V_k
\frac{\pi^2}{m \mid {\bf \vec{k}}\mid }
 \nonumber \\
\end{equation} 
where 
\begin{equation}V_k=\alpha^2(\frac{C_F^2}{2}-C_AC_F),
\end{equation}
and I have projected the potential onto the color singlet channel since that is all
we need when calculating the production current.
We must include one insertion of this potential in addition to
insertions of the order $v$ potentials already discussed in the
previous section. The relevant diagrams are shown in
(\ref{twolooprun}).

In addition there are corrections which contain Coulomb singularities
$1/v$, i.e. those contributions whose net scaling has inverse powers
of $v$. However, these won't contribute to the matching since they will
cancel in the full and effective theory. The purpose of this example
was just to illustrate that when you match you must remember that
sub-leading operators can be enhanced due to the $1/v$ nature of the
Coulomb potential. To my knowledge, as of the time of this writing,
this subtlety is unique to NRQCD.
It is interesting to note that had we done the two loop calculation using HQET
propagators we would get a vanishing anomalous dimension. This is most easily seen
in $A_0=0$ gauge.

\vskip.3in
{\bf Exercise 2.4:} Prove that in the $ {\overline{MS}}$ scheme the anomalous dimensions of the production current is gauge invariant. To prove this write down two equations for the current. One imposing
its independence from $\mu$ and the other its independence of the gauge fixing parameter $\alpha$.
Then commute the two equations to derive a set of relations which allows you to complete the
proof.
If you get stuck consult the original reference \cite{CW}.
\vskip.3in

The reason HQET gets it wrong lies in the theory's inability to properly describe fermion-fermion propagation. What happens is that HQET
misses out on the Coulomb singularity. This fact emphasizes an important
point about NRQCD. You can't treat the scales $mv$ and $mv^2$ independently. 
Usually we would argue that between the scales $m$ and $mv$, $mv^2$ should be treatable
as a perturbation, much as we would treat a light quark mass when working at a scale $E\gg m_q$.
But the fact that we need to keep the kinetic piece of the action at leading order is telling
us that both $mv$ and $mv^2$ must exist simultaneously in the theory.
After a brief diversion we will see the consequences of this correlation of scales.

\subsection{Summing Logs: The Velocity Renormalization Group (VRG)}
You have probably noticed that I have been careful not to specify the scale at which 
to evaluate the running couplings $\alpha_s(\mu)$. Such a specification implies we have
control of the logarithmic corrections, which I will only now discuss.
As shown in the previous lecture the recipe for summing large logs is typically straightforward.
Suppose we have a theory with two ``large" (relative to the scale at which experiments are to be performed $E$) scales $m_1\gg m_2$. In a general diagram we would expect to generate logs of the form $\ln(m_2/E),~\ln(m_1/E), ~\ln(m_1/m_2)$.
We would sum these large logs in steps: 1) Match at the scale $m_1$ onto a
theory where $\phi_1$ has been integrated out. 2) Use the RG to run
down to the scale $m_2$. 3) Match onto a theory where $\phi_2$ has been
integrated out. 4) Run down to the scale of the experiment.  We will call
this procedure ``two stage running''.
In the
case of NRQCD, where we have two large scales $m$ and $mv$ and a low
energy scale of $mv^2$, we will get logs of the form $\ln(m/mv^2)$ and
$\ln(m/mv)$. So we might be tempted to follow the standard machinery
and turn the crank. Match at $m$, run to $mv$ integrated modes of
momenta $mv$ and then run down to the scale $mv^2$. However, NRQCD is
a lot more interesting than that. The basic point is that the two scales
$mv$ and $mv^2$ are correlated. They shouldn't be treated
independently, since they coexist in the fermionic propagator.
If we were to follow the two stage running in NRQCD, we would first
match at $m$ onto an HQET like theory by rescaling the fields only by their
masses. In this case we should allow for quark virtualities
$q^2<m_Q^2$ and one might wish to treat  
${{\bf \vec{p}^2}\over{2m}}$ as a perturbation 
leaving the HQET form of the propagators. But as we discussed above, we
know that this will give the wrong anomalous dimension for the
current \cite{lmr}. Instead, in NRQCD what we must do is match
directly, at the scale $m$, onto the theory  in which 
$E\approx {{\bf \vec{p}^2}\over{2m}}$. So we can see immediately that
the two stage running will not be appropriate. Instead we do what
is known as ``one stage running''.


Given that we have to match directly onto a theory in which two scales
exist simultaneously, how can we sum the large logs?  In general,
potential and soft loops will generate $\ln(\sqrt{mE}/\mu)$, while US
loops will introduce logs of the form $\ln(E/\mu)$.  So in one-stage
running we introduce two different $\mu$ parameters, $\mu_S$ and
$\mu_{US}$. However, we are not free to choose $\mu_S$ and $\mu_{US}$
independently, since $\mu_S^2= m \mu_{US}$. Instead we should write
$\mu_S=m \nu$ and $\mu_{US}=m\nu^2$, and think of a subtraction
velocity $\nu$ instead of a subtraction scale $\mu$. This is the basis
of what is known as the ``velocity renormalization group"
(VRG)\cite{lmr}. The renormalization group equation is then derived
using the fact that bare quantities are independent of the subtraction
velocity, i.e.
\begin{equation}
\nu \frac{d}{d\nu}G_B=0.
\end{equation}
 Then by solving these equations and running from $\nu=1$ to $\nu=v$, we can sum all the large logs. How does the VRG differ from the standard RG \cite{MSS}? In VRG we have
\begin{equation}
\nu \frac{d}{d\nu}=\nu (\frac{d\mu_S }{d\nu}\frac{d}{d\mu_S}+\frac{d\mu_{US}}{d\nu}\frac{d}{d\mu_{US}})=\mu_S \frac{d}{d\mu_S}+2\mu_{US}\frac{d}{d\mu_{US}}.
\end{equation}
Note, the  crucial factor  of  $2.$
In the two-stage method one would first integrate from $m$ to
$mv$ using, $\gamma_S+\gamma_{US}$, then continue down to
the scale $mv^2$, using only $\gamma_{US}$. Whereas in one-stage one integrates
$\gamma_S+2\gamma_{US}$ from $\nu=1$ to $\nu=v$. If we assume that the anomalous dimensions are constant and don't run, then in the two stage running we would find for the running of
some generic Wilson coefficient
\begin{equation}
C(mv^2)-C(m)=(\gamma_S +\gamma_{US})\ln(mv/m)+\gamma_{US} \ln(mv^2/(mv)),
\end{equation}
while in one-stage we have
\begin{equation}
C(mv^2)-C(m)=(\gamma_S +2 \gamma_{US})\ln(v).
\end{equation}
So we have agreement. However, in general the anomalous dimensions depend upon
some couplings which are scale dependent. This scale dependence will lead to
a discrepancy between the two methods at higher orders.

Suppose we are interested in the  anomalous dimensions $\gamma_i$ 
of some set of Wilson coefficients $U_i$ \footnote{Operators and Wilson coefficients have opposite anomalous dimension  since their product is
RG invariant.}. These anomalous dimensions will, in general, depend on $U_i$ themselves.
\begin{equation}
\gamma_i\equiv \gamma_i(U_i(\mu)).
\end{equation}
$\gamma_i$ will be some function of $t\equiv \ln(\mu)=\ln(v)$, which we can
expand \footnote{This expansion is only for the purposes of illustrating the
difference between one-stage and two-stage running.}
\begin{eqnarray}
\gamma_i(t)&=&\gamma_i(0) +t\left(\frac{\partial \gamma_i}{\partial U_k}\frac{\partial U_k}{\partial t}\right)_{t=0}+ ... \nonumber \\
&=&\gamma_i(0) +t\left(\frac{\partial \gamma_i}{\partial U_k}\gamma_k\right)_{t=0}+ ...~~~~.
\end{eqnarray}
Suppose the $U_i$ have both soft and ultra-soft anomalous dimension (this
typically will occur at two loops). Then in one-stage running, integrating the
anomalous dimensions will give
\begin{eqnarray}
&&\int_0^t \gamma_i(t^\prime)dt^\prime=
\nonumber \\ & &t (\gamma_i^S+2 \gamma_i^{US})+\frac{t^2}{2}\left( 
\frac{\partial \gamma_i^S}{\partial U_k}\gamma_k^S+\frac{\partial \gamma_i^S}{\partial U_k}(2 \gamma_k^{US})+\frac{\partial (2\gamma_i^{US})}{\partial U_k}\gamma_k^S+\frac{\partial (2\gamma_i^{US})}{\partial U_k}(2 \gamma_k^{US})\right).\nonumber \\
\end{eqnarray}
Whereas two-stage running yields
\begin{eqnarray}
&&\int_{0}^{\ln(v)} \gamma_i^{(S,US)}(t^\prime)dt^\prime+
\int_{\ln(v)}^{\ln(v^2)} \gamma_i^{(S)}(t^\prime)dt^\prime
= \nonumber \\
& & 
t (\gamma_i^S+2 \gamma_i^{US})+\frac{t^2}{2}\left( 
\frac{\partial \gamma_i^S}{\partial U_k}\gamma_k^S+\frac{\partial \gamma_i^S}{\partial U_k}(\gamma_k^{US})+\frac{\partial (\gamma_i^{US})}{\partial U_k}\gamma_k^S+4 \frac{\partial (\gamma_i^{US})}{\partial U_k}( \gamma_k^{US})\right),\nonumber \\
\end{eqnarray}
and we see that the cross terms differ by a factor of two. 

Let us now return to the question of the running of the Coulomb potential.
This running is particularly simple
because the renormalization only involves soft loops. The matching coefficient
is simply $4 \pi \alpha_s(m)$, and  the running is generated
by figure (15), with each vertex giving a factor of $V_c$
\begin{equation}
\label{Coulanomdim}
Fig.15=-i\frac{\alpha_s^2}{ {\bf \vec{k}}^2}
(T^A\otimes \bar{T}^A)
\left(\frac{11}{3}C_A \left(\frac{1}{\bar{\epsilon}}+\ln\left(\frac{\mu_S^2}{{\bf \vec{k}}^2}\right)\right)+\frac{31 C_A}{9}\right). \end{equation}
There are also corrections stemming from US temporal gluon loops, but these vanish for reasons which were previously discussed.
Calling the coefficient of the Coulomb operator in the action $V_c$,  and using the fact that  its bare value $(V_c^B$) is independent of $\nu$, we have 
\begin{equation}
\nu \frac{d}{d\nu}V_c^B=0=\nu \frac{d}{d\nu}\left( \mu_S^{-2 \epsilon}( V_c-\frac{11}{3}C_A\frac{\alpha_s^2}{\epsilon})\right),
\end{equation}
and we are left with
\begin{equation}
\nu \frac{d}{d\nu}V_c=-2 \beta_0 \alpha_s^2(\nu).
\end{equation}
Where I've used $\beta_0=\frac{11}{3}C_A$ to make it clear that $V_c$ obeys the same equation as the gauge coupling.
Using our matching result for an initial condition, along with the explicit form for $\alpha_s(\nu)$, we find
\begin{equation}
V_c(m\nu)=4 \pi \alpha_s(m\nu).
\end{equation}
At this order the running of the potential is identical to the running of the beta function. The large logs can be avoided by choosing
$\alpha({\bf \mid \vec{k} \mid })$ in the potential. 
This agrees with our intuitive picture of the force law (Wilson loop) getting quantum corrections due to
the running of the coupling. However, this correspondence does not persist at higher orders
as the running of the Coulomb potential includes effects from higher order potentials which
vanish in the static limit \cite{BPSV,HMS}.

\vskip.3in
{\bf Exercise 2.5}: Show that the $\frac{11}{3}C_A$ gets replaced by the
 the full beta function $\beta_0=\frac{11}{3}C_A-\frac{2}{3}n_f$,
once $n_f$ soft light quark fields, $\phi_q$, are introduced.
To begin this calculation first show that the soft Lagrangian
gets an additional piece given by
\begin{equation}
L_S^\prime=-g^2 \sum_{{\bf \vec{p}, \vec{p}^\prime},q,q^\prime}
\left( 
\frac{1}{(\bf{\vec{p}^\prime-\vec{p}})^2}(\psi^\dagger_{\bf \vec{p}^\prime}T^A\psi_{\bf \vec{p}})
(\bar{\phi}_{q^\prime}\gamma_0 T^A \phi_q)\right)+(\psi \rightarrow \chi).
\end{equation}

There is a simple way to determine when, where and how the $\mu_S$ and $\mu_{US}$ show up.
We just make sure that in $4-2\epsilon$ dimensions, the operators have the same $v$ scalings
as they would in $4$ dimensions. For instance, in the
Coulomb potential the matching comes with a factor of $g^2$, which arises from an exchange of a
Coulomb gluon. In $4-2\epsilon$ dimensions, should we write this as $g^2 \mu_S^{2\epsilon}$ or
$g^2\mu_{US}^{2 \epsilon}$? 
The coupling should still have units $(mass)^\epsilon$ but the $v$ scaling appears ambiguous until we impose the fact that 
the operator should scale as $v^{-1}$. The fermion fields now scale
as $v^{3/2-\epsilon}$ and we have a factor of $1/v^2$ from the Wilson (label) coefficient. But now the
measure scales as $v^{-2} v^{-3+2\epsilon}$. So to insure that the operator scales as $v^{-1}$
we should write $g^2\rightarrow g^2 \mu_S^{2 \epsilon}$. If we consider an operator where a US gluon attaches to the potential \cite{MSI}, the scaling of the measure won't change (see section  2.4) but since the US gluon scales as $v^{2-2\epsilon}$ then we should include a factor of $\mu_{US}^{\epsilon}\mu_S^{2 \epsilon}$.

\subsection{The Lamb Shift  in QED}
I would now like to utilize this formalism in a simple setting,
namely, the Lamb shift in QED.  This is a beautiful example because,
not only does the formalism simplify the calculation, but it also
provides one with immediate information about sub-leading orders, as we
shall see.  Let me remind you that the Lamb shift is the splitting
between the $2S_{J=1/2}$ and $2P_{J=1/2}$ levels of Hydrogen. Recall
that these levels are ``accidentally" degenerate up to $\alpha^2$
corrections. Their splitting comes in at relative order $\alpha^3$,
corresponding to a over all energy shift of order $\alpha^5$. This
splitting is dominated by the so-called ``Bethe log", $\alpha^5\ln
v$. From an effective field theory point of view, all logs are RG
logs. If they look like ``long distance" logs, its only because the
scales have not been properly separated, i.e. you haven't written down
the proper effective field theory.  Thus, we should be able to also
resum these logs if there are more of them. In the case of the ``Bethe log" one
can ask the question, is this log just a first term in a series? i.e.
\begin{equation}
\Delta E \approx \alpha^5 \ln v(1+\alpha \ln v +\alpha^2 \ln^2 v+...)
\end{equation}

\begin{figure}[ht]
\centerline{\scalebox{0.75}{\includegraphics*[130,400][370,570]{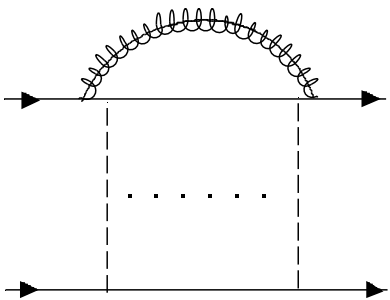}}}\caption{
The QED diagram responsible for the Lamb shift. The photon over the top is non-instantaneous, i.e. ultra-soft, while the ladder gluons are Coulombic and must be resummed.
}\label{lamb}
\end{figure}
Typically the Lamb shift comes about from a low energy matrix element involving diagrams
as shown in (\ref{lamb}). 
But in NRQCD we know that there should be no large logs in any low energy matrix element. The log should arise from running. Let's see how this comes about.

Recall that the leading order energy levels scale as $v^2$, or
using the Coulombic counting, $\alpha \propto v$, $\alpha^2$. 
So one would expect at order $v^4$ there would be a splitting due to the 
subleading contribution to the Hamiltonian  (\ref{V0}) and (\ref{Vk}). 
However, we must account for the fact that
in Hydrogen  the masses of the constituents are not equal. In fact, we may take the mass of the proton to infinity. 
\vskip.3in
{\bf Exercises 2.6}: Show that for hydrogen we have
\begin{eqnarray} \label{V0}
 V^{(1)} &=&   
\left[ 
{V_{so}
 \over m^2}\, \Lambda({\bf \vec{p}^\prime ,\vec{p}}) + V_2
\right] \nonumber \\
\end{eqnarray}
where $V_2$ is a contact delta function potential which vanishes in the 
equal mass case (considered in (\ref{V1})). Matching at $\nu=1$ we find 
\begin{equation}
V_2(\nu=1)=\frac{\pi \alpha}{2}\left(\frac{1}{m_e}-\frac{1}{m_P}\right)^2 \approx
\frac{\pi \alpha}{2m_e^2}.
\end{equation}

\begin{figure}[ht]
\centerline{\scalebox{0.75}{\includegraphics*[112,415][260,550]{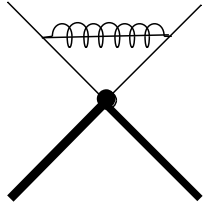}}}\caption{
The one loop diagram which generates an anomalous dimension for
$V_2$. The heavy lines are protons while the photon line with a line
through it is ultra-soft. In the limit where the proton mass goes to infinity
there is no soft anomalous dimension for $V_2$. 
 The filled circle is a Coulomb vertex,
while the US gluon couples to the electron via the order $v$ dipole
operator (\ref{dip}).}\label{lambanomdim}
\end{figure}

In addition, at this order, we should include corrections from the
operator \begin{equation}
L^{(2)}_{kin}=\sum_{\bf{\vec{p}}}\psi^\dagger_{\bf{\vec{p}}} (\frac{\bf{\vec{p}}^4}{8m^3})\psi_{\bf{\vec{p}}}~+~(\psi\rightarrow \chi).
\end{equation}
Thus the mass splitting is given by
\begin{equation}
\label{split}
\Delta E=\langle ^2\!S_{1/2} \!\mid L^{(2)}_{kin} +V_2 \mid \!^2\!S_{1/2} \rangle-\langle ^2\!P_{1/2} \mid  L^{(2)}_{kin}+{V_{so}
 \over m^2}\, \Lambda({\bf \vec{p}^\prime ,\vec{p}}) \mid \! ^2\!P_{1/2} \rangle.
\end{equation}
Calculating these matrix elements is nothing more than solving the
Coulomb problem to first order in perturbation theory, which we can
solve quantum mechanically\cite{TY}.  That is why it is simple to see
that $V_2$ does not contribute to the energy shift in the $P$ wave
since its wave function vanishes at the origin.  
 The degeneracy is only split once we consider the running of
$V_2$ and $V_\Lambda$.  $L^{(2)}_{kin}$ is not renormalized due to
reparameterization invariance but $V_2$ gets an anomalous dimensions
from the diagram shown in (\ref{lambanomdim}).
A short calculation yields the result
\begin{equation}
\gamma_{2}=\frac{2 \alpha}{3 \pi m_e^2} 
\end{equation}
where $V_c=-4\pi \alpha$.  Because we are looking at a one loop
diagram involving only $US$ loops the VRG reduces to the usual RG
(except of course for the factor of two arising from the fact
that $\mu_{US}=m\nu^2$) and we have
\begin{equation}
\nu \frac{d}{d\nu}V_2=\gamma_{2} 
\end{equation}
since in QED the coupling doesn't run below the scale $m_e$, the
solution to the RG equation is trivial
\begin{equation}
V_2(\nu=v)=\gamma_{2}V_c\ln(v)+V_2(1).
\end{equation}
We thus see that the series does indeed terminate. QED has a hard time
generating logs because the coupling freezes below the scale $m_e$.
At leading order the spin orbit potential has a vanishing anomalous
dimension, so the energy splitting can be determined
by taking the matrix element (\ref{split}), but this time renormalized
at the scale $\nu=v$, 
using
\begin{equation}
\langle V_2 \rangle =\gamma_2 V_c \ln(\nu) \mid \! \psi(0)\! \mid^2.
\end{equation}
and
\begin{equation}
 \mid \! \psi(0)\! \mid^2=\frac{(m \alpha)^3}{\pi n^3}
\end{equation}
we find the famous result
\begin{equation}
\Delta E= -\frac{m \alpha^5}{3 \pi}\ln(v),
\end{equation}
and we see that the series does indeed terminate at order $\alpha^5 \ln(v)$.
Using effective field theory we can go a step further and ask questions
about the next series of logs,
\begin{equation}
\alpha^5(\alpha \ln(v) +\alpha^2 \ln^2(v) + . . . )
\end{equation}
It turns out \cite{MSQED}, that this series terminates as well at order
$\alpha^8 \ln^3(v)$. The reason the series terminates, is due the fact that the coupling
in QED does not run below the scale of the electron mass. In QCD the series is not
so simple.
\subsection{States in the Effective Theory}
Before closing out this lecture I would like to emphasize an important point about the states
in any effective theory. If we wish to keep our ducks in a row we should be sure to
treat the states of the effective theory as  eigenstates of the {\it lowest} order
piece of the action. Power corrections to the states come from taking time ordered products
in the low energy matrix elements. This is exactly what we do in quantum mechanics. We perturb
around the lowest order wave functions. In this way we can keep track of various power law
corrections.  It is a simple exercise to see exactly how this works generically.
Suppose we are considering a correction to the expectation value for some transition operator
\begin{equation}
\mbox{\textsc {T}}=\int d^4y \langle \Phi_1(p_1) \mid T(O(0)H_1(y))\mid \Phi_2(p_2) \rangle
\end{equation}
where $\Phi_i$ are eigenstates of $H_0$.
The two time orderings correspond to  corrections to $\Phi_1$ and $\Phi_2$.
Let's consider the contribution  from $y_0<0$. Using the representation of the theta function
\begin{equation}
\theta(-y_0)=\frac{1}{i \pi}\int^\infty_{-\infty}d\tau\frac{ e^{-i y_0 \tau}}{\tau+i \epsilon}
\end{equation}  
and inserting a complete set of states yields
\begin{eqnarray}
\mbox{\textsc{ T}}&=&\sum_n \int d^4y\frac{1}{i \pi}\int^\infty_{\infty}d\tau e^{iy_0(p_2-p_n-\tau)} \langle \Phi_1(p_1) \mid O(0)\mid n^{(0)} \rangle \frac{\langle n^{(0)} \mid H_1(0)\mid \Phi_2 (p_2)\rangle }{E_2-E_n}.\nonumber \\
\end{eqnarray}
Then recalling from basic
time independent perturbation theory that the first correction ($\psi^{(1)}$) to the zeroth order  eigenstate  ($\psi^{0}$) of the full  Hamiltonian
$H=H_0+H_1$ is given by
\begin{equation}
\mid \psi^{(1)}\rangle =\sum_{n\neq 0} \mid n^{(0)} \rangle \frac{\langle n^{(0)} \mid H_1 \mid  \psi^{(0)} \rangle}{E_0-E_n}
\end{equation}
we explicitly see that the time ordered product accounts for corrections to the wave function.

Let me emphasize that there is nothing wrong with treating the physical states as eigenstates of the
exact Hamiltonian. This is often done in the literature. One often reads about
``higher Fock states" of a bound state. For instance in the case of NRQCD one often hears
the term ``dynamical gluons" as being part of a higher Fock state of the onium. 
 Again, there is nothing wrong with this, it just
goes against the spirit of effective field theory as espoused in these lectures.
But there are advantages to treating the states as lowest order eigenstates.
For instance, in the case of HQET the leading power corrections to the meson masses coming
from the operators 
\begin{equation}
O_1=\bar{h}_v\frac{\vec{D}^2}{2m_Q}h_v ~~~~O_2=\bar{h}_v\frac{g\sigma_{\mu \nu}G^{\mu \nu}}{4m_Q}h_v
\end{equation}
are cleanly separated from higher order power corrections. This would not be true if
we took the meson to be an eigenstate of the full Hamiltonian since the expectation values of
$O_1$ and $O_2$ would contain contributions of order $1/m^2$. Of course this would only be an issue if we were interested in $1/m^2 $ corrections.

\subsection{Summary}
In this second lecture I have introduced some new concepts into the game.
In particular, the idea that one can integrate out massless particles by
breaking up the field into various components. In the case of NRQCD
we split up the gluon field into four types of modes: hard $(m,m)$, though 
I didn't use this term previously. These are just the gluons which
we integrate out when we match at the scale $m$. 
Potential $(mv^2,mv)$, soft $(mv,mv)$ and ultra-soft $(mv^2,mv^2)$.
These last 3 modes are believed to be sufficient to correctly reproduce
all the non-analytic structure of non-relativistic Greens function.
To date this has only been proven up to two loops. Once we determined
the modes, we fixed their couplings by matching, although gauge invariance
and RPI fix many such couplings without doing any work. The way we arrived
at the proper mode decomposition was far from systematic. Indeed, finding
the proper modes is more of an art form at this point. The potential and
US modes fell out directly from matching, while we deduced the need for
the soft modes from physical argument. Of course had we carried out the
matching in the two quark sector we  would have immediately seen the
need for a soft mode. In particular, if you calculate the full QCD box
diagram you will see that just including the potential and the US modes
is not sufficient to reproduce the full result \cite{MoR}. 

There is no general rule for determining what modes are
necessary. However, there are some guiding principles. First we must
determine the scales involved in the external states of the theory.
In the case of NRQCD, those would be $mv$ and $mv^2$. Then there are some
obvious modes that one can pick out that are necessary. Those are
modes which leave the external lines on shell\footnote{By ``on-shell'' I mean its virtuality is on the same scale at the residual momentum.}. 
So for a fermion with
momentum $(mv,mv^2)$ we know that US and potential gluons should contribute 
to the non-analytic behavior of the Greens functions. 
After that, finding more modes starts to get harder 
and you have to do some matching
calculations to be sure. But its pretty clear that the other modes should
also be formed from these scales. So the soft $(mv,mv)$ modes are an
obvious choice. Since these modes throw quarks off-shell, its clear that
you will need more than one interaction vertex to generate an operator
in the effective theory. Then what about $(mv,mv^2)$ modes? Physically,
these modes would mediate interactions which are non-local in time
but local in space, which is clearly unphysical, and in fact a simple
calculation shows that such modes can't contribute. What about modes
with higher powers of $v$ such as $mv^3$? It is simple to see that
these modes would give zero since this scale can be dropped in all
propagator, and so loop integrals would be independent of the loop
variable itself. The interested reader should study SCET\cite{SCETI,SCETII} where
similar mode deconstructions are performed.

\section{Lecture III: Effective Theories and Extra Dimensions}
I would now like to discuss the application of effective field
theories to the  speculative topic of extra dimensions. The basic
idea of using extra dimensions to unify forces goes back to the early
part of the twentieth century with the work of Kaluza and Klein.  But
while extra dimensions have always played a role in string theory,
from a phenomenological stand point the topic was dormant until a few
years back.  The reason for this dormancy was that the size of the
extra dimensions was always assumed to be of order the Planck scale,
and therefore, the measurable effects of the extra dimensions are so suppressed that they are  irrelevant for all intents and purposes.
 However, with the advent of
D-brane ideas, it was realized that the extra dimensions can be of
arbitrarily large size (up to the experimental limits for gravity at
short distances\cite{will})  since string theory provides a mechanism for
field localization. That is, the usual problems involving 
observable effects of large
extra-dimensions are eliminated because the standard model fields live
on a fixed three dimensional space-like hyper-surface (the D-brane) and are thus
impervious to the extra dimensions.  This led to a flurry of model
building, including new solutions to the hierarchy problem. Many of
these models were discussed by other lecturers at this school\cite{Dienes,Csaki}. My
purpose here will be to treat these models from the point of view of
effective field theories. Please note that I will be abusing the term $D-brane$ throughout the rest of these lectures. 
You many substitute the term $D-brane$ with
a ``domain wall of Planck scale thickness'' for the purposes of these lectures. 

Let us consider a generic theory on a manifold $M$ which can be written as a direct product
\begin{equation} 
{\it M}=M_4\times M_d
\end{equation}
where $M_d$ is some $d$ dimensional compact manifold. We will assume for 
simplicity that there is only one relevant scale which characterizes $M_d$, 
namely $R$. So we would expect that for energies less than $1/R$, the effects
of the compact extra dimensions should be suppressed by $ER\ll 1$. 
Let us study a $4+d$ dimensional gauge theory on this manifold where the coupling
will have dimensions 
\begin{equation}
[g]=-d/2.
\end{equation}
 Furthermore, naturalness tells us that $g\simeq R^{d/2}$.
Since for $d>0$ the coupling is dimensionful, we would call this theory
``non-renormalizable'' in the old fashioned language. In our parlance we
would simply say that its only an effective field theory, with limited
predictive power. We would expect that this description of the theory would only correctly
describe the physics up to a scale of order $\Lambda \equiv R^{-1}$.  Typically the theory should contain 
a whole tower of higher dimensional operators which are suppressed by powers of $\Lambda$, which
all become equally important when we have external momenta on this scale.

\subsection{UV Fixed Points: A Theory Can Be More Than Effective}
Now before going on I 
should say that the above conclusions  need not necessarily hold. In particular,
there could be a strongly coupled fixed point, where our power 
counting would break down.  Indeed, as I previously mentioned, it is believed
that there are five and six dimension supersymmetric Yang-Mills theories with 16 supercharges
which posses non-trivial fixed points, although it is not known how to give Lagrangian descriptions of these theories.   Our argument in the previous paragraph seems to exclude this possibility.
So where is the loophole?
The answer, is that our power counting arguments were all based upon reasoning
which only necessarily holds at weak coupling. At strong coupling the power counting
can change. While its not clear how this happens in the higher dimensional SUSY theories,
there is a nice example 
in three dimensions \cite{BWP} which illustrates a mechanism by which naive power counting breaks down.

 Consider the theory of $N$ fermions in three dimensions
interacting via a four Fermi interaction
\begin{equation}
\label{2+1}
L=\bar{\psi}_i i\partial \!\!\!\! \slash \psi_i+ \frac{g^2}{2 N}(\bar{\psi}_i \psi_i)^2,
\end{equation}
In three dimensions the coupling $g^2/N$ has mass dimensions $-1$. So naively
this theory is only an effective theory and breaks down at the scale
$N/g^2$. However, it has been shown that in the large $N$ limit this
is not the case. The conclusion that the theory is non-renormalizable
was deduced from a weak coupling analysis. We know that quantum corrections
can lead to scaling modifications. Typically, we study weakly coupled theory,
where the quantum corrections only change the power counting by logs. 
But in strong coupling it can happen that this naive power counting breaks
down, and a theory which we thought was non-renormalizable can indeed
be renormalizable. This is exactly what happens if you study (\ref{2+1}) in the
large $N$ limit. The way this comes about is seen by re-writing (\ref{2+1}) in terms of an auxiliary scalar field
\begin{equation}
\label{aux}
L=\bar{\psi}_i i\partial \!\!\!\! \slash \psi_i+\sigma \bar{\psi}_i\psi_i-\frac{N}{2 g^2}\sigma^2.
\end{equation}
Now naively, the $\sigma$ propagator is constant.
However, in the large $N$ limit we are forced to resum sets of
diagrams, and in doing so the $\sigma$ propagator scales as
$1/p$, if the bare coupling satisfies a certain bound. Given this scaling
it is not hard, using textbook arguments, to show that all the divergences
can be absorbed using only  the operators shown in (\ref{aux}).
This means that we can naturally set all the higher dimension operators in the
action to zero without worrying that we will regenerate them via quantum corrections.

The point of this little diversion was just to let the reader know that
sometimes field theories can actually be more than just effective, they may
be complete. The trouble is that the strong coupling, which is necessary to generate large
anomalous dimensions, makes it difficult to perform controlled approximations. So finding strongly
coupled fixed points is a difficult task. I should also point out, that even without such
a fixed point, there could well be a field theoretic description of the theory above
the cut-off in terms of a different set of variables. For instance, the underlying theory
could consist of a quiver gauge theory \cite{Cohen}.
\subsection{Naive Dimensional Analysis}
Returning now to our non-renormalizable effective gauge theory, 
it actually turns out that there are  simple arguments which lead to the
conclusion that the scale at which the  effective theory  breaks down is numerically enhanced relative to the compactification scale by a geometrical factor. These arguments are based on what is 
called ``naive dimensional analysis" which was first discussed in the context of chiral
perturbation theory \cite{MG}. A discussion of NDA within the context of super-symmetric theories can be found in \cite{NDASUSY}. The basic argument goes as follows.
Suppose we wish to calculate some n-point function ($G^{(n)}_i$) in a non-renormalizable field theory.  Generically
there will be a tree level contribution from some unknown counter-term ($C_i$), as well as some calculable (in terms of lower order coefficients) non-analytic terms
which arise from loop corrections.  Both contributions will be
suppressed by powers of the UV
scale $\Lambda$. Furthermore, dimensional analysis tells us
that there is a correlation between the order in the loop expansion at which the observable
is generated and the dimensionality of the operator which contributes to the observable at tree
level.

 Schematically the amplitude $M$ can be written as
\begin{equation}
M=(p^{2n}/\Lambda^{2n})\Big(C_i(\mu)+( C(\frac{1}{\epsilon}+\ln(p^2/\mu^2))+\kappa)\Big)
\end{equation}
Here the piece proportional to $C_i$ is the contribution from some local operator, while the
log piece comes from a loop integral involving lower order interactions whose couplings are related to $C$ by a geometric factor generated by the loop integration ($\rho$). That is, if $C_0$ is the lower
order coupling then $C\simeq C_0^2 \rho$.
The divergence will get absorbed into $C_i$. The momentum $p$ represents some generic set of
external momenta.
Varying the renormalization scale on a interval of order one generates an effective change 
in result of order $C (p^{2n}/\Lambda^{2n})$.
This is roughly  equivalent to the statement that even if we set $C_i$ to zero at tree level,  it will be
generated at the loop level.
Thus, we would expect that the size of the counter-term
is thus naturally of order $Cp^{2n}/\Lambda^{2n}$. Now given that the log arises from a loop, 
$\rho$ is just a geometric factor which comes about from the angular integration and is of the form $1/(\gamma \pi^r)$ where $\gamma$ and $r$ are a function of the number of dimensions. Thus NDA tells us that the cut-off is more like
$\rho^{-1} \Lambda$ instead of $\Lambda$. While this does not seem like a huge distinction, the chiral Lagrangian description of low energy QCD would
be far less powerful if it
were no for this factor of $\rho=4 \pi$. It is important to realize that NDA is a naturalness argument and assumes that couplings are all of order one.
As such, it  should only be taken as a guiding principle.

Let us now see how this works for a $4+d$ dimensional Yang-Mills theory.
The leading interaction is suppressed by $\Lambda^{-d/2}$, so that any  loop contribution to the
effective action will be suppressed by $\Lambda^d$. Following the argument in the previous paragraph, let us compare the tree level contribution to the photon two point function
arising from the higher dimensional operator
\begin{equation}
\label{O_2}
O_2=\frac{C_2}{\Lambda^2}F_{AB}\partial^2 F^{AB}
\end{equation}
to the contribution arising from loop corrections. 
 The one loop correction to the two point function arising from interaction with a massless field is just the usual vacuum polarization
 graph, which in $4+d$ dimensions is given by
 \begin{equation}
 \pi  (q^2)\propto \frac{g^2}{(4\pi)^{2+d/2}} (-q^2)^{d/2}\Gamma[-d/2].
 \end{equation} 
 where, as usual $\pi^{\mu \nu}(q^2)=(q^2 g^{\mu \nu}-q^\mu q^\nu)\pi(q^2)$.
 First we notice that the result only diverges in even dimensions. We could  have
 guessed that since in odd dimensions there are no allowed (i.e. analytic) counter-terms which could
 absorb the divergence. So in odd dimensions we cannot use this particular calculation to learn anything from NDA. In even dimensions, e.g. $d=2$, we would conclude that the true cut-off is not $g^{-1}$ but
$(g(4\pi)^{3/2})^{-1}$. Notice that I didn't worry about the overall constants in 
(\ref{O_2}). The full result  could include group theory factors or  gamma functions which might arise 
from doing Feynman parameter integrals. However, as I said before, NDA is only supposed
to give you a rough guess of the true cut-off. I would not wager much on its exactness, but
it does make qualitative sense that the cut-off will in general be enhanced by a geometric
factor, which grows with the dimensionality. 
\vskip.3in
{\bf Exercise 3.1} Using a physical argument determine whether one would expect
the NDA guess for the strong coupling scale to change due to compactification.
\vskip.3in
 
\subsection{The force law in $d$ dimensional Yang Mills Theory}
It is interesting to try to understand the force law in a compactified version of a $d$ dimensional
Yang-Mills theory. Let us calculate $V(r)$ a la exercise 1.6 for a five dimensional gauge
theory compactified on an $S^1$ with radius $R$. If we KK reduce the theory (I am assuming here that the reader is familiar with KK reduction, otherwise see \cite{Csaki}), then the force will
get contributions from zero mode as well as KK mode exchange between our heavy sources.
For distances much larger than $R$ we would expect that only the zero mode force is non-negligible since the KK modes, whose masses scale with $n/R$, mediate a Yukawa force which dies off exponentially fast $e^{-rn/R}$. Thus, in this regime,  we reproduce the four dimensional result, with a rescaled   coupling $g/\sqrt{R}$. The rescaling arises from the
the norm of the zero mode.
On the other hand, for $r\ll R$, we can consider a whole tower of effectively massless $(R/n\ll r)$ KK modes each of which will contribute to the force.
Thus, qualitatively we would expect the following behavior for the one loop force law at tree level
\begin{eqnarray}
\label{DA}
r\gg R&:& ~~~~~~~~V(r)\propto \frac{g^2}{R r}
\nonumber \\
r\ll R&:& ~~~~~~~~V(r)\propto \frac{g^2}{ r^2}
\end{eqnarray}
How will this result be modified by quantum corrections?
At large $r\gg R$ we know that we should be dominated by external zero modes as well
as zero modes in the loop.  Whereas at small $r\ll R$ we will get contributions not only
from corrections to zero mode exchange but also from corrections to the exchange of external
KK modes as well. 
%

At large $r$ the dominant correction will arise from the pure zero
mode vacuum polarization diagram which   generates a correction of the form
\begin{equation}
\delta V(r)\propto \frac{g^4}{r R^2}\ln(r),
\end{equation}
whereas dimensional analysis tells us that the correction to the full five
dimensional potential, at small $r$, will be of the form
\begin{equation}
\delta V(r) \propto \frac{g^4}{r^3},
\label{5dQM}
\end{equation}
so that at small $r$ the quantum corrections dominate the classical result as
per our previous discussion.

 Now let's do an honest calculation to see what the exact result looks
like. Instead of calculating the full five dimensional potential, I will concentrate
on the potential generated by external zero modes and its correction.
The reason being that this will be the relevant object to calculate when we discuss grand unification. I will do the calculation in two ways. First by working exactly in five dimensions, and then
by matching and running, as discussed in the first lecture.

I  will specialize to the case of scalar QED for simplicity. In this case there are
two contributions to the vacuum polarization. The result for a massless scalar can be written as \cite{GnRII}
\begin{eqnarray}
\pi_{\mu \nu}(p^2)=\frac{g^2}{R}\sum_n \int
\frac{d^dk}{(2\pi)^4}\frac{1}
{(k^2-m_n^2)}\left[ \frac{(2k+p)_\mu (2k+p)_\nu}{(p+k)^2-m_n^2}-2g_{\mu \nu}\right]\equiv (p_\mu p_\nu-p^2 g_{\mu \nu})\pi(p^2).
\nonumber \\
\end{eqnarray}
\begin{equation}
\label{S1}
\pi(p^2)=\frac{g^2}{R}\left[\frac{\sqrt{p^2} R}{256}-\frac{1}{8\pi^2}\int_0^1x dx \sqrt{1-x^2}\ln(1-e^{-\pi x R \sqrt{ p^2 }})\right].
\end{equation}
\vskip.3in
{\bf Exercise 3.2}: Prove equation ({\ref{S1}})
\vskip.3in
 A few points are worth noticing here. First off the result is finite. This is because dimensional
 regularization sets the power divergence to zero. 
 This finiteness may seem strange given that each mode gives a divergent contribution.
 What happens is that the sum of the poles cancels.
Also, we would expect that at long distances we
 should exactly reproduce the four dimensional non-analytic behavior. So if we expand for
 small values of the momentum we should get a log and nothing else non-analytic.
Defining $\delta=R \sqrt{ p^2 }$, and expanding the log we find
\begin{equation}
\ln(1-e^\delta)=-\frac{\delta}{2}+\ln(\delta)+a_1 \delta^2+a_2 \delta^4 +....
\end{equation}
where the rest of the series contains only even powers of $\delta$. The first term in the expansion
cancels the power dependence, the second term is the 4-D log and the higher order terms are 
analytic. At $pR\gg 1$, the result is  dominated by the power dependence and taking the Fourier transform of the first term in the Dyson series we get back the radius independent result we would expect from dimensional analysis (\ref{5dQM}). We can also see from dimensional analysis that
in the regime  $pR\gg 1$ the theory becomes strongly coupled as all loops will be of the same order. 

We should be able to reproduce the force law  $p$ dependence by the running and matching procedure. For long distances $p<<R^{-1}$, all of the effects of the KK modes can be absorbed
into local counter-terms and we simply reproduce the four dimensional result from the zero mode contribution.
The higher KK modes just go into making up the low energy (measured) coupling.
It is interesting though to see how we can reproduce the short distance behavior as well.
If we are interested in the two point function at the scale $p$, then we should sum over all
the KK modes with masses less than that scale. Each mode will contribute a $log(p/\mu)$ along with
some scheme dependent constant. It seems clear how the linear dependence on $p$ will arise
from summing over the constant pieces, but how does the $log(p)$ dependence becomes power-like?

Suppose we have measured the coupling  at some scale below the first KK mass $M_1$. To predict
the coupling at $P\simeq M_1$, we run the coupling to the scale $M_1$ using the four dimensional
beta function, then we match, at the scale $M_1$, onto a theory which contains the zero mode as well as the first KK mode. In fact, let's generalize this procedure. Suppose we know the value of
the coupling at the scale $M_n$ and we wish to predict its value at some higher scale. We just match in steps as we discussed in the first lecture. Since we are only interested in the log piece
let us ignore additive constants in the matching (even though in this particular case the constants are of the same order as the logs, so this calculation does not follow the standard systematics). Thus, at the order we are
working we may use the relation
\begin{equation}
\alpha(\mu)=\frac{\alpha(\mu_0)}{1-\beta_0 n\alpha(\mu_0)\ln(\mu/\mu_0)},
\end{equation}
where $\beta_0$ is the beta function for one massless particle, and $n$ counts the number
of massless particles in the theory \footnote{Remember once we are
above threshold for a given particle it is treated as massless in the
running. The mass effects  are power suppressed.}.
So the value of the coupling in a theory with $n$ KK modes, $\alpha_n$, at the scale $M_{n+1}$ is given
by
\begin{equation}
\alpha_n(M_{n+1})=\alpha_n(M_n)+n \beta_0\alpha_n^2(M_n)\ln(\frac{n+1}{n})
\end{equation}
where I have used the fact that $M_n\propto n$ and dropped higher order terms in the couplings
since $\ln(\frac{n+1}{n})$ is not a large log. Now we match onto a theory with $n+1$ KK modes with
a coupling we label $\alpha_{n+1}$. To determine $\alpha_{n+1}$, we equate the potentials
of the two theories
\begin{equation}
V_n(M_{n+1})=V_{n+1}({M_{n+1}})
\end{equation}
Note that, in general, this relation does $\it{not}$ mean we equate the couplings.

Now we may use this result to derive the value of the coupling at the scale $M_n$ in terms
of $\alpha(1/R)$.
\begin{equation}
\alpha(n/R)=\alpha(1/R)+\sum_{n=1}^N\alpha^2(1/R)n\beta_0 \ln(\frac{n+1}{n}),
\end{equation}
taking $N\gg 1$ we have
\begin{equation}
\alpha(n/R)\approx \alpha(1/R)+\alpha^2(1/R)\beta_0 N
\end{equation}
Then using 
\begin{equation}\label{match}
N\simeq pR
\end{equation}
 we see that we have reproduced the power behavior we found when doing the
full five dimensional calculation (\ref{S1}). But we also notice that the result is linearly sensitive to the matching scale. That is, an order one  variation in the matching scale 
(\ref{match}) changes the result by an amount of order one. Usually such a variation would lead to a subleading correction, but in this case  the power law behavior leads to
linear sensitivity to the UV physics. 

\subsection{Grand Unification}
One of the first uses of effective field theories was in the context of grand unification
\cite{WeinbergII,Hall}. In the standard model the couplings approach each other near the
scale $10^{15-16}$ GeV, and in the minimal super-symmetric standard model (MSSM) the coupling
are within errors of unifying. What unification of the couplings really means is that in the context of a GUT we get the correct post-diction for the weak mixing angle $\sin^2\theta_W$.
This post-diction is a compelling reason to believe that there
is a desert of physics between the weak scale and the GUT scale\footnote{ Another hint that this might be the correct scenario is the long life time of the proton. Since  higher dimensional baryon number violating operators need to be sufficiently suppressed. This is an additional  challenge to extra dimensional model builders.}. This desert would
seem to exclude the existence of any extra dimensions whose scale is less than the usual ($10^{16}$ GeV) GUT scale.

It is thus interesting see whether or not we can retain the grand unified prediction for
$\sin^2\theta_W$ within the confines of a theory of ``large" ($R^{-1}\simeq$ TeV) extra dimensions \cite{DDG}.
Let us see what a compactified five dimensional GUT would predict for $\sin^2\theta_W$.
I will discuss this calculation in the context of a simplified Abelian model. 

Suppose we have a unified 5-D theory with gauge group $U(1)_1\times U(1)_2 \times U(1)_3$,  which is compactified on a circle of radius $R$. We will furthermore impose a permutation symmetry to 
insure that the bare couplings for the three gauge groups are equal.
We will assume that each gauge group has a matter content composed of two scalars.
We will choose the first gauge group to have all its scalars massless. The second group will have one massless and one of mass $M_2$, while the third group will have all its matter content of mass $M_3$. All the masses for the scalars will be of the same order which we will call $M_{GUT}$. These masses will mimic the effect of GUT symmetry breaking.
For instance, they could be generated by choosing a Higgs potential in an appropriate way.
Well above the scale of the masses, the values of the two point functions will be identical.

Let us first review how the prediction for $\sin^2\theta_W$ comes about in four dimensions in this
model.
Suppose that we've measured the low energy (say at the scale $M_Z$) values of two of the couplings, $g_1(M_Z)$ and $g_2(M_Z)$.
For each gauge group we may write down a relation between the
$\overline{MS}$ coupling, $\bar{g}_i$, and the physical coupling, $g(q)_i$,  (the
object sitting in front of the $F^2$ piece of the effective action)
 \begin{eqnarray}
\frac{1}{g_1^2(q)}&=&\frac{1}{\bar{g}_1^2(\mu)}+ \frac{\beta_0}{\pi^2}
\ln(q/\mu) \\
\frac{1}{g_2^2(q)}&=&\frac{1}{\bar{g}^{2}_2(\mu)}+ \frac{\beta_0}{2\pi^2}
\ln(q/\mu)+ \frac{\beta_0}{2\pi^2} \ln(M_2/\mu) \\
\frac{1}{g_3^2(q)}&=&\frac{1}{\bar{g}^{2}_3(\mu)}+ \frac{\beta_0}{\pi^2}
\ln(M_3/\mu). 
\end{eqnarray}
In writing these results I have ignored constants of order one, as they
are subleading in the leading log approximation. Furthermore, I have
dropped the $q$ dependence in the contributions from massive scalars,
assuming that $q<M_{GUT}$. When $q$ is order $M_{GUT}$ this log will
be small and sub-leading, so the $q$ dependence will be irrelevant.
We have at our disposal two measurements as well as the relation
$g_1(q)=g_2(q)=g_3(q)$ for $q\geq M_{GUT}$. Naively, this is not
enough
information to make any predictions given the number of unknowns.
However, it is the magic of the log
that will save the day. First we solve for $\bar{g}_1$ and $\bar{g_2}$ in
terms  of the measured couplings
\begin{eqnarray}
\frac{1}{g_1^2(q)}&=&\frac{1}{g_1^2(M_Z)}+ \frac{\beta_0}{\pi^2}
\ln(q/M_Z)\\
\frac{1}{g_2^2(q)}&=&\frac{1}{g_2^2(M_Z)}+ \frac{\beta_0}{2\pi^2}
\ln(q/M_Z)  
\end{eqnarray}
Using this result we may solve for
$M_{GUT}$
\begin{equation}
M_{GUT}=M_Z
\exp\left(
\frac{2\pi^2}{\beta_0} 
\left[ \frac{1}{g_1(M_Z^2)}
-\frac{1}{g_1(M_Z^2)} \right] 
\right).
\end{equation} 
Now $g_3$ is frozen at the GUT scale, so we can determine its value by
solving for $g_1(M_{GUT})$
\begin{equation}
\frac{1}{g_3^2}=\frac{1}{g_1^2(M_Z)}+ \frac{\beta_0}{\pi^2}
\ln(M_{GUT}/M_Z).
\end{equation}
Notice that in making this prediction it is crucial that $M_3 \simeq
M_{GUT}$, otherwise $g_3$ would run between those two scales.
Perhaps more importantly the prediction relies upon the fact that
the sensitivity to the threshold mass $M_3$ is only logarithmic, so
that even if $M_3$ differs from $M_{GUT}$ by a number of order one,
the running between these two scales is negligible.
If we only consider {\bf{flat}} manifolds), then this property is
unique to four  dimensions.

{\bf Exercise 3.3:} This calculation was done in the context of a
non-decoupling scheme. Redo the calculation in terms of an effective
theory calculation by matching and running to each threshold.

Let us now see how this prediction runs into trouble in five
dimensions.  Notice that the couplings do not ``run" in the five dimensional theory since there are no log divergences.  That is, there is no
dimensional transmutation, the theory was not conformal classically.  Thus it is always true that $g_1=g_2=g_3$,
but is not true that the force between two charges will be the same
for all the gauge groups at all distances. Thus instead of a running
coupling I will define an ``effective coupling", $g_i^{eff}$ which
will represent the strength of the force.  Calculationally $g_i^{eff}$
is sum of the bare coupling $g_i$ plus radiative corrections.  We will
assume that we have compactified on a circle of radius $R$, and that
the symmetry breaking masses are such that $M_iR\geq 1$. For smaller
masses we would expect that we would just reproduce the four
dimensional log running which we know gets the answer for the weak angle wrong when the
GUT scale is near a TeV.  Consider the prediction
for $\sin^2\theta_W$ in this model. Assume we have measured the force
between two charges at some low scale $r^{-1}=M_Z$ for $g_1$ and
$g_2$.  For the group with only massless fields in the loops the
strength of the force at the scale $E=M_Z$ is
\begin{equation}
\frac{1}{(g_1^{eff})^2}=\frac{1}{g_1^2}+\frac{\beta_0}{\pi^2} \ln(2\pi M_Z R),
\end{equation}
This result is what we would expect. At low energies the corrections to the potential are
governed by loops of zero modes of massless fields and we get back the four dimensional result. For the second gauge
group we get
\begin{equation}
\frac{1}{(g_2^{eff})^2}=\frac{1}{g_2^2} +\frac{\beta_0}{2\pi^2} \ln(2\pi M_Z R)+C_2\frac{M_2 R}{\pi}.
\end{equation}
The term proportional to $\beta_0$ is the usual massless result. The second term arises from the massive scalar field. 
Using the same logic as in the four dimensional case, we would like to predict $\frac{1}{(g_3^{eff})^2}$ via
\begin{equation}
\frac{1}{(g_3^{eff})^2}=\frac{1}{g_3^2} +C_2\frac{2 M_3 R}{\pi}.
\end{equation}
In these equations I have assumed that the power like piece dominates in (150), otherwise we would just reproduce the
four dimensional result.
We can see that when we try to solve for $R$ using the first
two equations, we won't be able to drop the term which goes as $M_2 R$,
since it's not small compared to the log term.  There are more
unknowns than equations and we have no predictive power. Notice that
this shortcoming is not related to calculability, but to the UV
sensitivity of the result. Even if you knew the ``beta function"
exactly it would not matter, you still would need to {\it know} the
values of the masses of the GUT particles in order  to accurately predict  the low
energy value of $\sin^2\theta_W$.
\subsection{Orbifolding}
As I mentioned at the beginning of this lecture, the phenomenological interest in extra dimensional theories
was resurrected when it was realized that the D-brane techniques discovered in string theory \cite{Pol}
could be  useful for model building. The world volume theory of a D-brane can be described by a gauge theory which lives on the
brane. Thus, D-branes allow for a natural way of ``trapping" fields onto a surface (these fields won't have KK modes). In this
way one can make the space as large as one wants without worrying about introducing phenomenologically meddlesome light KK modes. However, we are not free to restrict gravity to
the brane, since by general covariance, all stress-energy couples to the graviton, and thus
gravity must propagate in the ``bulk" of space-time. 
Let us explore the effective field theories of
theories with D-branes. 

If all the fields of our theory are trapped to the D-brane, then there really  isn't anything interesting going on, since
the theory is just a normal four dimensional theory up to the scale of quantum gravity\footnote{I will assume here that we are talking about $D3$ branes which have four dimensional world volumes.}. So to make things interesting we will allow at least some of our fields to traverse
the bulk, and discuss the description of the theory in terms of an effective field theory \cite{Sundrum}.
Let us assume, for the moment, we are working in a flat spacetime \footnote{In general the brane tension itself will
curve the space, but we will assume that the tension is small enough so as not to
significantly perturb the theory away from Minkowski space.}. Then in order to keep the KK modes
sufficiently massive to conform to accelerator limits, we must also compactify the space with a radius at least as big as a $TeV$.
The scenario I shall consider here is a compactification on the five dimensional manifold
\begin{equation}
M=M^4\times S^1/Z_2.
\end{equation}
Where the $Z_2$ defines an equivalence class via a reflection across the $S^1$.
Note that $Z_2$ acts (K) non-trivially on both the coordinates and the fields
\begin{eqnarray}
K&:&~~~~~(x,y)\rightarrow (x,-y)\\
K&:&~~~~~\Phi \rightarrow R(k) \Phi
\end{eqnarray}
where $R(k)$ furnishes a representation of the $Z_2$. I have chosen the circle to have radius $R$ and coordinatized it in the
 two dimensional plane $(x,y)$. The points $y=0$ and $y=\pi R$ are {\it fixed points } meaning they are left invariant under the action of the $Z_2$ and  are singular. The Einstein equations require that we place a singular sources of stress energy at those points, so that is where we must put 
a pair of D-branes. Such singular spaces are technically not manifolds, and are instead called
orbifolds. The thickness of the D-branes are of order the string scale, so as far as our effective field theory description is concerned, they are infinitely thin in the limit where $E/M_s$ goes to
zero.

How would the force law in this space differ from the case of the pure
 $S^1$ we previously discussed? We would expect that at long distances the force laws should also  be identical, as long as the massless spectra are the same. Suppose we allow for our charged scalar to propagate into the bulk, as in the previous
case. There is a simple way to use our result for $S^1$ to calculate the vacuum polarization 
on $S^1/Z_2$. First we must decide how the bulk fields transform under the $Z_2$. If we want the massless spectrum to be the same then we must choose the field to be even under the $Z_2$ otherwise
the theory would posses no zero mode. We then KK decompose the field in the usual way
\begin{equation}
\Phi=\sum_{n=-\infty}^{\infty} \Phi_{(n)}(x)\frac{e^{\frac{i n z}{R}}}{\sqrt{2 \pi R}},
\end{equation}
and $0 \geq z \leq \pi$ coordinatizes the fifth dimension. 
We then further decompose the field as
\begin{equation}
\Phi=\Phi_+ +i \Phi_-
\end{equation}
where
\begin{eqnarray}
\Phi_+^{(n)}=\frac{1}{\sqrt{2\pi R}} (\Phi_{(n)}(x)+\Phi_{(-n)}(x))\cos\left( \frac{n z}{R}\right) \nonumber \\
\Phi_-^{(n)}=\frac{1}{\sqrt{2\pi R}} (\Phi_{(n)}(x)-\Phi_{(-n)}(x))\sin\left( \frac{n z}{R}\right)
\end{eqnarray}
Then on $S_1$ the action can be written as
\begin{equation}
S=(\partial_\mu \Phi_+^{(0)})^2+\sum_{n=1}^\infty \int d^4x \left[ (\partial_\mu \Phi_+^{(n)})^2 +(\partial_\mu \Phi_-^{(n)})^2-
\frac{n^2}{R^2}\left((\Phi_+^{(n)})^2 +( \Phi_-^{(n)})^2\right)\right].
\end{equation}
Orbifolding by $Z_2$ eliminates the $\Phi_-$ component so that we may write
\begin{equation}
\pi_{S^1}=\pi_{S^1/Z_2}+\pi^{(-)},
\end{equation}
where $\pi^{(-)}$ is the contribution from $\Phi_-$ . Recall that in the case of $S_1$, we found a finite result, and we concluded that the UV divergence from the zero mode must have cancelled with the pole from the $KK$ sum. But since $\pi^{(-)}=\overline{\pi}^{(+)}$ (where the bar indicates that we don't include the zero mode)  we can see that for
the orbifold we  no longer expect a finite result since
\begin{equation}
\pi_{S^1/Z_2}=\pi_{S^1}-\frac{1}{2}(\overline{\pi}^{(+)}+\pi^{(-)})=\frac{1}{2}\pi_{S^1}+\frac{1}{2}\pi^0,
\end{equation}
and we know that the KK mode sum $(\overline{\pi}^{(+)}+\pi^{(-)})$ is divergent, since its divergent piece is minus the zero mode divergence.

 At first this may seems rather mysterious given that we would not expect that a non-local operation like a change in boundary conditions
could affect a local structure like a UV counter-term. How can the local bulk physics be sensitive to what's going on at the boundaries? The answer is, it is not, the divergence corresponds to a local
divergence on the boundary. Introducing the boundary allows for the existence of new counter-terms, since we can now write down localized interactions on the boundary,
\begin{equation}
S_B=\int d^5x (\delta(z)L_0+\delta(z-R) L_R)
\end{equation} 
where $L_0$ and $L_R$ are the most general actions allowed by the symmetries of the theory.

Returning to our calculation, we find 
\begin{equation}
\label{S1Z2}
\pi_{S^1/Z_2}
\approx \frac{1}{2}\pi_{S^1}+\frac{1}{48 \pi^2}(\frac{1}{\bar{\epsilon}}+C)
-\frac{1}{96\pi^2}\ln(q^2/\mu^2).
\end{equation}
As I said previously we would expect that at long distances one cannot distinguish between
$S^1$ and $S^1/Z_2$, so  we can immediately read off the $q$ dependence of $
\pi_{S^1/Z_2}$ from (\ref{S1Z2}), when $qR \ll 1$ without calculation
\begin{equation}
{\pi}(q^2)_{S^1/Z_2}={1 \over 2}{\pi}(q^2)_{S^1}-\frac{1}{96 \pi^2}\ln(q^2/\mu^2)=-\frac{1}{\\48 \pi^2}\ln(q^2/\mu^2),
\end{equation}
and in fact this agrees with the expansion of the result we found previously in (\ref{S1}).
Note again that we do not expect constants to agree as these are local short distance contributions. The $\frac{1}{\epsilon}$ divergence can be subtracted by putting
a $FF$ counter-term on the boundary.

Before closing this section I would like to discuss a common mistake you might find in the
literature. Suppose we have a theory on $S^1/Z_2$, where we have a certain set of fields ($\Phi_0$)
which are localized to a boundary at one boundary at $z=0$,    another set $\Phi_R$ on the second boundary which we will place at $z=R$,  and finally yet another set which propagate in the bulk $\Phi_B$.
At low energies we may write down an effective field theory consisting of the low energy modes of $\Phi_0$ and $\Phi_R$ and a few ($N_{KK}$) of the low lying KK modes of the bulk fields \footnote{The number of modes we choose to keep is fixed by the maximal energy allowed in the theory.}, such
that 
\begin{equation}
S^{EFT}=\int d^4 x L^{EFT}=\int d^4x \sum_{n=0}^{N_{KK}}(L(\Phi_0)+L(\Phi_R)+L(\Phi_B^{n})).
\end{equation}
At this point I have not allowed for interactions between the various fields \footnote{ Strictly speaking
this isn't possible since they must interact at least gravitationally given that the fields
are sources of stress energy. Such interactions terms will be described in the effective
theory via higher dimensional (i.e. $d>4$) operators which are suppressed by the Planck scale.}. Whose effective theory is this? An observer in the bulk? Or on one of the
boundaries? The answer is that the question is not well posed. At low energies we cannot
probe distances short enough to talk about positions in the fifth dimension. The action
correctly describes the physics for all the observers. The Lagrangian will correctly describe
all low energy experiments. However, the possible experiments which can be performed will
depend upon the location of the observer, because an observer on the brane at $z=R$ will
not have access to a beam of particles which are interpolated by $\Phi_0$ and visa versa.
In fact, matter on the brane at $z=R$  will just look like dark matter for an observer 
on the brane at $z=0$. 

\subsection{Classical Divergences as UV Ignorance}
Let me now utilize an extra dimensional theory to emphasize one of the key ideas that
I have tried to convey in these lectures. The UV divergences we find in field theory
calculations are not exotic, nor do they imply the theory is sick. They just tell
us that our theory has limited predictive power, because they do not correctly describe
the UV. To hammer this point home, I would now like to give an example where the
UV divergences show up at the classical level\cite{GoW,GG}. Suppose I have a six dimensional
theory with a 3-brane located at the origin, around which I impose a $Z_2$ symmetry, so that the space is $M^4 \times R^2/Z_2.$
Strictly speaking for arbitrary brane tension, the true geometry is conical \cite{DtHJ} and
not a flat orbifold, but I can always fine tune the tension so as to get a solution to
Einstein's equation which yields a $Z_2$ flat orbifold.
To simplify matters I will consider a simple theory of a real scalar field in Euclidean space such that
\begin{equation}
S=\int d^6x \left(\frac{1}{2}(\partial_A \phi)^2+\frac{1}{2}m^2 \phi^2+\lambda \phi^4 +\delta^{(2)}(x)\frac{1}{2}m_b^2\phi^2\right).
\end{equation}
where for simplicity I have only included a mass term on the brane. 
Now let's calculate the two point function for $\phi$. Ignoring for the moment the
brane mass term we find
\begin{equation}
\label{prop}
G_0(x,x^\prime)=\int \frac{d^4 k}{(2\pi)^4}\int \frac{d^2q}{(2\pi)^2}e^{ik\cdot(x-x^\prime)}\frac{\left(e^{ikq\cdot(y-y^\prime)}+
e^{ikq\cdot(y+y^\prime)}\right)}{k^2+q^2}
\end{equation}

\vskip.3in
{\bf Exercise 3.4} Prove equation (\ref{prop}).
\vskip.3in
To calculate the complete two point function we include insertions of the brane mass so that 
\begin{eqnarray}
G(x,x^\prime)&=&G_0(x,x^\prime)-m_b^2 \int d^6z G_0(x,z)\delta^2(z)G_0(z,x^\prime)\nonumber \\&+& m_b^4 \int d^6z d^6z^\prime G(x,z)\delta^2(z)G(z,z^\prime)\delta^2(z^\prime)G(z^\prime,x^\prime)+...
\end{eqnarray}
Since we have four dimensional Poincare symmetry it is simpler to work in the
mixed momentum configuration space representation of the Greens function $G_k(w)$, where
$w$ are the transverse coordinates. Then

\begin{equation}
G_k(w,w^\prime)=G^0_k(w,w^\prime)-m_b^2 G^0_k(w,0)G^0_k(0,w^\prime)+m_b^4 G^0_k(w,0)G^0_k(0,0)G_k^0(0,w^\prime)+....
\end{equation}
Notice that $G^0_k(0,0)$ is divergent.  This divergence arises as a consequence of the
fact that we have approximated the brane as being infinitely thin. But we know this is not
a problem, we simply parameterize our ignorance of the correct brane structure in a local
counter-term.

\vskip.3in
{\bf Exercise 3.5:} Show that in the $\overline{MS}$ scheme the beta function for $m_b$ is given by 
\begin{equation}
\mu \frac{ d m_b^2}{d\mu}=\frac{m_b^2}{\pi}.
\end{equation}
\vskip.1in

 Suppose we knew the true structure of the brane, and that it can be described by some profile $F(w)$, which in the limit where the UV scale (which well call the Planck mass) goes to infinity approaches a delta function,
\begin{equation}
\lim_{M_{Pl}\rightarrow \infty}F(w)=\frac{1}{2}m_b^2 \delta^2(w)
\end{equation}
This information can then be encoded in the action
\begin{equation}
S=\int d^6x \left(\frac{1}{2}(\partial_A \phi)^2+\frac{1}{2}m^2 \phi^2+\lambda \phi^4 +F(w) \phi^2\right).
\end{equation}
Then the divergent integrals gets damped out by $F$ and the result is finite.
We have one less counter-term to fix and thus increased predictive power, but only
because we claim to know the underlying theory. 
Of course, there is an even simpler example of UV divergences arising classically.
The total energy in the Coulomb field of a charge particle diverges when the electron is
treated as point-like.

\subsection{Effective Field Theory in Curved Space}
Adding curvature to the space-time manifold forces one to carefully inspect many of our starting assumptions in building a quantum field theory. For instance, even  the notion of what is a particle 
must be reconsidered \cite{BD} since if the space-time does not admit a  time-like killing vector there is no notion of a conserved energy. However, things simplify considerably  as the space-time becomes more symmetric. So we will limit our discussion to a specific set of theories
which posses four dimensional Poincare symmetry. These theories have preferred vacua
around which to quantize. Such space-times have metrics which can be written as
\begin{equation}
ds^2=dz^2+g(z)\eta_{\mu \nu}dx^\mu dx^\nu,
\end{equation}
for simplicity I've chosen to work in five dimensions, generalization
to higher dimensional spaces is straightforward. Such space-times are
often called ``warped".  We can use quantum field theory on this, as well as any other, manifold as long as the curvature scale is small compared to the Planck scale.

Much, though not all, of our flat space effective field theory intuition carries over into curved space.
Most importantly, the counter-term structure of the theory does not change since for momentum much
shorter than the curvature the space looks flat \footnote{Here I'm ignoring counter-terms involving powers of the curvature which will in general be generated.}
. So, as in flat space, we write down
all possible counter-terms which are consistent with the symmetries of the theory 
under consideration. However, since momentum scales are now  dependent upon the coordinate in the
fifth direction, the momentum at which the loop expansion breaks down will depend upon
the choice of observable. 

Let me specialize to a particular space which has received much attention in the past few years.
Namely, five dimensional anti-deSitter space compactified on $S^1/Z_2$.
anti-deSitter space is a maximally symmetric space with negative cosmological constant \cite{HE}, 
and its compactification on $S^1/Z_2$ was used in \cite{RSI} to solve the hierarchy problem.
The warped metric can be written in the conformally flat form
\begin{equation}
ds^2=\frac{1}{(kz)^2}(dz^2+\eta_{\mu \nu}dx^\mu dx^\nu)
\end{equation}
where $k$ is the inverse radius of curvature.
The two branes necessary to account for curvature singularities are positioned at
$z=1/T$ (the ``TeV brane") and $z=1/k$ (the ``Planck brane"). $T$ is taken to be near the TeV scale while the bulk cosmological
constant ($\Lambda$) is related to the curvature scale via
\begin{equation}
k^2=-\frac{\Lambda}{24M^3}
\end{equation}
where $M$ is the five dimensional fundamental scale, i.e. it is the scale in front of the
Einstein term in the action. Note, that $\Lambda$ will not be the four dimensional cosmological
constant, nor will $M$ be the effective Planck mass.
This theory (sometimes called RSI)  solves the hierarchy problem \footnote{Technically speaking the hierarchy problem is not
really solved until the brane separation is naturally stabilized as in \cite{GWII}.} in a very elegant way.
A simple way to see this is to note that if the standard model fields are localized to the
TeV brane, then the masses of the fields are naturally of order $M T/k$ (see \cite{Csaki}), while
gravity remains weak. Let us study the low energy effective field theory derived from this model.

Now we would like to ask, what observables are calculable in an effective field theory?
If we perform a KK decomposition, the action looks like
\begin{equation}
S=\int d^4x \sum_n L(\phi_T,\phi_P,\phi_n),
\end{equation}
where the fields $\phi_T$ ($\phi_P$) are fields localized to the TeV (Planck) brane, and $\phi_n$ are the KK modes of the bulk fields. The KK masses correspond to zeros of Bessel functions
(see \cite{Csaki}), and the lowest lying masses are near a TeV (note 
that there is a vanishing effective
cosmological constant. This only comes about after fine tuning a
linear combination of the brane tensions.).
From a low energy view point
this looks like a theory with an effective compactification scale of a TeV, since the first
KK mode has a mass of that order.
So if we dimensionally reduce to a four dimensional effective theory, we will have, in general
a set of zero modes and a tower of KK modes. Additionally we could have zero mass ($m\ll TeV$)
particles which are restricted to live on the brane.

Presumably, in any such model, we would be composed of the set of zero mass fluctuations.
To make things interesting we will assume that all the gauge fields are bulk fields.
So if we want to determine the potential between charge objects, we need to calculate  the
corrections to the zero mode propagator. Given that the effective compactification scale is
around a $TeV$ we would conclude that the potential is only calculable up to that scale, just as in the flat case.
Indeed, it is simple to see that this is true even without looking at quantum corrections.
All we need to do is to consider the effect of higher dimensional operators.
Suppose we  add a term to the action, such as
\begin{equation}
S=\frac{\lambda}{2}\int d^5x \sqrt{g}F_{MN}\partial^2 F^{MN}.
\end{equation}
where $\lambda$ is a dimensionful quantity of order $1/M$ in our
normalization where the gauge field kinetic term is accompanied by a
factor
of $1/g^2$.
The contribution of this term to the zero mode gauge field two point function will be of the
form
\begin{equation}
\Pi(q^2)=\lambda k \frac{q^2}{T^2},
\label{FFop}
\end{equation}
so we can see explicitly that we have no hope of calculating the gauge
field zero mode two point function for $q^2$much  larger than $T^2$.

\vskip.3in
{\bf Exercise 3.6} Prove equation (\ref{FFop}). Be sure to remember that the gauge potentials are the
components of a one-form, so its index is naturally down.
\vskip.3in

At first this result is rather disheartening, since it appears to mean that 
extra dimensions with size of order the $TeV$ scale have no hope of reproducing the
four dimensional prediction for the weak mixing angle. Is there no way to have higher dimensional
physics which is accessible at the LHC without the loss of this beautiful prediction \cite{pomarol}?
The answer is yes, and it is through the magic  of AdS.

\subsection{Unification  in $AdS_5$}

Suppose we calculate the low energy ($q<$TeV) potential for the gauge zero mode, ignoring for the
moment the issues of the breakdown of the effective theory. I will work in the usual toy scalar
electro-dynamics  to simplify matters, and I will take the massive bulk scalar to transform
evenly under the $Z_2$
\begin{equation}
\label{Q}
\pi(q^2)  =  -{1\over 48\pi^2}\left[{1\over{\bar\epsilon}}+{1\over 2}\ln\left({\mu^2\over 4kT}\right)\right]+{1\over 16\pi^2} \int_0^1 dx  x \sqrt{1-x^2} \ln N\left({x\sqrt{q^2}\over 2}\right),
\end{equation}
where 
\begin{equation}
\label{eq:ppsum}
\ln N(p)  = -\ln\left|i_\nu(p/T) k_\nu(p/k)-i_\nu(p/k) k_\nu(p/T)\right|,
\end{equation}
\begin{eqnarray}
i_\nu(z)=(2-\nu) I_\nu(z) + z I_{\nu-1}(z),\nonumber \\ 
k_\nu(z)=(2-\nu) K_\nu(z) - z K_{\nu-1}(z),
\end{eqnarray}
and $\nu=\sqrt{4+m^2/k^2}$. Let's study this result in some detail. We first notice the UV pole ${1\over{\bar\epsilon}}$. This pole is identical
to the pole we found in flat space orbifold, and can be subtracted away by a boundary counter-term. This fact also tells us that the $\mu$ dependence in (\ref{eq:ppsum}) determines the running of the
boundary couplings, one of which lives at $z=1/T$ ($\lambda_T$) the other at $z=1/k$ ($\lambda_k$).  
In the $\overline{MS}$ scheme, the effective dimensionally reduced coupling looks like
\begin{eqnarray}
\label{eq:coup}
{1\over g^2(q^2)} = {R\over g_5^2} +\lambda_k(\mu) + \lambda_T(\mu) -{1\over 96\pi^2}\ln\left({\mu^2\over 4kT}\right) \nonumber \\
+{1\over 16\pi^2} \int_0^1 dx  x \sqrt{1-x^2} \ln N\left({x\sqrt{q^2}\over 2}\right),
\end{eqnarray}
where $R=\ln(k/T)/k$, is the effective compactification radius.
The first three terms result from the tree level bulk gauge coupling and boundary couplings, respectively.  The remaining terms  encode all the information about the KK modes. In flat space this term
would be linearly sensitive to the bulk mass so in AdS we would expect that for $m$ larger than
the curvature scale $k$ we would again have linearly sensitivity. The interesting case is $m<k$.
Using the expansions
\begin{eqnarray}
i_\nu(z) &\simeq& {2+\nu\over \Gamma(\nu+1)} \left({z\over 2}\right)^\nu\left[1+{ O}(z^2)+\cdots\right],\\
k_\nu(z) &\simeq& {\Gamma(\nu)\over 2}(2-\nu)\left({z\over 2}\right)^{-\nu}\left[1+{ O}(z^2)+\cdots\right],
\end{eqnarray}
 for $p\ll T$, we have
\begin{equation}
\ln N(p)\simeq -\ln\left[{\nu^2-4\over\nu}\right] - \nu\ln{k\over T} + \cdots,
\end{equation}
and thus, for $m\neq 0$
\begin{equation}
\label{eq:adsmpi}
{1\over g^2(q^2)} \simeq {R\over g_5^2}+ \lambda_k(\mu)+\lambda_T(\mu) -{1\over 96\pi^2}\ln\left({\mu^2\over 4kT}\right) - {1\over 48\pi^2}\left[\ln\left({m^2\over T^2}\right) \right] + \mbox{analytic in $q^2$} .
\end{equation}
This is a remarkable result, in that the five dimensional vacuum polarization looks just like
what we would expect from a four dimensional calculation. The result (\ref{eq:adsmpi}) is only
logarithmically sensitive to the masses, so there is hope that a GUT theory with a large mass scale (by large I mean large compared to the TeV scale, but still smaller than $k$) will still lead to the correct
value of the weak mixing angle. 

\vskip.3in
{\bf Exercise 3.7} Show that in the limit where $m\gg k$ the inverse coupling is linearly
sensitive to the mass.
\vskip.3in

Let us try to get a better understanding of how the logarithmic sensitivity to
the symmetry breaking scale arises. It seems rather strange that we can
make a prediction for a lower energy observable in terms of high energy parameters (the symmetry breaking scale $m$)
when the observable under consideration is not well defined at scales above $T$.
Clearly we can't get the result (\ref{eq:adsmpi}) from an
effective field theory calculation where we run the zero mode correlator down from
the scale $m$. Instead of tracking the zero mode correlator
we will calculate another observable, calculable at higher scales,
which  
matches onto the zero mode correlator at the scale $T$ \cite{GnRI}. 
This other observable is the Planck brane
gauge field correlator, which is defined as the gauge field two point
function with its end point on the Planck brane. This correlator is 
calculable up to the scale $k$. 
In fact, the cut-off scale
for any correlator restricted to a hyper-surface of fixed $z$ is $1/z$.
The tree level Planck brane correlator, in mixed momentum coordinate space representation, can be written in the   $A_z=0$ gauge, as (here $\eta_{\mu\nu}$ is the flat Euclidean metric)
\begin{equation}
D^{\mu\nu}_q(z,z') = \eta^{\mu\nu} D_q(z,z') + {q^\mu q^\nu\over q^2} H_q(z,z'),
\end{equation}
where we won't need the form of $H_q$. 
If we take both end points on the Planck brane ($z=1/k$) one finds
(for $q> T$)
\begin{equation}
\label{eq:planck} 
\int d^4 x e^{iq\cdot x}\langle A_\mu(x,1/k) A_\nu(0,1/k)\rangle \propto 
{g_5^2 \over q^2\ln(q)} \eta_{\mu\nu}+\cdots,
\end{equation}
whereas in the case where only one end-point is on the Planck brane (this
is needed for the one loop calculation) we find \cite{GnRII}
\begin{equation}
D_q(z,1/k)=\frac{kz}{q} \frac{K_1(qz)I_0(q/T)+K_0(q/T)I_0(qz)}{I_0(q/T)K_0(q/k)-I_0(q/k)K_0(q/T)}.
\end{equation}
At $q\ll T$ this expression approaches the zero mode propagator so
that for small momenta the one loop Planck brane correlator will match
onto the zero mode (external zero modes, but with internal KK modes)  result.

Note that at the scale $q=1/T$ (\ref{eq:planck}) matches onto the first term in
(\ref{eq:adsmpi}).  This is an interesting result, because 
 a Planck brane observer would see a
logarithmic running of the coupling at tree level due just to the
gravitational effects arising from the bulk. 
(\ref{eq:planck}) only freezes to $g_5^2/R$ at 
scales below $T$.
This log has a simple interpretation in terms of the 
AdS/CFT correspondence \cite{Mald} as understood within a cut-off AdS
space \cite{ADSk}.  As far as the weak mixing angle is
concerned, this log  is irrelevant, as it is universal (meaning
it contributes to all the couplings identically) since it depends only
upon the geometry. So that is a first step in understanding the result
(\ref{eq:adsmpi}), but the real challenge in understanding the
logarithmic mass dependence. From a four dimensional perspective this term looks like
it's coming from a heavy scalar which decouples at its mass scale (compare to the result in section 3.4). In
terms of the Planck brane correlator we can understand this from the
following remarkable fact. All of the KK modes with masses much less than $k$, save for one, are 
localized to the TeV brane while the external gauge boson lines in the
self energy die off exponentially fast as one moves away from the
Planck brane. So the net contribution of the KK modes to the symmetry breaking effects is suppressed
relative to contribution of a single mode. The modes with masses larger than $k$ contribute in a symmetric way and thus only affect the unmeasurable counter-term. The one special mode has a mass near the value of the bulk mass.
 In the massless case this
mode is the zero mode, which is flat, while in the massive case there
is a ``magic mode'' which has support near the Planck brane \cite{Agashe}. Let us
see how this happens quantitatively. 
 In the limit  $T\ll\sqrt{q^2}\ll k$
\begin{equation}
D_q(z,1/k)\sim {k\over q}\sqrt{\pi z\over 2 q} {1\over K_0(q/k)}e^{-qz}.
\end{equation}
we see the exponential suppression  of the gauge field propagator. 
This suppression will kill the contribution of the TeV brane localized
KK modes.
However, one must be careful to note that there is regulator
dependence in this statement. A more precise statement would be that the
non-analytic momentum dependence generated by the KK modes is exponentially small.
I will refer the reader to reference \cite{GnRI,GnRII,GnRIII} for more
details.

For the case of gauge fields and fermions, the suppression of the KK modes
does not follow this argument, as they are not in general TeV brane localized. 
In fact, the contributions from these 
modes are only logarithmically suppressed. That is, they look like
they contribute to the two loop beta function. This can be seen from studying
the non-decoupling calculation  for gauge field  \cite{choi}.
But given that these effects are of the same order as the usual unknown threshold effects, they won't change the
prediction for the weak mixing angle \cite{GnRIII}. 
To understand this behavior from a non-decoupling 
argument one can use arguments based on AdS/CFT, but these are beyond the scope of these lectures.
At this point I should also mention that the gauge fields differ from scalar fields in
another important way. The one loop calculation for a gauge
field with mass $m_X$
gives
a contribution of the form
\begin{equation}
\pi(q^2) \propto \ln(\frac{m_X^2}{\mu k})+ . ~ .~.
\end{equation}
so that the effective mass is {\it parametrically} enhanced \cite{GnRIII}, though
numerically, this effect is small unless we are 
willing to tolerate a large hierarchy between
$k$ and $m_X$. 

Let's now  consider the effects
of the higher dimensional operators 
on our prediction for the weak mixing angle, 
as they could perhaps ruin our predictive power. 
Any higher dimensional operator which is insensitive 
to the symmetry breaking masses will obviously not
affect the prediction. These higher dimensional 
operators would all correspond to some universal contribution to the
couplings and thus be irrelevant for the difference 
which determines the mixing angle. 
It is also possible that there are higher dimensional operators such as $\phi F^2$ which
would be sensitive to the symmetry breaking if $\phi$ got an
expectation value.
However, since $m\ll k \ll M$ these terms will be suppressed by powers of $m/M$ and thus
negligible. So higher dimensional operators are not a problem, and it seems that we will
be able to predict the weak mixing angle.

The net result of this discussion is that one can understand how AdS
can lead to a successful prediction of the weak mixing angle. The high
energy behavior of the Planck brane correlator looks four
dimensional. In addition to the logs which arise from zero modes of
bulk fields there is a ``tree-level'' log which is purely geometric
and does not contribute to the mixing angle prediction. Scaling down
the Planck brane correlators, one then matches onto the zero mode
correlators which are related to the physical measurement.

It is quite exciting to think about unification in $AdS$ 
because one would expect to
see $GUT$ symmetry violating processes at the TeV scale.
Recent progress in model building in this 
direction can be found \cite{models}. 

\section*{Acknowledgments}
I would like to thank the TASI organizing committee, Howie Haber, Ann Nelson and especially  K.T. Manhanthappa for organizing yet another
fruitful TASI meeting. I would also like to thank the UCSD theory group for its hospitality, as well as Iain Stewart and especially Walter Goldberger for comments on the manuscript.


\begin{thebibliography}{0}
\bibitem{Wilson} The basic ideas behind effective field theory go back
to Wilson's original work.  The intuition gained by understanding the
renormalization group in position space in terms of block spins makes
it well worth the time to study this topic. I suggest the following
review to the interested reader.  K.~G.~Wilson and J.~B.~Kogut,
Phys.\ Rept.\  {\bf 12}, 75 (1974).
\bibitem{Peskin}
M.~E.~Peskin and D.~V.~Schroeder,
``An Introduction To Quantum Field Theory,'',  Reading, USA: Addison-Wesley (1995)
\bibitem{Weinberg}
S.~Weinberg,
``The Quantum Theory Of Fields. Vol. 2: Modern Applications",
Cambridge, UK: Univ. Pr. (1996)
\bibitem{Georgi}
H.~Georgi,
``Weak Interactions And Modern Particle Theory,''
Menlo Park, USA: Benjamin/Cummings (1984)
\bibitem{decoup}
T.~Appelquist and J.~Carazzone,
Phys.\ Rev.\ D {\bf 11}, 2856 (1975).



\bibitem{Sterman} For a discussion see
G.~Sterman,
``An Introduction To Quantum Field Theory,'', Cambridge, UK: Univ. Pr. (1993) 

\bibitem{equiv}
S.~Kamefuchi, L.~O'Raifeartaigh and A.~Salam,
Nucl.\ Phys.\  {\bf 28}, 529 (1961).


R.~E.~Kallosh and I.~V.~Tyutin,
Yad.\ Fiz.\  {\bf 17}, 190 (1973)
[Sov.\ J.\ Nucl.\ Phys.\  {\bf 17}, 98 (1973)].


\bibitem{DIS} The use of the OPE in DIS is discussed in \cite{Peskin}.
The B decay calculation is discussed in \cite{MW}.


\bibitem{GeorgiI}
H.~Georgi,
Nucl.\ Phys.\ B {\bf 361} (1991) 339.


\bibitem{WZ}
J.~Wess and B.~Zumino,
Phys.\ Lett.\ B {\bf 37}, 95 (1971).

\bibitem{Witten2}
E.~Witten,
Nucl.\ Phys.\ B {\bf 223}, 422 (1983).



\bibitem{Witten}
E.~Witten,
Phys.\ Lett.\ B {\bf 117}, 324 (1982).
\bibitem{DF}
E.~D'Hoker and E.~Farhi,
Nucl.\ Phys.\ B {\bf 248}, 59 (1984),
Nucl.\ Phys.\ B {\bf 248}, 77 (1984).

\bibitem{GW}
J.~Goldstone and F.~Wilczek,
Phys.\ Rev.\ Lett.\  {\bf 47}, 986 (1981).



\bibitem{nuc} For a review and references, see 
P.~F.~Bedaque and U.~van Kolck,
arXiv:nucl-th/0203055.


\bibitem{Seiberg}
N.~Seiberg,
Phys.\ Lett.\ B {\bf 390}, 169 (1997)
[arXiv:hep-th/9609161].
N.~Seiberg,
Phys.\ Lett.\ B {\bf 388}, 753 (1996)
[arXiv:hep-th/9608111].



\bibitem{buras}
A.~J.~Buras,
arXiv:hep-ph/9806471.




\bibitem{MW}
A.~V.~Manohar and M.~B.~Wise,
Cambridge Monogr.\ Part.\ Phys.\ Nucl.\ Phys.\ Cosmol.\  {\bf 10}, 1 (2000).
Also see M. Luke, these proceedings, as well as
M.~Neubert,
Phys.\ Rept.\  {\bf 245}, 259 (1994)
[arXiv:hep-ph/9306320].




\bibitem{GeorgiII}
H.~Georgi,
Phys.\ Lett.\ B {\bf 240}, 447 (1990).



\bibitem{BL}
The theory was originally formulated in,
W.~E.~Caswell and G.~P.~Lepage,
Phys.\ Lett.\ B {\bf 167} (1986) 437. The formulation discussed in these
lectures was given in \cite{lmr}. For a review see \cite{hoang}.
NRQCD in the context of production and decay was discussed in \cite{BBL}.
\bibitem{lmr}
M.~E.~Luke, A.~V.~Manohar and I.~Z.~Rothstein,
Phys.\ Rev.\ D {\bf 61}, 074025 (2000)
[arXiv:hep-ph/9910209].
\bibitem{hoang}
A.~H.~Hoang,
arXiv:hep-ph/0204299. I.~Z.~Rothstein,
arXiv:hep-ph/9911276.


\bibitem{BBL}
G.~T.~Bodwin, E.~Braaten and G.~P.~Lepage,
Phys.\ Rev.\ D {\bf 51}, 1125 (1995)
[Erratum-ibid.\ D {\bf 55}, 5853 (1997)]
[arXiv:hep-ph/9407339].
\bibitem{SCETI}
The first attempt at such a theory was written down in: 
M.~J.~Dugan and B.~Grinstein,
Phys.\ Lett.\ B {\bf 255}, 583 (1991).
This theory turned out to be incomplete as clarified in :
C.~W.~Bauer, S.~Fleming and M.~E.~Luke,
[arXiv:hep-ph/0005275].
The full effective theory was formalized in: C.~W.~Bauer, S.~Fleming,
D.~Pirjol and I.~W.~Stewart,
Phys.\ Rev.\ D {\bf 63}, 114020 (2001)
[arXiv:hep-ph/0011336],
C.~W.~Bauer and I.~W.~Stewart,
Phys.\ Lett.\ B {\bf 516}, 134 (2001)
[arXiv:hep-ph/0107001],
C.~W.~Bauer, D.~Pirjol and I.~W.~Stewart,
Phys.\ Rev.\ D {\bf 65}, 054022 (2002)
[arXiv:hep-ph/0109045].
C.~W.~Bauer, D.~Pirjol and I.~W.~Stewart,
Phys.\ Rev.\ D {\bf 67}, 071502 (2003)
[arXiv:hep-ph/0211069].

\bibitem{SCETII}
C.~W.~Bauer, S.~Fleming, D.~Pirjol, I.~Z.~Rothstein and I.~W.~Stewart,
Phys.\ Rev.\ D {\bf 66}, 014017 (2002)
[arXiv:hep-ph/0202088].
I.~Z.~Rothstein,
arXiv:hep-ph/0301240.

\bibitem{SCETNRQCD}

S.~Fleming and A.~K.~Leibovich,
Phys.\ Rev.\ Lett.\  {\bf 90}, 032001 (2003)
[arXiv:hep-ph/0211303].

S.~Fleming, A.~K.~Leibovich and T.~Mehen,
arXiv:hep-ph/0306139.
S.~Fleming and A.~K.~Leibovich,
Phys.\ Rev.\ D {\bf 67}, 074035 (2003)
[arXiv:hep-ph/0212094].




\bibitem{WittenEFT}
E.~Witten,
Nucl.\ Phys.\ B {\bf 122}, 109 (1977).


\bibitem{Grinstein}
B.~Grinstein,
Nucl.\ Phys.\ B {\bf 339}, 253 (1990).

\bibitem{AE}
S.~G.~Gorishnii,
Nucl.\ Phys.\ B {\bf 319}, 633 (1989),
K.~G.~Chetyrkin,
Theor.\ Math.\ Phys.\  {\bf 75}, 346 (1988)
[Teor.\ Mat.\ Fiz.\  {\bf 75}, 26 (1988)],
K.~G.~Chetyrkin,
Theor.\ Math.\ Phys.\  {\bf 76}, 809 (1988)
[Teor.\ Mat.\ Fiz.\  {\bf 76}, 207 (1988)],
F.~V.~Tkachov,
Int.\ J.\ Mod.\ Phys.\ A {\bf 8}, 2047 (1993)
[arXiv:hep-ph/9612284],
V.~A.~Smirnov,
Commun.\ Math.\ Phys.\  {\bf 134}, 109 (1990).
\bibitem{MoR}
M.~Beneke and V.~A.~Smirnov,
Nucl.\ Phys.\ B {\bf 522}, 321 (1998)
[arXiv:hep-ph/9711391].

\bibitem{BSS}
M.~Beneke, A.~Signer and V.~A.~Smirnov,
Phys.\ Lett.\ B {\bf 454}, 137 (1999)
[arXiv:hep-ph/9903260].

\bibitem{Fischler}
W.~Fischler,
Nucl.\ Phys.\ B {\bf 129}, 157 (1977).
\bibitem{Kogut}
J.~B.~Kogut,
Rev.\ Mod.\ Phys.\  {\bf 55}, 775 (1983).
\bibitem{ADM}
T.~Appelquist, M.~Dine and I.~J.~Muzinich,
Phys.\ Rev.\ D {\bf 17}, 2074 (1978).


\bibitem{NRQCDc}
M.~Beneke,
arXiv:hep-ph/9703429,
N.~Brambilla, A.~Pineda, J.~Soto and A.~Vairo,
Phys.\ Rev.\ D {\bf 63}, 014023 (2001)
[arXiv:hep-ph/0002250],
S.~Fleming, I.~Z.~Rothstein and A.~K.~Leibovich,
Phys.\ Rev.\ D {\bf 64}, 036002 (2001)
[arXiv:hep-ph/0012062].
\bibitem{HMST}
A.~H.~Hoang, A.~V.~Manohar, I.~W.~Stewart and T.~Teubner,
Phys.\ Rev.\ D {\bf 65}, 014014 (2002)
[arXiv:hep-ph/0107144].
\bibitem{LS}
M.~E.~Luke and M.~J.~Savage,
Phys.\ Rev.\ D {\bf 57}, 413 (1998)
[arXiv:hep-ph/9707313].
\bibitem{Labelle}
P.~Labelle,
Phys.\ Rev.\ D {\bf 58}, 093013 (1998)
[arXiv:hep-ph/9608491].
\bibitem{GR}
B.~Grinstein and I.~Z.~Rothstein,
Phys.\ Rev.\ D {\bf 57}, 78 (1998)
[arXiv:hep-ph/9703298].
\bibitem{Griesshammer}
H.~W.~Griesshammer,
Phys.\ Rev.\ D {\bf 58}, 094027 (1998)
[arXiv:hep-ph/9712467].
\bibitem{MSI}
A.~V.~Manohar and I.~W.~Stewart,
Phys.\ Rev.\ D {\bf 63}, 054004 (2001)
[arXiv:hep-ph/0003107].
\bibitem{ML}
M.~E.~Luke and A.~V.~Manohar,
Phys.\ Lett.\ B {\bf 286}, 348 (1992)
[arXiv:hep-ph/9205228].
\bibitem{CW}
W.~E.~Caswell and F.~Wilczek,
Phys.\ Lett.\ B {\bf 49}, 291 (1974).
\bibitem{MSS}
A.~V.~Manohar, J.~Soto and I.~W.~Stewart,
Phys.\ Lett.\ B {\bf 486}, 400 (2000)
[arXiv:hep-ph/0006096].
\bibitem{MSII}
A.~V.~Manohar and I.~W.~Stewart,
Phys.\ Rev.\ D {\bf 62}, 014033 (2000)
[arXiv:hep-ph/9912226].
\bibitem{TY} For a discussion of the calculation of these, as well as other
matrix elements see, 
S.~Titard and F.~J.~Yndurain,
Phys.\ Rev.\ D {\bf 49}, 6007 (1994)
[arXiv:hep-ph/9310236].
\bibitem{HMS}
A.~H.~Hoang, A.~V.~Manohar and I.~W.~Stewart,
Phys.\ Rev.\ D {\bf 64}, 014033 (2001)
[arXiv:hep-ph/0102257].
\bibitem{BPSV}
N.~Brambilla, A.~Pineda, J.~Soto and A.~Vairo,
Phys.\ Rev.\ D {\bf 60}, 091502 (1999)
[arXiv:hep-ph/9903355].
\bibitem{MSQED}
A.~V.~Manohar and I.~W.~Stewart,
Phys.\ Rev.\ Lett.\  {\bf 85}, 2248 (2000)
[arXiv:hep-ph/0004018].
\bibitem{Dienes} K. Dienes, these proceedings.
\bibitem{Csaki} C. Csaki, these proceedings.
\bibitem{will}
C.~M.~Will,
Living Rev.\ Rel.\  {\bf 4}, 4 (2001)
[arXiv:gr-qc/0103036].

\bibitem{BWP}
For a review of this theory see, 
B.~Rosenstein, B.~Warr and S.~H.~Park,
Phys.\ Rept.\  {\bf 205}, 59 (1991).
\bibitem{Cohen} See A. Cohen's lectures, in these proceedings.
\bibitem{MG}
A.~Manohar and H.~Georgi,
Nucl.\ Phys.\ B {\bf 234}, 189 (1984).
\bibitem{NDASUSY}
M.~A.~Luty,
Phys.\ Rev.\ D {\bf 57}, 1531 (1998)
[arXiv:hep-ph/9706235].
A.~G.~Cohen, D.~B.~Kaplan and A.~E.~Nelson,
Phys.\ Lett.\ B {\bf 412}, 301 (1997)
[arXiv:hep-ph/9706275].
Z.~Chacko, M.~A.~Luty, A.~E.~Nelson and E.~Ponton,
JHEP {\bf 0001}, 003 (2000)
[arXiv:hep-ph/9911323].
\bibitem{WeinbergII}
S.~Weinberg,
Phys.\ Lett.\ B {\bf 91}, 51 (1980).
\bibitem{Hall}
L.~J.~Hall,
Nucl.\ Phys.\ B {\bf 178}, 75 (1981).

\bibitem{DDG}
K.~R.~Dienes, E.~Dudas and T.~Gherghetta,
Phys.\ Lett.\ B {\bf 436}, 55 (1998)
[arXiv:hep-ph/9803466].
K.~R.~Dienes, E.~Dudas and T.~Gherghetta,
Nucl.\ Phys.\ B {\bf 537}, 47 (1999)
[arXiv:hep-ph/9806292].
\bibitem{Pol}
J.~Polchinski,
``String Theory. Vol. 2: Superstring Theory And Beyond,''
 Cambridge, UK: Univ. Pr. (1998) 531 p. 
\bibitem{Sundrum}
R.~Sundrum,
Phys.\ Rev.\ D {\bf 59}, 085009 (1999)
[arXiv:hep-ph/9805471].
\bibitem{GG}
H.~Georgi, A.~K.~Grant and G.~Hailu,
Phys.\ Lett.\ B {\bf 506}, 207 (2001)
[arXiv:hep-ph/0012379].
\bibitem{GoW}
W.~D.~Goldberger and M.~B.~Wise,
Phys.\ Rev.\ D {\bf 65}, 025011 (2002)
[arXiv:hep-th/0104170].
\bibitem{DtHJ}
S.~Deser, R.~Jackiw and G.~'t Hooft,
Annals Phys.\  {\bf 152}, 220 (1984).
\bibitem{BD} For an introduction to quantum fields in curved spaces see, 
N.~D.~Birrell and P.~C.~Davies,
``Quantum Fields In Curved Space,''
 Cambridge, Uk: Univ. Pr. ( 1982) 340p. 
\bibitem{HE} S.W. Hawking and G.F.R. Ellis,
``The Large Scale Structure of Space-Time'',
Cambridge, Uk: Univ. pr. (1975). 
\bibitem{RSI}
L.~Randall and R.~Sundrum,
Phys.\ Rev.\ Lett.\  {\bf 83}, 3370 (1999)
[arXiv:hep-ph/9905221].


\bibitem{GnRI}
W.~D.~Goldberger and I.~Z.~Rothstein,
Phys.\ Rev.\ Lett.\  {\bf 89}, 131601 (2002)
[arXiv:hep-th/0204160].
\bibitem{GnRII} A technique for calculating loop diagrams in higher dimensional spaces, including warped spaces, can be found in the appendix of, 
W.~D.~Goldberger and I.~Z.~Rothstein,
arXiv:hep-th/0208060. For results including higher spin fields see \cite{choi}.
A method of calculating vacuum bubbles is given in \cite{GnR0}.
\bibitem{GnR0}
W.~D.~Goldberger and I.~Z.~Rothstein,
Phys.\ Lett.\ B {\bf 491}, 339 (2000)
[arXiv:hep-th/0007065].
\bibitem{Mald}
J.~M.~Maldacena,
Adv.\ Theor.\ Math.\ Phys.\  {\bf 2}, 231 (1998)
[Int.\ J.\ Theor.\ Phys.\  {\bf 38}, 1113 (1999)]
[arXiv:hep-th/9711200].
S.~S.~Gubser, I.~R.~Klebanov and A.~M.~Polyakov,
Phys.\ Lett.\ B {\bf 428}, 105 (1998)
[arXiv:hep-th/9802109].
E.~Witten,
Adv.\ Theor.\ Math.\ Phys.\  {\bf 2}, 253 (1998)
[arXiv:hep-th/9802150].
\bibitem{ADSk}
E.~Witten,~remarks~at~ITP~Santa~Barbra~conference,
~``New~Dimensions~in~Field~Theory~and~String~Theory''.
S.~S.~Gubser,
Phys.\ Rev.\ D {\bf 63}, 084017 (2001)
[arXiv:hep-th/9912001].
N.~Arkani-Hamed, M.~Porrati and L.~Randall,
JHEP {\bf 0108}, 017 (2001)
[arXiv:hep-th/0012148].
R.~Rattazzi and A.~Zaffaroni,
JHEP {\bf 0104}, 021 (2001)
[arXiv:hep-th/0012248].

\bibitem{choi}
K.~w.~Choi and I.~W.~Kim,
Phys.\ Rev.\ D {\bf 67}, 045005 (2003)
[arXiv:hep-th/0208071].
\bibitem{GnRIII}
W.~D.~Goldberger and I.~Z.~Rothstein,
arXiv:hep-ph/0303158.

\bibitem{GWII}
W.~D.~Goldberger and M.~B.~Wise,
Phys.\ Rev.\ Lett.\  {\bf 83}, 4922 (1999)
[arXiv:hep-ph/9907447].
\bibitem{pomarol}
A.~Pomarol,
Phys.\ Rev.\ Lett.\  {\bf 85}, 4004 (2000)
[arXiv:hep-ph/0005293].
L.~Randall and M.~D.~Schwartz,
Phys.\ Rev.\ Lett.\  {\bf 88}, 081801 (2002)
[arXiv:hep-th/0108115].

\bibitem{Agashe}
K.~Agashe, A.~Delgado and R.~Sundrum,
Nucl.\ Phys.\ B {\bf 643}, 172 (2002)
[arXiv:hep-ph/0206099].
\bibitem{models}
K.~Agashe, A.~Delgado and R.~Sundrum,
Annals Phys.\  {\bf 304}, 145 (2003)
[arXiv:hep-ph/0212028].
W.~D.~Goldberger, Y.~Nomura and D.~R.~Smith,
Phys.\ Rev.\ D {\bf 67}, 075021 (2003)
[arXiv:hep-ph/0209158].
M.~Carena, A.~Delgado, E.~Ponton, T.~M.~Tait and C.~E.~Wagner,
arXiv:hep-ph/0305188.
L.~J.~Hall, Y.~Nomura, T.~Okui and S.~J.~Oliver,
arXiv:hep-th/0302192.
Z.~Chacko and E.~Ponton,
arXiv:hep-ph/0301171.
\end{thebibliography}
\end{document}